\documentclass[a4paper]{article}
\usepackage[perpage]{footmisc}
\usepackage{chapterbib}
\usepackage{titling}
\usepackage{blindtext}
\usepackage{graphicx}
\usepackage{authblk}
\usepackage{sectsty}
\usepackage{lipsum}
\usepackage{titlesec}
\usepackage{ifthen}
\usepackage{subfigure}
\usepackage{amsmath,amssymb,amsfonts}                                                 
\usepackage{feynmp-auto}%%%{feynmp}
\usepackage[utf8]{inputenc}
\usepackage[T1]{fontenc}
\usepackage[margin=1.0in]{geometry}
\usepackage{tikz}
\usepackage{slashed}
\usepackage{hyperref}

\newboolean{articletitles}
\setboolean{articletitles}{false} % False removes titles in references
\newboolean{uprightparticles}
\setboolean{uprightparticles}{false} %True for upright particle symbols
\newboolean{inbibliography}
\setboolean{inbibliography}{false} %True once you enter the bibliography
\newcommand{\ptvecmiss}{\ensuremath\vec{p}_{\scriptstyle\mathrm{T}}^{\scriptstyle\mathrm{miss}}}
\newcommand{\ETm}{\ensuremath{{E}_{\textrm{T}}^{\textrm{miss}}} }
\newcommand{\sessionday}[1]{\iffalse#1\fi} %% please leave that line

\newcommand{\pt}{\ensuremath{p_\mathrm{T}}}
\newcommand{\mt}{\ensuremath{m_{\mathrm{T}}} }
\newcommand{\gsim}{\!\mathrel{\hbox{\rlap{\lower.55ex \hbox{$\sim$}} \kern-.34em \raise.4ex \hbox{$>$}}}}                                     

\newcommand{\MET}{${\mathrm{E_{T}^{miss}}}$}
\newcommand{\welkept}{${\mathrm{p_{T}}}$}
\newcommand{\ttbar}{${\mathrm{t\bar{t}}}$}
\newcommand{\HT}{$\mathrm{H_T}$}
\newcommand{\MJ}{$\mathrm{M_J}$}
\newcommand{\MT}{$\mathrm{M_T}$}
\newcommand{\MTtwo}{$\mathrm{M_{T2}}$}
\newcommand{\LT}{$\mathrm{L_T}$}
\newcommand{\dphiwl}{$\mathrm{\Delta\Phi(W, \ell)}$}
\newcommand{\njets}{$\mathrm{N_{jets}}$}
\newcommand{\nbtags}{$\mathrm{N_{b-tags}}$}
\newcommand{\mll}{$\mathrm{M_{\ell\ell}}$}

\newcommand{\Bto}[1]{
	\ifnum #1 = 521343 \ensuremath{B^+ \to K^+ e^+ e^- \xspace}\fi
	\ifnum #1 = 521347 \ensuremath{B^+ \to K^+ \mu^+ \mu^- \xspace}\fi
	\ifnum #1 = 521345 \ensuremath{B^+ \to K^\ast(892)^+ e^+ e^- \xspace}\fi
	\ifnum #1 = 521349 \ensuremath{B^+ \to K^\ast(892)^+ \mu^+ \mu^- \xspace}\fi
	\ifnum #1 = 511332 \ensuremath{B^0 \to K^0 e^+ e^- \xspace}\fi
	\ifnum #1 = 511336 \ensuremath{B^0 \to K^0 \mu^+ \mu^- \xspace}\fi
	\ifnum #1 = 511335 \ensuremath{B^0 \to K^\ast(892)^0 e^+ e^- \xspace}\fi
	\ifnum #1 = 511339 \ensuremath{B^0 \to K^\ast(892)^0  \mu^+ \mu^- \xspace}\fi
}
\newcommand{\Kto}[1]{
	\ifnum #1 = 313421 \ensuremath{K^{\ast 0} \to K_S \pi^0}\fi
	\ifnum #1 = 313532 \ensuremath{K^{\ast 0} \to K^+ \pi^-}\fi
	\ifnum #1 = 323432 \ensuremath{K^{\ast +} \to K^+ \pi^0}\fi
	\ifnum #1 = 323521 \ensuremath{K^{\ast +} \to K_S \pi^+}\fi
}

\newcommand*{\factorh}{0.48}

\newcommand{\yfs}{$\Upsilon(4S)$\xspace}

\newcommand{\bkllzero}{\ensuremath{B^0 \to K^\ast(892)^0  \ell^+ \ell^-\xspace}}

\newcommand{\kast}{$K^\ast$\xspace}

\titleformat*{\section}{\LARGE\bfseries}
\titleformat*{\subsection}{\Large\bfseries}

\author[1,2]{\bf Editors\thanks{lhcski2016@oeaw.ac.at}: Jochen Schieck}
\author[1]{\bf Wolfgang Adam}
\author[1]{\bf Josef Pradler}
\author[1]{\bf Christoph Schwanda}
\author[1]{\bf Wolfgang Waltenberger}
\affil[1]{Institute for High Energy Physics, Austrian Academy of Sciences, Nikolsdorfergasse 18, 1050 Wien, Austria}
\affil[2]{Atominstitut, Vienna University of Technology, Stadionallee 2, 1020 Wien, Austria}

\title{\vspace*{-2cm} \Huge \bf LHCSki 2016 -- \\ A First Discussion of 13 TeV Results \\ \vspace*{2cm} }
\date{April 10-15, 2016, Obergurgl Universit\"atszentrum, Tirol, Austria}

\begin{document}

%% \pagenumbering{arabic}

\begin{titlingpage}
\maketitle
\begin{figure}[h!t]
\begin{center}
\includegraphics[width=\textwidth]{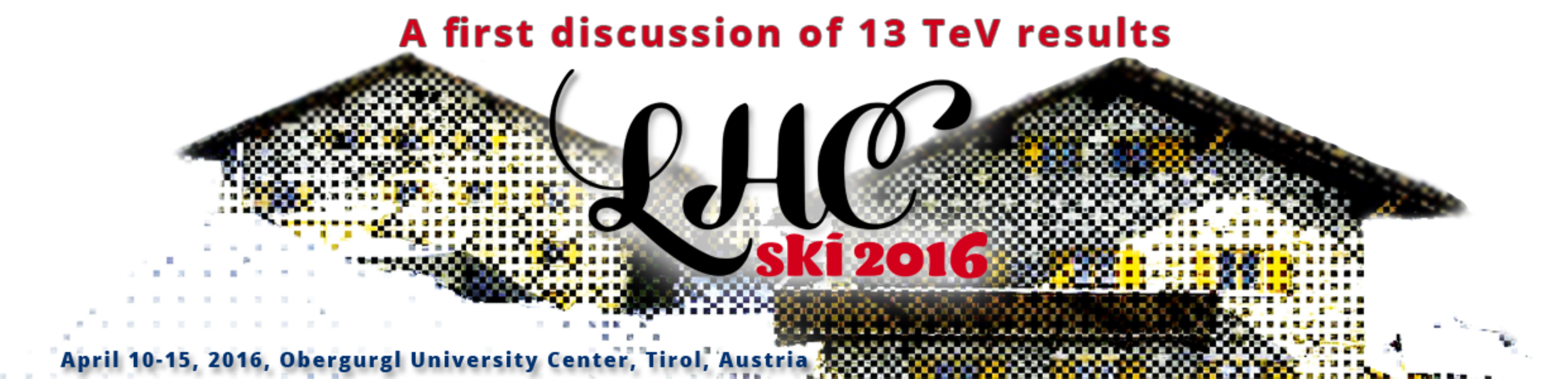}
%\caption{The working principle of SModelS}
%\label{fig:workingscheme}
\end{center}
\end{figure}
\begin{abstract}
These are the proceedings of the LHCSki 2016 workshop --
``A First Discussion of 13 TeV Results'' -- that has been held
at the Obergurgl Universit\"atszentrum, Tirol, Austria, April 10 -- 15, 2016.
In this workshop the consequences of the most recent results from the LHC 
have been discussed, with a focus also on the interplay with dark matter physics,
flavor physics, and precision measurements.
Contributions from the workshop speakers have been compiled into this document.

\end{abstract}
\end{titlingpage}

\newpage

\section{Introduction}
\label{sec:introduction}
\begin{figure}[h!]
\begin{center}
\includegraphics[width=\textwidth]{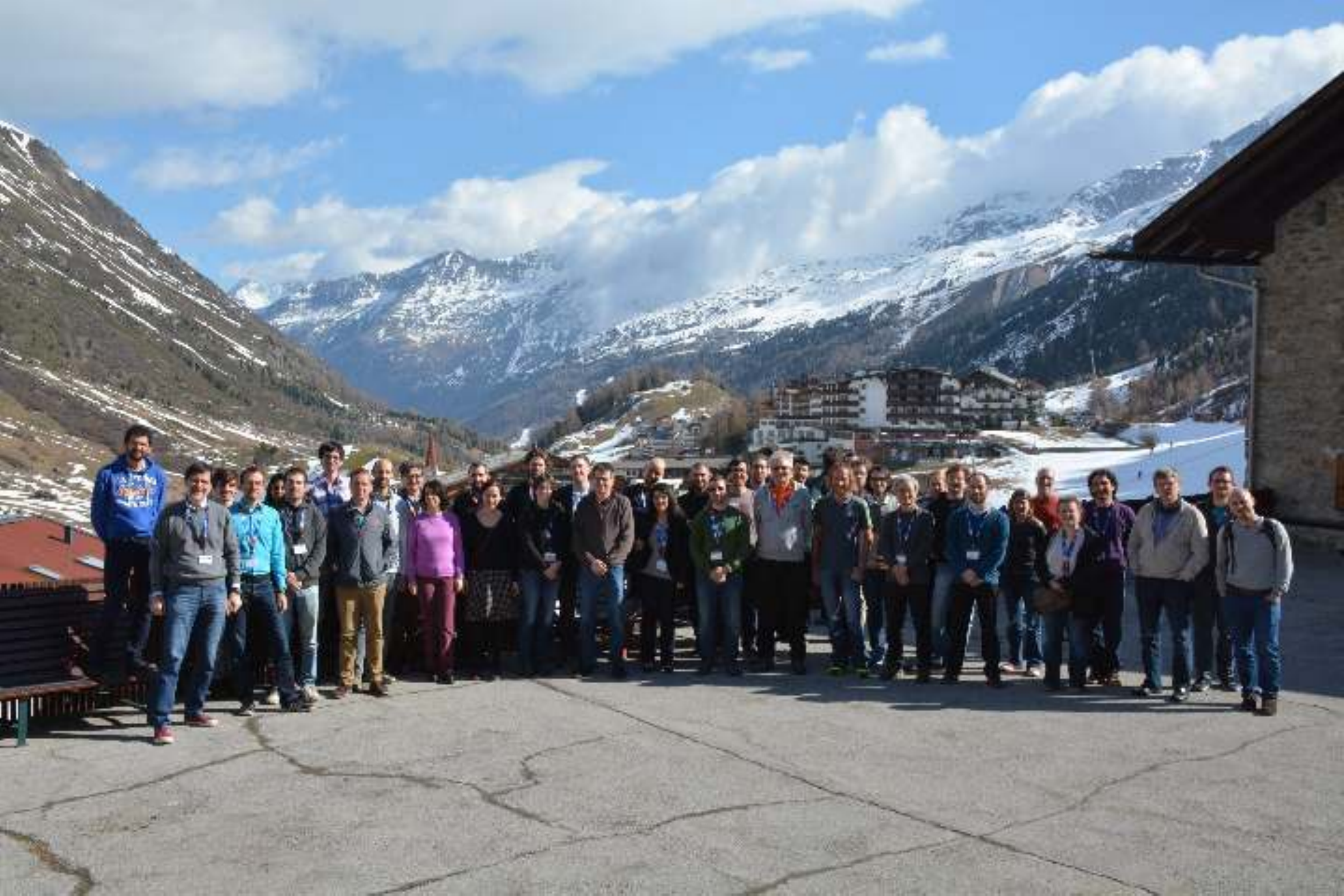}
\caption{ Group photo. }
%\label{fig:workingscheme}
\end{center}
\end{figure}
The workshop ``LHCSki 2016 -- A First Discussion of 13 TeV Results'' was held at the Obergurgl Universit\"atszentrum in 
the Austrian Alps in Tirol in the time period from April 10 until April 15, 2016.
Its date has been chosen to follow the traditional winter conferences.  
%The idea behind that choice is that, whereas winter conferences such as the
%``Rencontres de Moriond'' {\it present} the latest results of the LHC
%experiments, it is in this workshop that a 
%discussion of the {\it implications} of the results should take place.
The idea behind that choice is that, whereas most winter conferences present the latest results of particle physics
experiments, it is in this workshop that a discussion of the implications of the results should take place.
To this end, it was intended to balance the talks between theorists
and experimentalists. The workshop aimed to facilitate not only
communication between theory and experiment, but also between the high
energy, the cosmological, and the high luminosity frontiers.
Reflecting this idea of enabling cross talk, each workshop day was dedicated
to one of these endeavours: the scalar sector was discussed on Monday, followed
by physics beyond the standard model on Tuesday, dark matter was the subject of 
Wednesday, precision measurements and the flavor sector were the topics of Thursday.
Finally summary and outlook talks were scheduled for Friday.
In addition, each day held a  ``young scientists' forum''. % in which our younger
%colleagues were given the opportunity to claim the stage in an international
%environment for a short time period to report on their own research.

In total, 60 participants attended the workshop, 51 talks were given, 28 of
which are represented in these conference proceedings.
The workshop website can be found at  \\
\vspace*{-3mm}
\begin{center}
	\href{http://lhcski2016.hephy.at}{ \texttt{http://lhcski2016.hephy.at} }
\end{center}

%% \newpage
Slides and background reference materials are also to be found at \\
\vspace*{-3mm}
\begin{center}
	\href{https://indico.cern.ch/event/351843/} { \texttt{ https://indico.cern.ch/event/351843/ } }
\end{center}

Next year’s workshop ``ALPS 2017 - An Alpine LHC Physics Symposium'' will be
held at the same location, from April 17 to 22, 2017. Detailed information
about the workshop will be made available on \href{http://alps.hephy.at}{\texttt{http://alps.hephy.at}} in due course.

%LHCSki 2016 is intended to be the beginning of the ``ALPS'' annual series of
%workshops: ``An Alpine LHC Physics Symposium''.
%Please note the name change: next year's workshop will be called ``ALPS
%2017''.  The dates for next year are already fixed for April 17 -- 22, 2017.
%We hope to see many of you again in Tirol in 2017!

\newpage

\subsection*{Speakers}
Antonio Augusto Alves Junior (University of Cincinnati), Brian Thomas Amadio (LBNL), Brian Thomas Batell (University of Pittsburgh), Florian Urs Bernlochner (Universit\"at Bonn), Michael Brodski (Rheinisch-Westfälische Tech. Hoch.), Giorgio Busoni (University of Melbourne), Alejandro Celis (LMU), Eung Jin Chun (Korea Institute for Advanced Study), Andreas Crivellin (CERN), Jeff Dandoy (University of Chicago), Michael Duerr (DESY), Georgi Dvali (NYU and LMU), Martin Flechl (Austrian Academy of Sciences), Timothy Gershon (University of Warwick), Kiel Howe (Fermi National Accelerator Laboratory), Tetiana Hryn'ova (Centre National de la Recherche Scientifique), Kalliopi Iordanidou (Columbia University), Manfred Jeitler (Austrian Academy of Sciences), Felix Kahlhoefer (DESY), Jernej F. Kamenik (Jozef Stefan Institute), Valentin Khoze (Durham University), Muhammad Bilal Kiani (Università e INFN Torino), Christian Kiesling (Werner-Heisenberg-Institut), Joachim Kopp (Johannes-Gutenberg-Universit\"at Mainz), Suchita Kulkarni (Austrian Academy of Sciences), Dan Levin (University of Michigan), Luigi Li Gioi (Max-Planck-Institut f\"ur Physik, Werner-Heisenberg-Institut), Jenny List (DESY), Emiliano Molinaro (CP3-Origins), Sebastian Ohmer (Max-Planck-Institut f\"ur Kernphysik, Heidelberg), Maurizio Pierini (CERN), Tilman Plehn (Heidelberg University), Maxim Pospelov (University of Victoria), Olaf Reimer (Uni Innsbruck), Alexei Safonov (Texas A \& M University), David Salek (NIKHEF), Daniel Salerno (Universit\"at Z\"urich), Francesco Sannino (CP3-Origins), Veronica Sanz Gonzalez (University of Sussex), Marc Schumann (University of Bern), Christian Schwanenberger (DESY), Sezen Sekmen (Kyungpook National University), Nikoloz Skhirtladze (Kansas State University), Peter Stangl (Excellence Cluster Universe, Munich), Barbara Storaci (Universit\"at Z\"urich), Carlos Tamarit (IPPP Durham), Zoltán Tulpiánt (University of Debrecen), Nicola Venturi (University of Toronto), Tomer Volansky (Tel Aviv University), Simon Wehle (DESY Hamburg), Hartger Weits (NIKHEF), Charles Vincent Welke (Univ. of California San Diego), Zinonas Zinonos (Georg-August-Universit\"at Göttingen)

\subsection*{Acknowledgements}
%First of all we would like to thank all attendants for taking the effort 
%of travelling to a fairly remote corner of the Alps to take part of this
%workshop. 
Thanks to all the speakers for contributing actively to
the scientific content of the workshop. We cordially thank
Juliane Mayer and the ``Transferstelle Universit\"at Innsbruck Wissenschaft –
Wirtschaft – Gesellschaft'' for taking care of all non-scientific 
aspects of the workshop, as well as the Universit\"atszentrum Obergurgl
for their hosting our workshop. 
Thank you all and looking forward to seeing you at ALPS 2017!
\vspace*{4mm}
\newline
Wien, April 25, 2016,
\newline
\begin{center}
Jochen Schieck\\
Wolfgang Adam\\
Josef Pradler\\
Christoph Schwanda\\
Wolfgang Waltenberger\\
\end{center}

\newpage

\section{Contributions}
\subsection*{Monday}
\hfill Page\newline 
K. Howe: {\it Induced Electroweak Symmetry Breaking and the Composite Twin Higgs} \dotfill\,\pageref{ssec:InducedElectrow}\,\\
M. Kiani: {\it New CMS Results on $\mathrm{H}\rightarrow\mathrm{Z}\mathrm{Z}\rightarrow4\ell$ at 13 TeV} \dotfill\,\pageref{ssec:NewCMSResultson}\,\\
M. Flechl: {\it The Scalar Sector of the Standard Model} \dotfill\,\pageref{ssec:TheScalarSector}\,\\
E. Molinaro: {\it The 750 GeV Diphoton Resonance in Scenarios of Minimal Composite Dynamics} \dotfill\,\pageref{ssec:The750GeVDiphot}\,\\
S. Kulkarni: {\it Making Sense of LHC Diboson and Diphoton Excesses} \dotfill\,\pageref{ssec:MakingSenseofLH}\,\\
D. Salerno: {\it Studies of Higgs Bosons Decaying to Fermions with CMS} \dotfill\,\pageref{ssec:StudiesofHiggsB}\,\\
N. Venturi: {\it h(125) Boson Measurements in ATLAS: Run-1 Legacy and Early Run-2 Results} \dotfill\,\pageref{ssec:h125BosonMeasur}\,\\
\subsection*{Tuesday}
\hfill Page\newline 
P. Stangl: {\it Constraining Composite Higgs Models with Direct and Indirect Searches} \dotfill\,\pageref{ssec:ConstrainingCom}\,\\
C. Welke: {\it Searches for Supersymmetry at CMS in Leptonic Final States with 13 TeV Data} \dotfill\,\pageref{ssec:SearchesforSupe}\,\\
S. Sekmen: {\it Low Scale Supersymmetry: R.I.P. or Resurrection?} \dotfill\,\pageref{ssec:LowScaleSupersy}\,\\
K. Iordanidou: {\it Searches for New Physics with Bosons at the ATLAS Detector in LHC Run II} \dotfill\,\pageref{ssec:SearchesforNewP}\,\\
Z. Zinonos: {\it Searches for Beyond-Standard-Model Higgs Bosons in ATLAS} \dotfill\,\pageref{ssec:SearchesforBeyo}\,\\
M. Brodski: {\it Search for New Physics in Z+MET channel at CMS} \dotfill\,\pageref{ssec:SearchforNewPhy}\,\\
T. Berger-Hryn'ova: {\it ``Exotica'' - Speaking up for Minorities} \dotfill\,\pageref{ssec:``Exotica''Spea}\,\\
\subsection*{Wednesday}
\hfill Page\newline 
M. Schumann: {\it Dark Matter: The Next 5 Years and Beyond} \dotfill\,\pageref{ssec:DarkMatterTheNe}\,\\
M. Duerr: {\it Baryonic Dark Matter at the LHC} \dotfill\,\pageref{ssec:BaryonicDarkMat}\,\\
F. Kahlhoefer: {\it How (not) to Use Simplified Models to Search for DM at the LHC} \dotfill\,\pageref{ssec:HownottoUseSimp}\,\\
M. Jeitler: {\it Dark Matter Searches with CMS} \dotfill\,\pageref{ssec:DarkMatterSearc}\,\\
\subsection*{Thursday}
\hfill Page\newline 
L. Li Gioi: {\it Belle II Studies of Missing Energy Decays and Searches for Dark Photon Production} \dotfill\,\pageref{ssec:BelleIIStudieso}\,\\
J. F. Kamenik: {\it CP Violation in Standard Model} \dotfill\,\pageref{ssec:CPViolationinSt}\,\\
S. Wehle: {\it Angular Analysis of $\bkllzero$} \dotfill\,\pageref{ssec:AngularAnalysis}\,\\
T. Gershon: {\it Precision Measurements in Heavy Flavour Physics} \dotfill\,\pageref{ssec:PrecisionMeasur}\,\\
A. Celis: {\it A New Class of Family Non-Universal $Z^{\prime}$ models} \dotfill\,\pageref{ssec:ANewClassofFami}\,\\
J. List: {\it Physics at Future Colliders} \dotfill\,\pageref{ssec:PhysicsatFuture}\,\\
A. Crivellin: {\it New Physics in the Flavour Sector} \dotfill\,\pageref{ssec:NewPhysicsinthe}\,\\
E. Chun: {\it LHC Probe of Leptophilic 2HDM for Muon g-2} \dotfill\,\pageref{ssec:LHCProbeofLepto}\,\\
\subsection*{Friday}
\hfill Page\newline 
T. Plehn: {\it The Rise of Effective Lagrangians at the LHC} \dotfill\,\pageref{ssec:TheRiseofEffect}\,\\
C. Kiesling: {\it Towards the Next Standard Model - Experimental Challenges} \dotfill\,\pageref{ssec:TowardstheNextS}\,\\

\newpage
\newpage
\setcounter{figure}{0}
\setcounter{table}{0}

%% sessionday{monday} 

\subsection*{\hfil Induced Electroweak Symmetry Breaking and the Composite Twin Higgs \hfil}
\label{ssec:InducedElectrow}
\vspace*{10mm}

\def\be{\begin{equation}}
\def\ee{\end{equation}}
\def\ie{{\it i.e.}}
\def\eg{{\it e.g.}}
\def\GeV{\text{ GeV}}
\def\TeV{\text{ TeV}}

\def\Mfive{{MCHM$_{5+1}$}}
\def\betaexp{\ensuremath{\beta_{\rm SM}}}

\def\beq{\begin{equation}}
\def\eeq{\end{equation}}

R.  Harnik, \underline{K.  Howe}, J.  Kearney\vspace*{4mm} 
 \\ Theoretical Physics Department, Fermi National Accelerator Laboratory\\ Batavia, IL 60510 USA \\  
\newline \noindent 
We investigate induced electroweak symmetry breaking\cite{Simmons:1988fu,Samuel:1990dq,Dine:1990jd,Kagan:1990az,Kagan:1991gh,Kagan:1992aq,Carone:1992rh,Carone:1993xc,Antola:2009wq,Antola:2011bx,kagantalk,Azatov:2011ht,Azatov:2011ps,Galloway:2013dma,Chang:2014ida} (EWSB) in models in which the Higgs is a pseudo-Nambu-Goldstone boson (pNGB) \cite{Kaplan:1983fs,Kaplan:1983sm}. In pNGB Higgs models, Higgs properties and precision electroweak measurements imply a hierarchy between the EWSB and global symmetry-breaking scales, $v_H \ll f_H$. When the pNGB potential is generated radiatively, this hierarchy requires fine-tuning to a degree of at least $\sim v_H^2/f_H^2$.
We show that if Higgs EWSB is induced by a tadpole arising from an auxiliary sector at scale $f_\Sigma \ll v_H$, this tuning is significantly ameliorated or can even be removed.
We present explicit examples both in Twin Higgs models \cite{Chacko:2005pe,Chacko:2005un,Barbieri:2015lqa,Low:2015nqa} and in Composite Higgs models based on $SO(5)/SO(4)$ \cite{Agashe:2004rs,Bellazzini:2014yua,Panico:2015jxa}. For the Twin case, the result is a fully natural model with $f_H \sim 1$~TeV and the lightest colored top partners at 2 TeV. These models also have an appealing mechanism to generate the scales of the auxiliary sector and Higgs EWSB directly from the scale $f_H$, with a natural hierarchy $f_\Sigma \ll v_H \ll f_H \sim{\rm TeV}$. The framework predicts modified Higgs coupling as well as new Higgs and vector states at LHC13 \cite{Chang:2014ida}.

\begin{figure}[h!]
\centering
\includegraphics[width=0.499\textwidth]{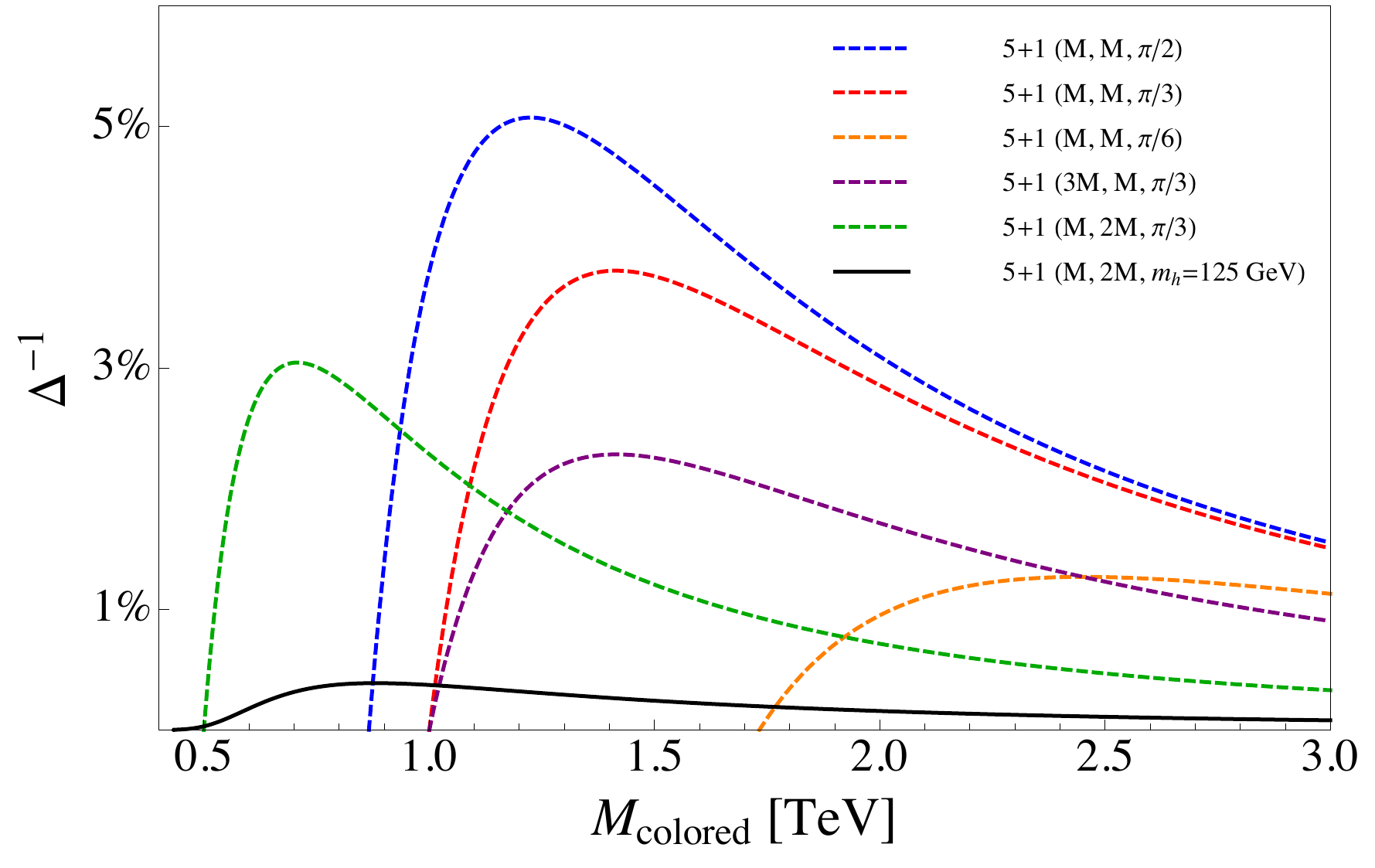}
\caption{
\label{fig:tadpoletuningSO5}
Top sector radiative tuning in the $SO(5)/SO(4)$ 5+1 model (or \Mfive) with a tadpole as a function of the lightest colored top partner mass $M_{\rm colored}$ for $f_H = 1\TeV$. Dashed curves correspond to different choices of $(M_1, m_4, \theta_R)$, as listed in the legend. For comparison, the black solid line corresponds to \Mfive~without a tadpole (\ie, with $\beta=\betaexp$ generated by large $q_L$ compositeness, determining $\theta_R$).
}
\end{figure}

Fig.~\ref*{fig:tadpoletuningSO5} shows the tuning for the $SO(5)/SO(4)$ minimal coset as a function of the lightest top partner mass for several different sets of parameters $(M_1, m_4, \theta_R)$ specifiying the top sector masses and degree of right-handed compositeness.
For comparison, we also show the tuning for the \Mfive~model without a tadpole in which the minimal top sector generates $\beta=\betaexp$ radiatively to give $m_h = 125 \GeV$. Achieving sufficiently large Higgs mass requires an increase in $q_L$ compositeness, such that the Higgs experiences more explicit breaking from $y_L > y_t$, similar to raising $m_h$ via large $A$-terms in the MSSM---the increased explicit symmetry breaking enhances the quartic, but also results in more tuning.
A model exhibiting top partners with masses $\gsim f_H$ and a tadpole contribution to the potential can be significant more natural (with tuning reduced by ${\cal O}(5-10)$) than the \Mfive~with  $m_h=125~\GeV$ generated by the minimal top sector.

\begin{figure}
\centering
\includegraphics[width=0.47\textwidth]{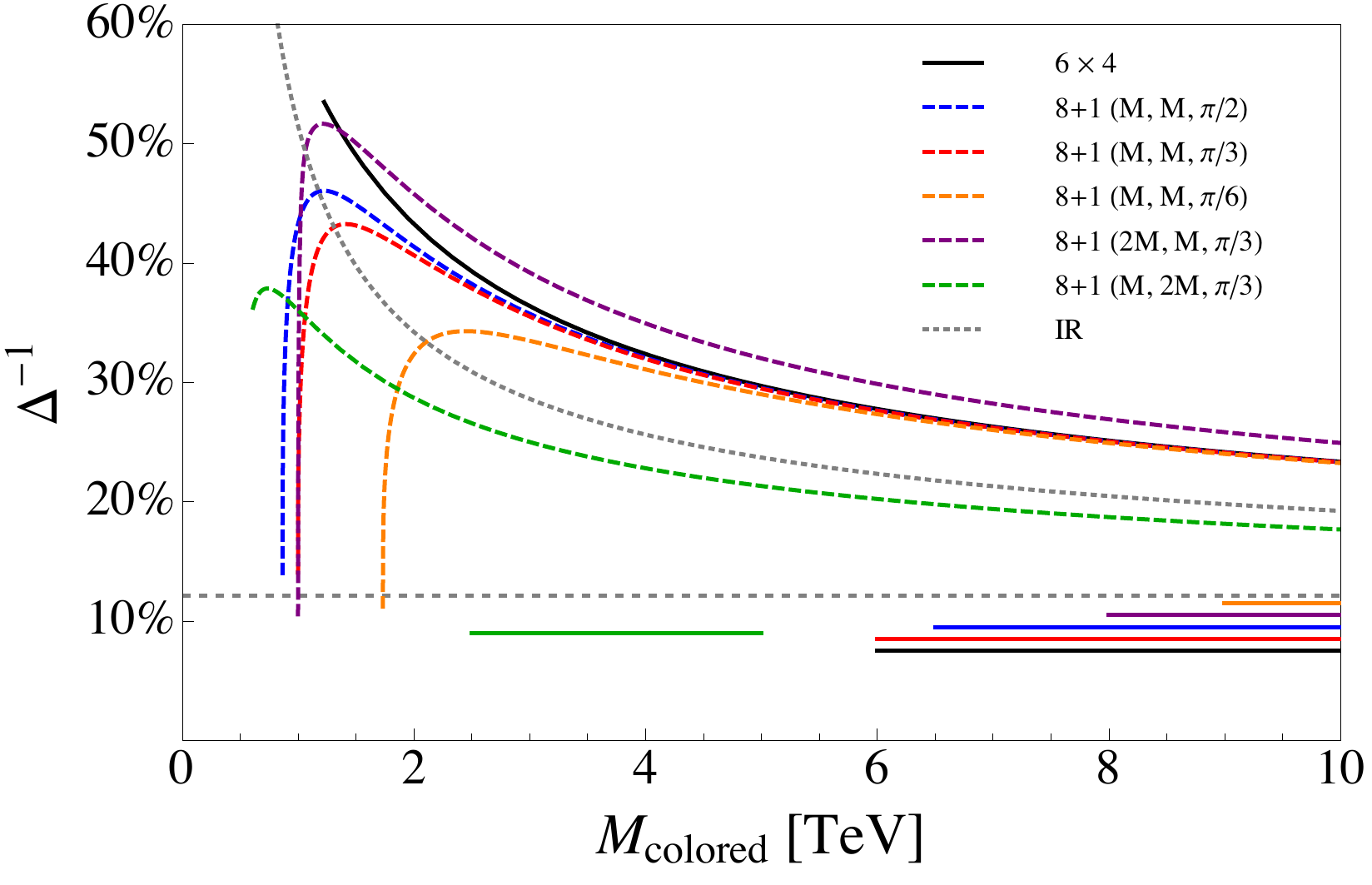}
\caption{
\label{fig:tadpoletuning}
Top sector radiative tuning in a Twin Higgs model with a tadpole as a function of the lightest colored top partner mass $M_{\rm colored}$ for $f_H=1\TeV$. Dotted gray is the estimated tuning from the pure $t_B$ IR contribution. Solid black is the $6\times4$ model, while dashed, colored curves correspond to the $8+1$ model with $(M_1, m_7, \theta_R)$ as listed. For comparison, the horizontal dotted gray line corresponds to the minimal tuning $\frac{2 v_H^2}{f_H^2} \simeq 10\%$ of the radiative quartic potential, with horizontal lines indicate the top partner mass range which can radiatively generate $\beta=\betaexp$ (saturating this tuning) within theoretical uncertainty.}
\end{figure}

Fig.~\ref*{fig:tadpoletuning} shows the tuning for the twin Higgs model for several choices of top sector.
The minimal tuning occurs for top partners with masses roughly just above the smallest possible value required to realize the top Yukawa, $M_T \simeq \sqrt{2} f_H \sim m_{t_B}$. For these values, induced EWSB can reduce tuning by a factor of $\sim 5$ relative to the minimal $\frac{2 v_H^2}{f_H^2} \sim 10\%$ tuning of the radiative quartic potential.

\begin{figure}
\centering
\includegraphics[width=0.4\columnwidth]{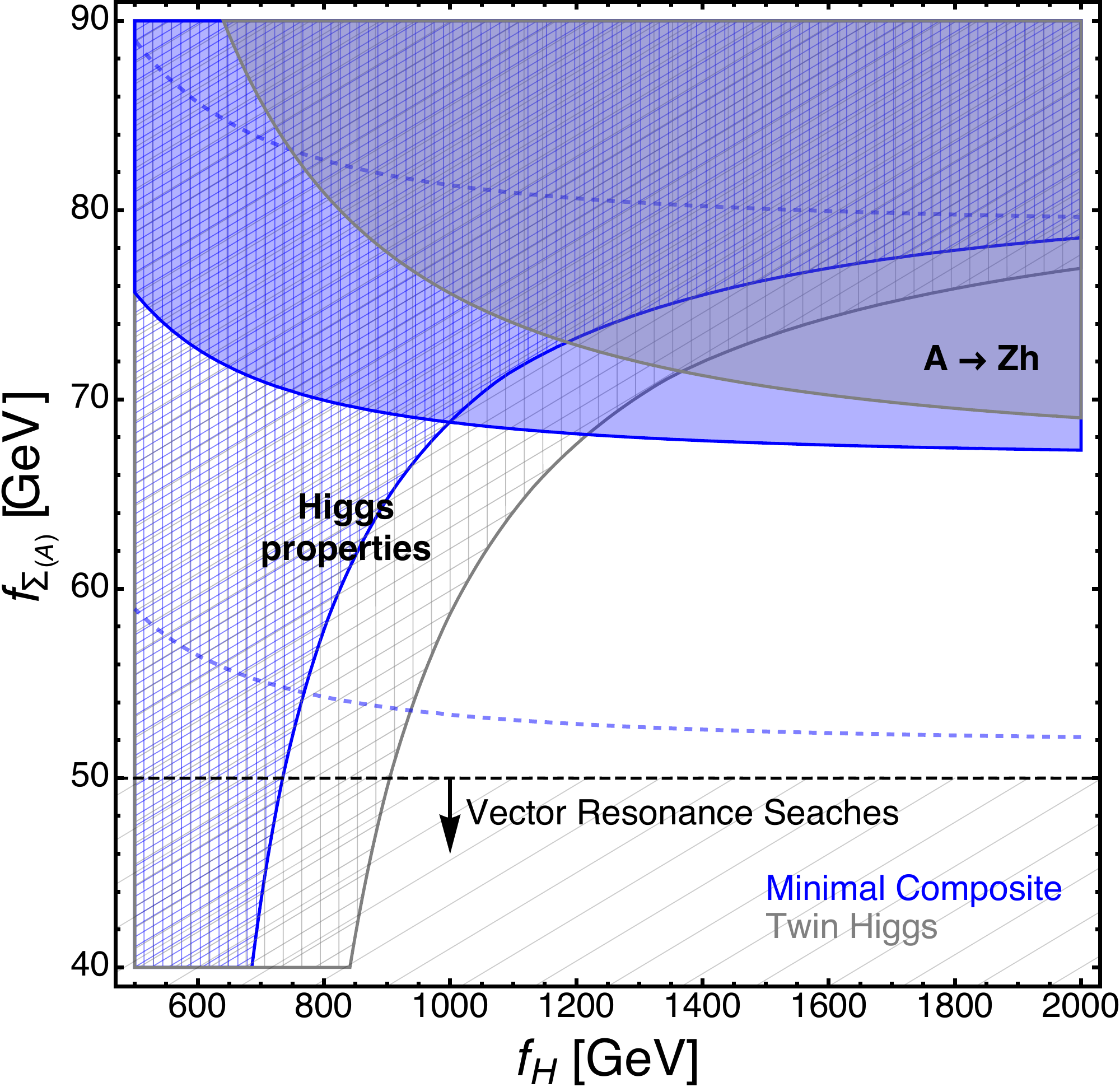}
\caption{
\label{fig:ffSigma}
Regions of $(f_H,f_\Sigma)$ excluded by Higgs coupling measurements (hatched) and direct $A \rightarrow Zh$ searches (solid) for Minimal Composite (blue) and Twin (gray) Higgs models.
Dashed blue contours represent the effect of rescaling $m_A^2$ by $0.6$ (lower) or $1.4$ (upper) and thus represent the theoretical uncertainty on the solid blue line.
The dashed black line denotes approximate lower bound $f_\Sigma \gsim 50 \GeV$ from vector resonance searches.}
\end{figure}

In Fig.~\ref*{fig:ffSigma}, we show approximate constraints from Higgs property measurements \cite{ATLAS-CONF-2015-044}, the extended Higgs sector $A \rightarrow Zh$ decay \cite{Khachatryan:2015lba},  and the strongly-coupled auxiliary sector vector resonance $\rho^\pm \rightarrow W^\pm Z$ \cite{Aad:2014pha} decay. 

Not only can tadpole-induced models feature a pNGB potential with a fully natural scale for EWSB, but in fact searches at LHC13 and future colliders will likely be able to probe the entire remaining range of viable models independent of any naturalness arguments.

\bibliographystyle{plain}
\bibliography{TadpoleRefs}

\newpage
\setcounter{figure}{0}
\setcounter{table}{0}

%% sessionday{monday} 

\subsection*{\hfil New CMS Results on $\mathrm{H}\rightarrow\mathrm{Z}\mathrm{Z}\rightarrow4\ell$ at 13 TeV \hfil}
\label{ssec:NewCMSResultson}
\vspace*{10mm}

\underline{M.  Kiani}\vspace*{4mm} 
 \\ Dipartimento di Fisica, Universit\`a degli studi di  Torino - Torino, Italy \\  
INFN, Sezione di Torino - Torino, Italy \\  
\newline \noindent Studies of Higgs boson production are presented using the $\mathrm{H}\rightarrow\mathrm{Z}\mathrm{Z}\rightarrow4\ell$ ($\ell=\mathrm{e},\mu$) decay. These studies are performed using a data sample corresponding to an integrated luminosity of $2.8~\mathrm{fb}^{-1}$ of pp collisions at a center-of-mass energy of $13~\mathrm{TeV}$ collected by the CMS experiment at the LHC during 2015. The observed significance for the standard model Higgs boson with $m_{\mathrm{H}}=125.09~\mathrm{GeV}$ is $2.5\sigma$, where the expected significance is $3.4\sigma$. The model independent fiducial cross section is measured to be $\sigma_{\mathrm{fid.}}=2.48^{+1.48}_{-1.14}(\mathrm{stat. \oplus sys.})^{+0.01}_{-0.04}(\mathrm{model~dep.})~\mathrm{fb}$

\section{Description}

The studies in various Higgs decay channels and production modes with the full LHC Run 1 ~\cite{prop} data set and combined measurements from ATLAS and CMS showed that the properties of the new boson are so far consistent with expectations for the SM Higgs.

The start of the LHC Run 2 in 2015, at an increased center-of-mass energy of $\sqrt{s}=$13  TeV, opens the way for an era of new precision measurements of the Higgs boson, which will involve the observation and study of its rare production modes such as vector boson fusion (VBF) and associated production with a vector boson (WH, ZH) and top pair production ttH.

\section{Results}

\begin{figure}[h!]
\vspace*{0.3cm}
\begin{center}
\includegraphics[width=0.4\textwidth]{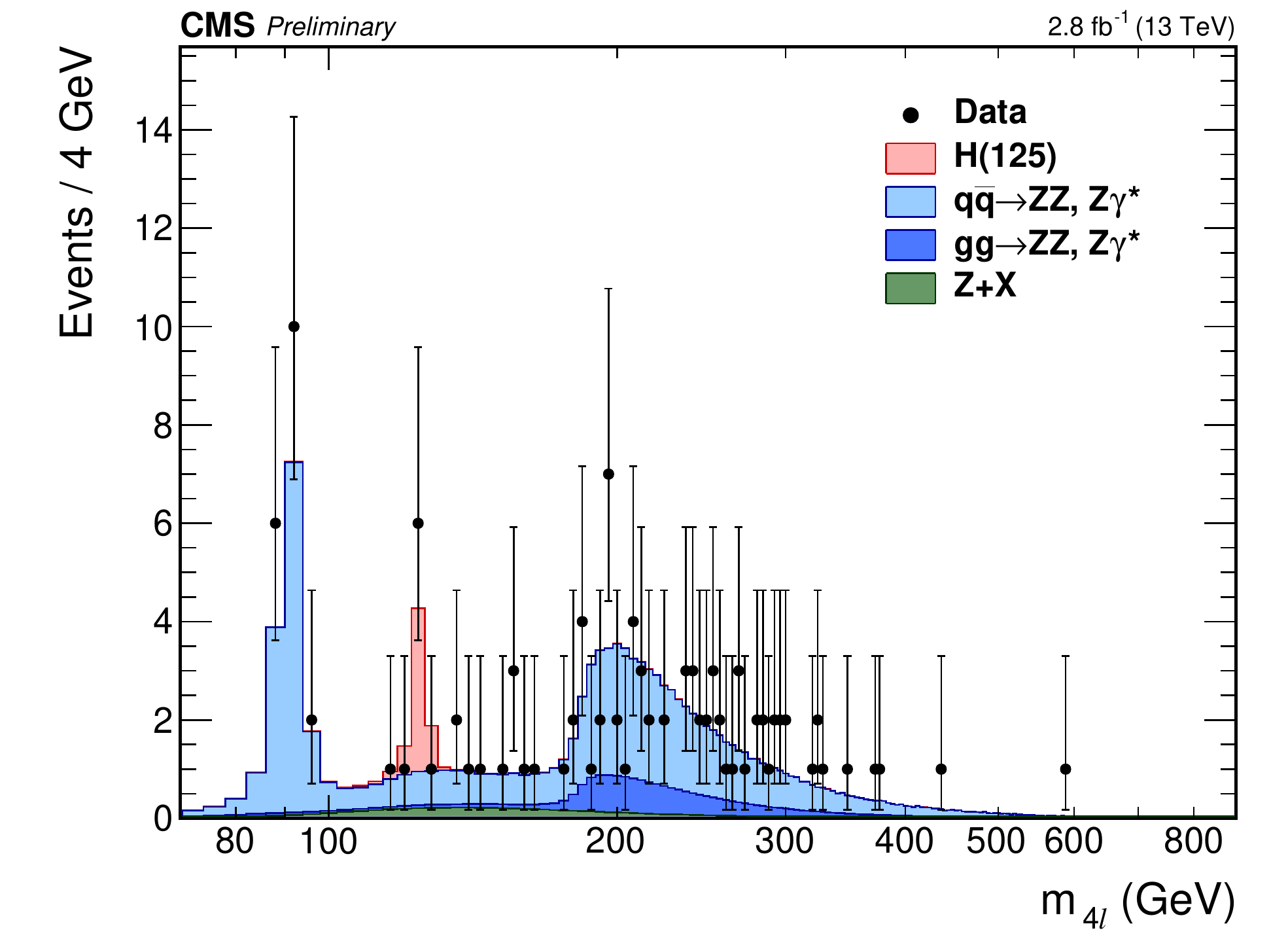} 
\includegraphics[width=0.3\textwidth]{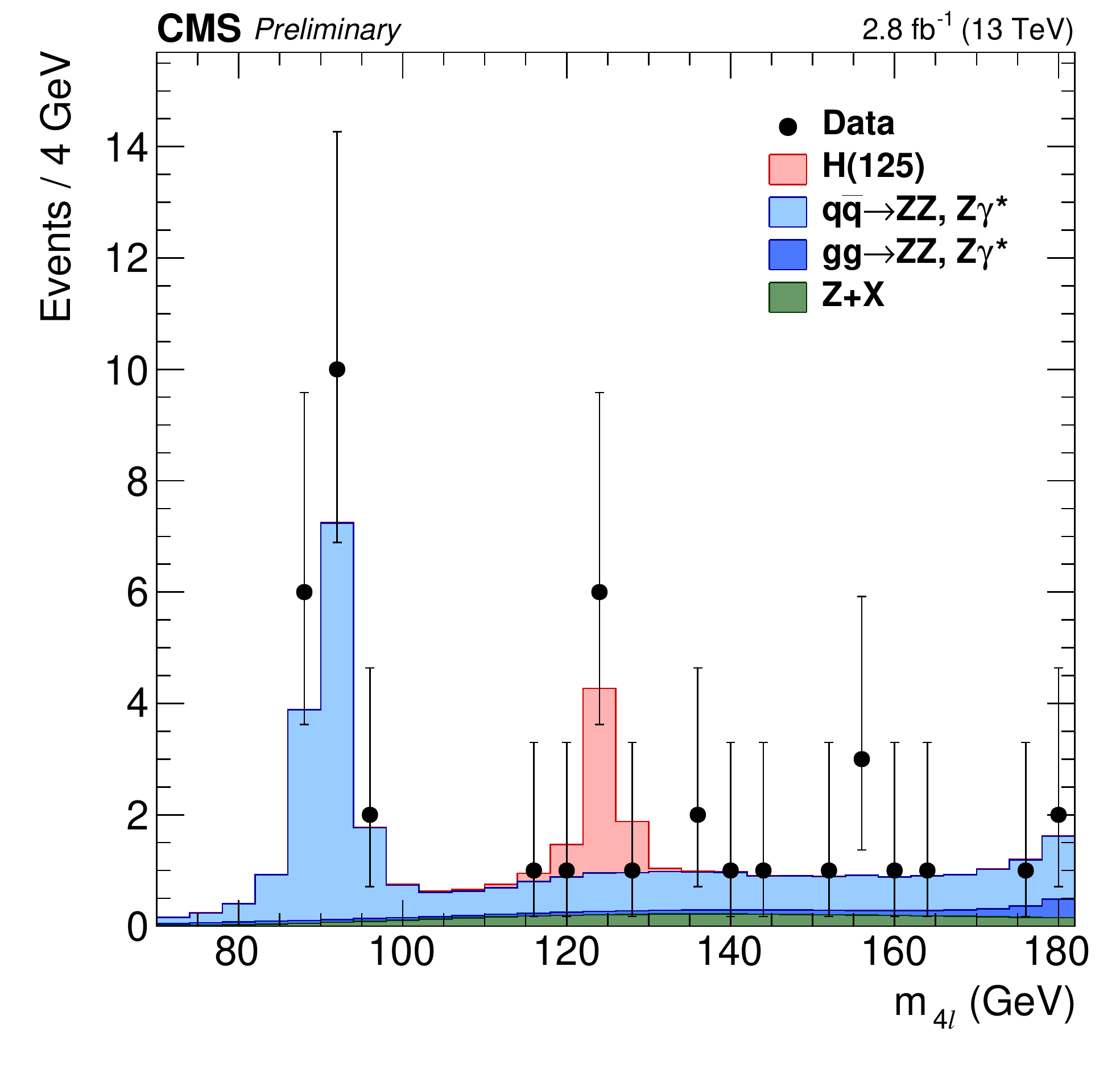}     \caption{Distribution of the four-lepton invariant mass $m_{4l}$ in the full mass range (left) and in the low-mass region (right). Points with error bars represent the data and stacked histograms represent expected distributions.}
\label{mass}
\end{center}
\end{figure}

\begin{figure}[h!]
\vspace*{0.3cm}
\begin{center}
\includegraphics[width=0.3\textwidth]{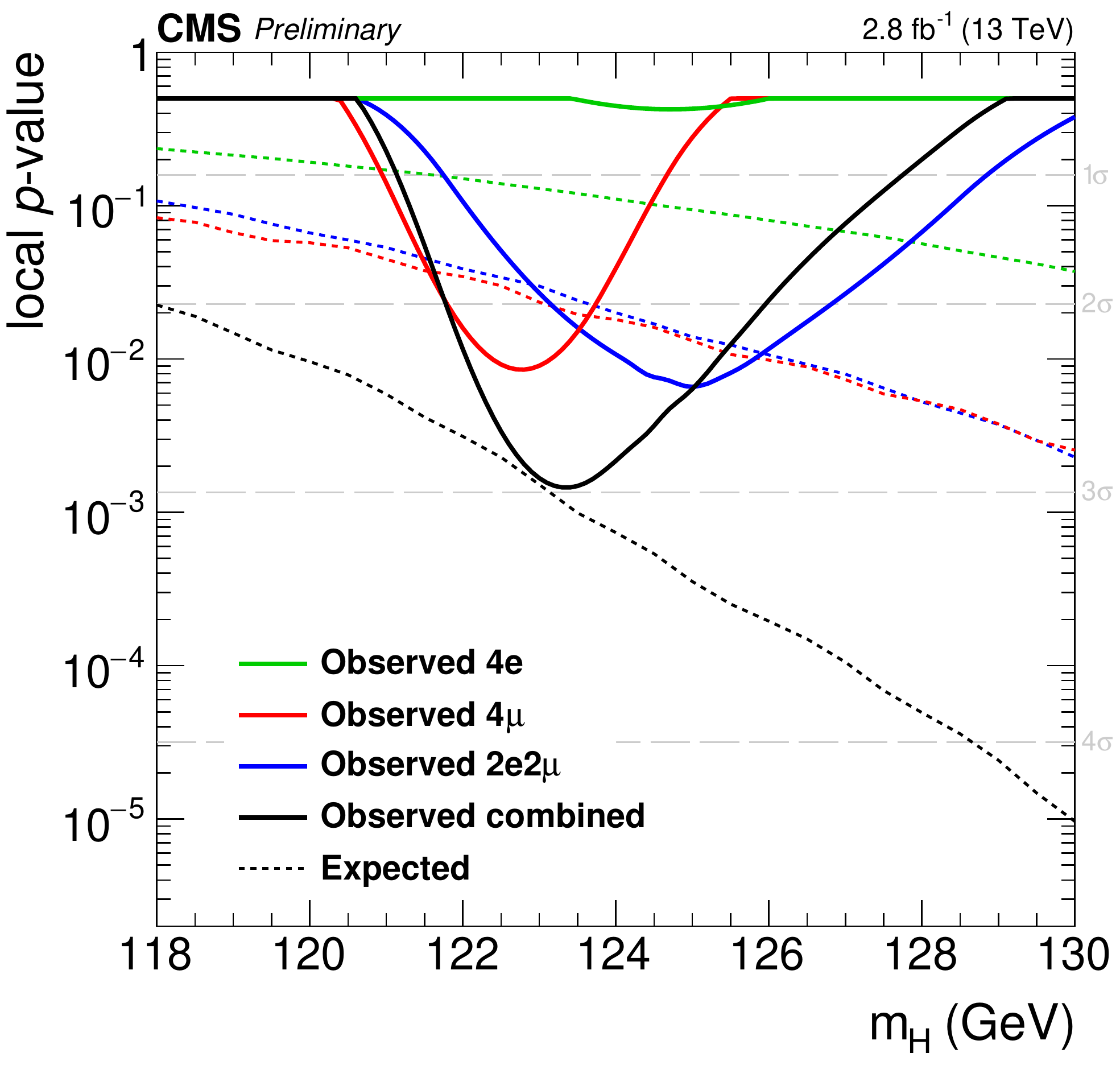} 
\includegraphics[width=0.3\textwidth]{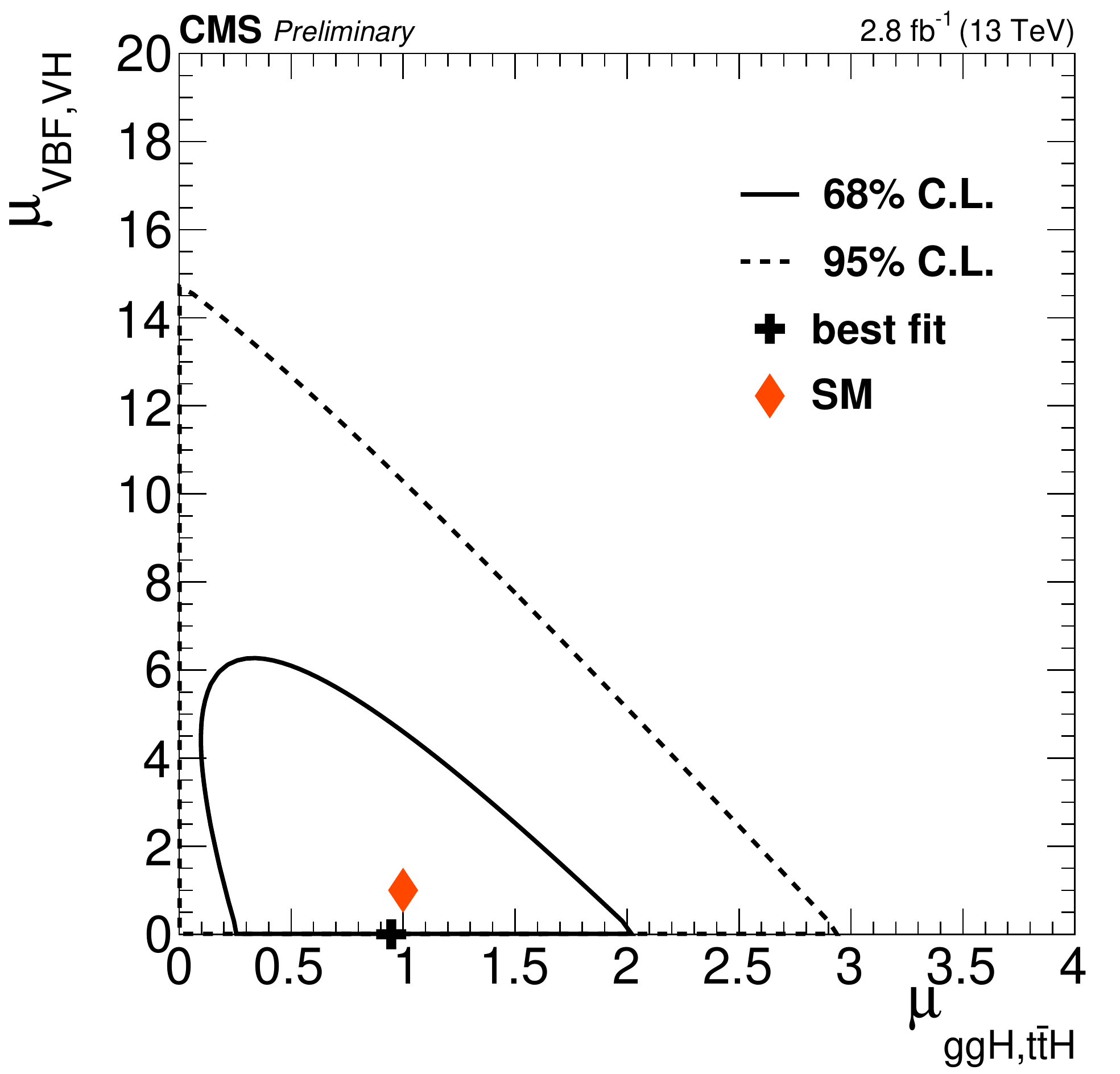} 
\caption{Significance of the local fluctuation with respect to the SM expectation as a function of the Higgs boson mass. Dashed lines show the mean expected significance of the SM Higgs boson for a given mass hypothesis.(right) Result of the 2D likelihood scan for the $\mu_{F}$ and $\mu_{V}$ signal-strength modifiers. The solid and dashed contours show the 68\% and 95\% CL regions, respectively. The cross indicates the best-fit values, and the diamond represents the expected values for the SM Higgs boson. }
\label{sig}
\end{center}
\end{figure}

The $\mathrm{H}\rightarrow\mathrm{Z}\mathrm{Z}\rightarrow4\ell$  analysis is based on the reconstruction, identification and isolation of leptons. The event selection is designed to extract signal candidates from events containing at least four well-identified and isolated leptons, each originating from the primary vertex.
 First, Z candidates are formed with pairs of leptons of the same flavor and opposite-charge  and required to pass $12 < m_{l^{+}l^{-}}  < 120GeV$.
They are then combined into ZZ candidates, wherein we denote as Z1 the Z candidate with an invariant mass closest to the nominal Z boson mass, and as Z2 the other one. The details of the event selection can be found at ~\cite{15pas004}. Figure ~\ref*{mass} shows the distributions obtained from this selection. 
To extract the signal significance for the excess of events observed in the Higgs peak region and estimate its signal strength, a multi-dimensional fit is  performed that relies on two variables: the four-lepton invariant mass $m_{4l}$ and the $D_{kin}^{bkg}$  discriminant(which separates the gluon fusion from production modes ) Figure ~\ref*{sig}

\begin{figure}[h!]
\vspace*{0.3cm}
\begin{center}
\includegraphics[width=0.45\textwidth]{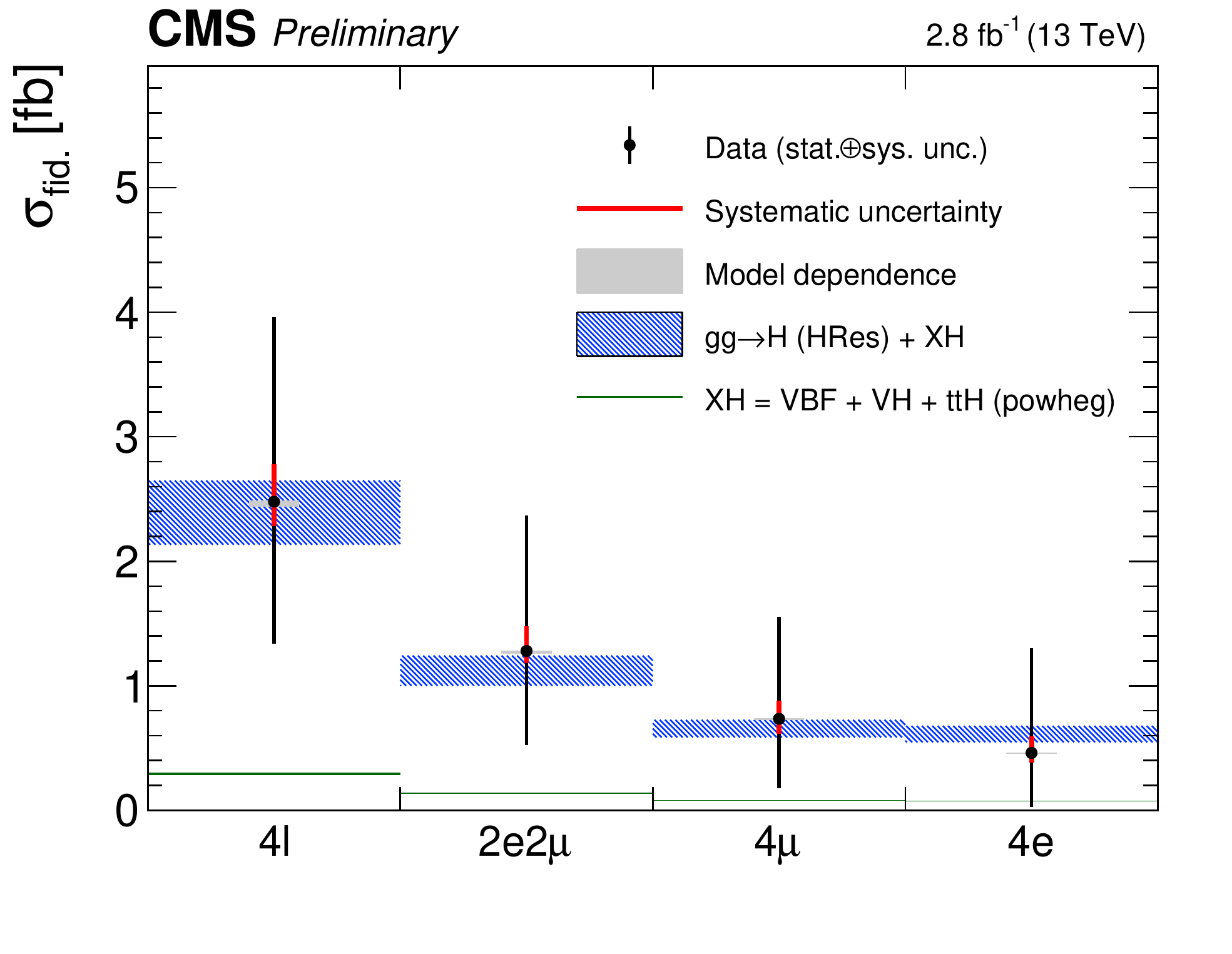} 
\includegraphics[width=0.45\textwidth]{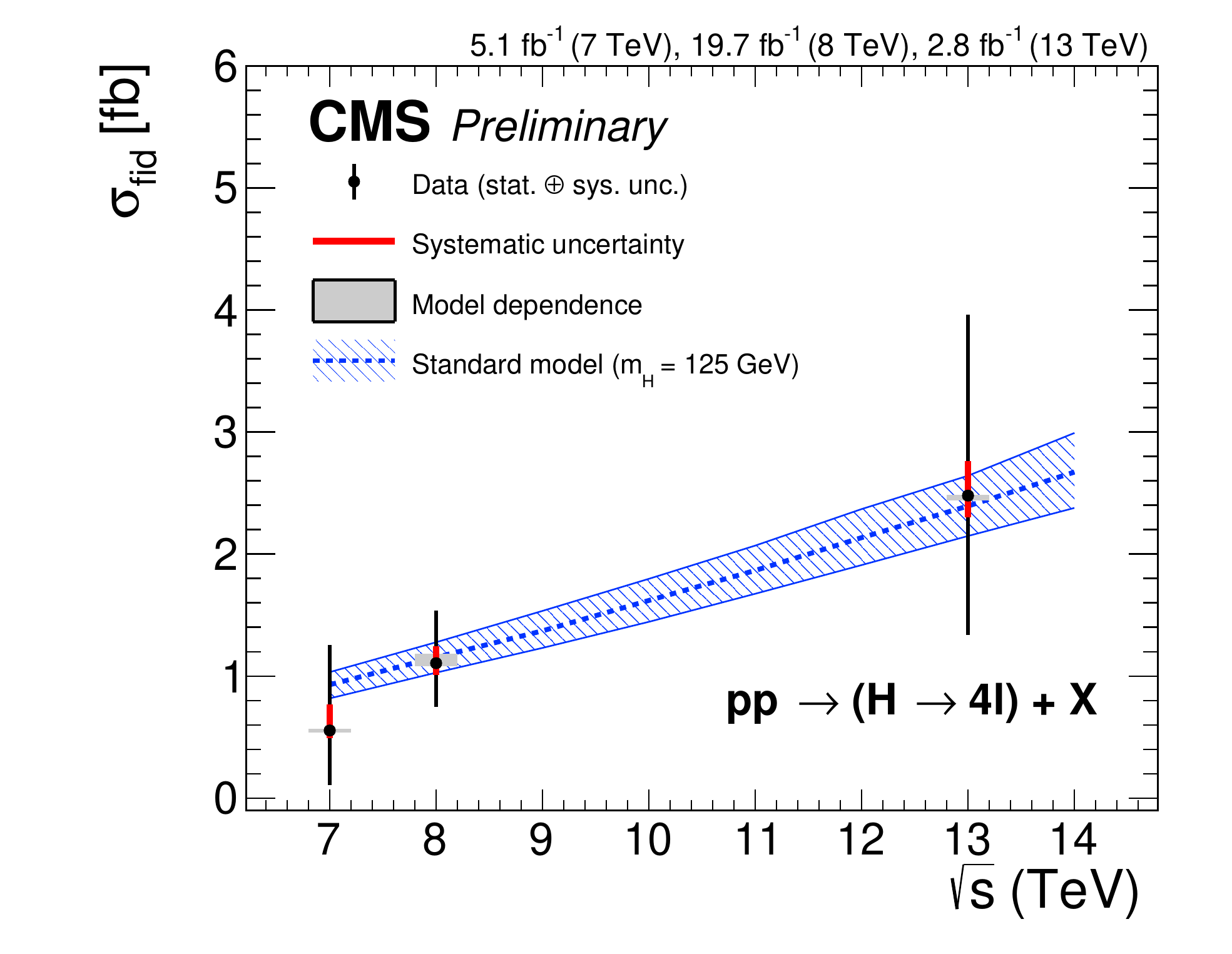}     \caption{Result of measured fiducial cross section in each final state (left). the measured fiducial cross section as a function of $\sqrt{s}$ (Right)}
\label{fid}
\end{center}
\end{figure}

The fiducial volume is defined in ~\cite{fidcross}. Integrated fiducial cross section for pp $\rightarrow$H$\rightarrow$4$\ell$ is obtained by performing a maximum likelihood fit of the signal and background parameterizations to the observed $4\ell$ mass distribution, . The results are shown in Figure ~\ref*{fid}.

\newpage
\setcounter{figure}{0}
\setcounter{table}{0}

%% sessionday{monday} 

\subsection*{\hfil The Scalar Sector of the Standard Model \hfil}
\label{ssec:TheScalarSector}
\vspace*{10mm}

\underline{M.  Flechl} (on behalf of the ATLAS and CMS Collaborations)\vspace*{4mm} 
 \\ Institute of High Energy Physics\\ Austrian Academy of Sciences\\ Nikolsdorfergasse 18, 1050 Vienna, Austria \\  
\newline \noindent 
In 2012, the ATLAS~\cite{atlas} and CMS~\cite{cms} experiments have discovered a 
Higgs boson~\cite{Chatrchyan:2012xdj,Aad:2012tfa} and since then 
started an industry of Higgs boson property measurements. Higgs bosons at 
the LHC are dominantly produced in gluon-gluon fusion (ggF), vector boson fusion (VBF), 
and associated production with a W or Z boson (VH), or a top quark pair (ttH)~\cite{yr3}.

Higgs boson property measurements are typically the result of global fits 
combining studies aimed at different production and decay modes. For several 
measurements, the ATLAS and CMS analyses of the years 2011 and 2012 have been 
combined. 

Higgs boson production in ggF and VBF as well as decays to $\gamma\gamma$, $ZZ$, $WW$ and 
$\tau\tau$ have been observed at the LHC with a significance of more than 5 Gaussian standard 
deviations, VH and ttH production with more than 3~\cite{atlas_kappa,cms_kappa}. 
The mass has been measured by ATLAS and CMS as $m_H=125.09 \pm 0.24$ GeV~\cite{mass_lhc}. 
The global signal strength $\mu$ is measured as  $1.09 \pm 0.11$ (ATLAS+CMS) and all 
results for $\mu$ for different production and decay modes are in good agreement 
with the standard model (SM) expectation, with only $\mu_{ttH}$ being about two standard deviations higher than 1~\cite{atlas_kappa,cms_kappa}. 

The combined ATLAS+CMS coupling strength modifier $\kappa$ fits~\cite{atlas_kappa,cms_kappa} are done for different 
assumptions and free parameters. All results agree with the SM expectation. Examples are 
universal modifiers for fermions and vector bosons in 2D fits or $\kappa$ ratios for down- and up-type fermions as well 
as leptons to quarks: All agree with the SM expectation within uncertainties of typically about 15\%.

Spin-parity properties have been measured separately by CMS~\cite{Khachatryan:2014kca} and ATLAS~\cite{Aad:2015mxa}. 
Among the tested alternatives, the data favor the SM hypothesis but there is still room for a sizable anomalous CP-odd or 
CP-even admixture.

Both ATLAS and CMS measure fiducial differential cross sections~\cite{diff_gg_atlas,diff_4l_atlas,diff_gg_cms}. 
They agree with the SM expectation within large statistical uncertainties.

While the overall agreement of measurements  with the SM is excellent, a few mild 
anomalies have been observed in the 2011+2012 data, e.g. in $ttH$~\cite{tth_cms,tth_atlas}, $WH \to e\nu bb$~\cite{wh_cms}, 
$HH \to bb\gamma\gamma$~\cite{hh_atlas} and $H\to\tau\mu$~\cite{lfv_cms,lfv_atlas1,lfv_atlas2}.

A list of most recent ATLAS and CMS Higgs physics results can be found on these TWikis~\cite{atlas_higgs_url,cms_higgs_url}. 
Prospects are discussed in Refs.~\cite{prosp_cms,tp,prosp_atlas,prosp_pair_atlas}.

\bibliographystyle{unsrt}
\bibliography{flechl_contr}

\newpage
\setcounter{figure}{0}
\setcounter{table}{0}

\def\Tr{\mbox{Tr}\,}

%% sessionday{monday} 

\subsection*{\hfil The 750 GeV Diphoton Resonance in Scenarios of Minimal Composite Dynamics \hfil}
\label{ssec:The750GeVDiphot}
\vspace*{10mm}

\underline{E.  Molinaro}\vspace*{4mm} 
 \\ CP$^3$-Origins, University of Southern Denmark, \\ Campusvej 55, DK-5230 Odense M, Denmark \\  
\newline \noindent

We explain the recent  excess in diphoton invariant mass searches reported by ATLAS \cite{ATLAS} and CMS \cite{CMS} under  the two currently viable hypotheses of dominant photon and gluon fusion production mechanisms. The two frameworks sensitively depend  on how the new physics couples to the Standard Model (SM) degrees of freedom. We encode the new physics in the effective Lagrangian
\begin{eqnarray*}\label{eq:L-eff}
\mathcal{L}_{\rm eff}&=& -i y_t \frac{m_t}{v} a\, \bar{t} \gamma_5 t - \frac{c_{GG}}{8v}  a\, \Tr\left[ G^{\mu\nu}\tilde{G}_{\mu\nu}\right]  - \frac{c_{AA}}{8v}  a\, A^{\mu\nu}\tilde{A}_{\mu\nu} \nonumber\\
&& - \frac{c_{AZ}}{4v}  a\, A^{\mu\nu}\tilde{Z}_{\mu\nu}- \frac{c_{WW}}{4v}  a\, W^{+ \mu\nu}\tilde{W}^{-}_{\mu\nu} - \frac{c_{ZZ}}{8v}  a\, Z^{\mu\nu}\tilde{Z}_{\mu\nu}\,,
\end{eqnarray*}
in which $a$ is a new pseudoscalar boson of mass $m_a\approx 750$ GeV,  $v=246$ GeV is the electroweak scale, and $\tilde{V}^{\mu\nu}=\epsilon^{\mu\nu\rho\sigma}V_{\rho\sigma}$.   We neglect in $\mathcal{L}_{\rm eff}$ any direct  coupling of $a$ to the SM fermions, except for the top quark $t$. 
 
\begin{figure}[t!]
\centering
\begin{tabular}{cc}
\includegraphics[width=0.47\textwidth]{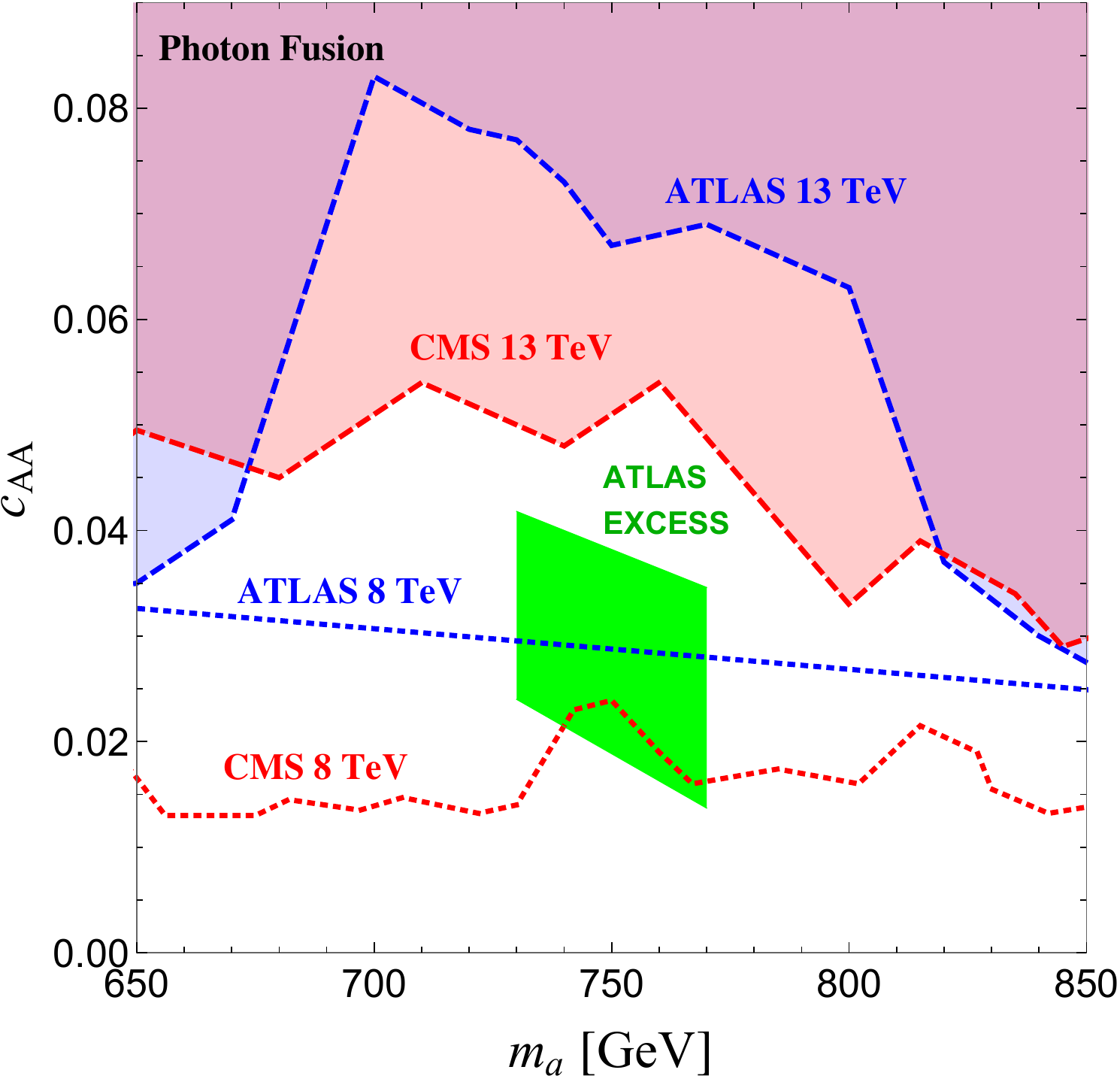} &
\includegraphics[width=0.47\textwidth]{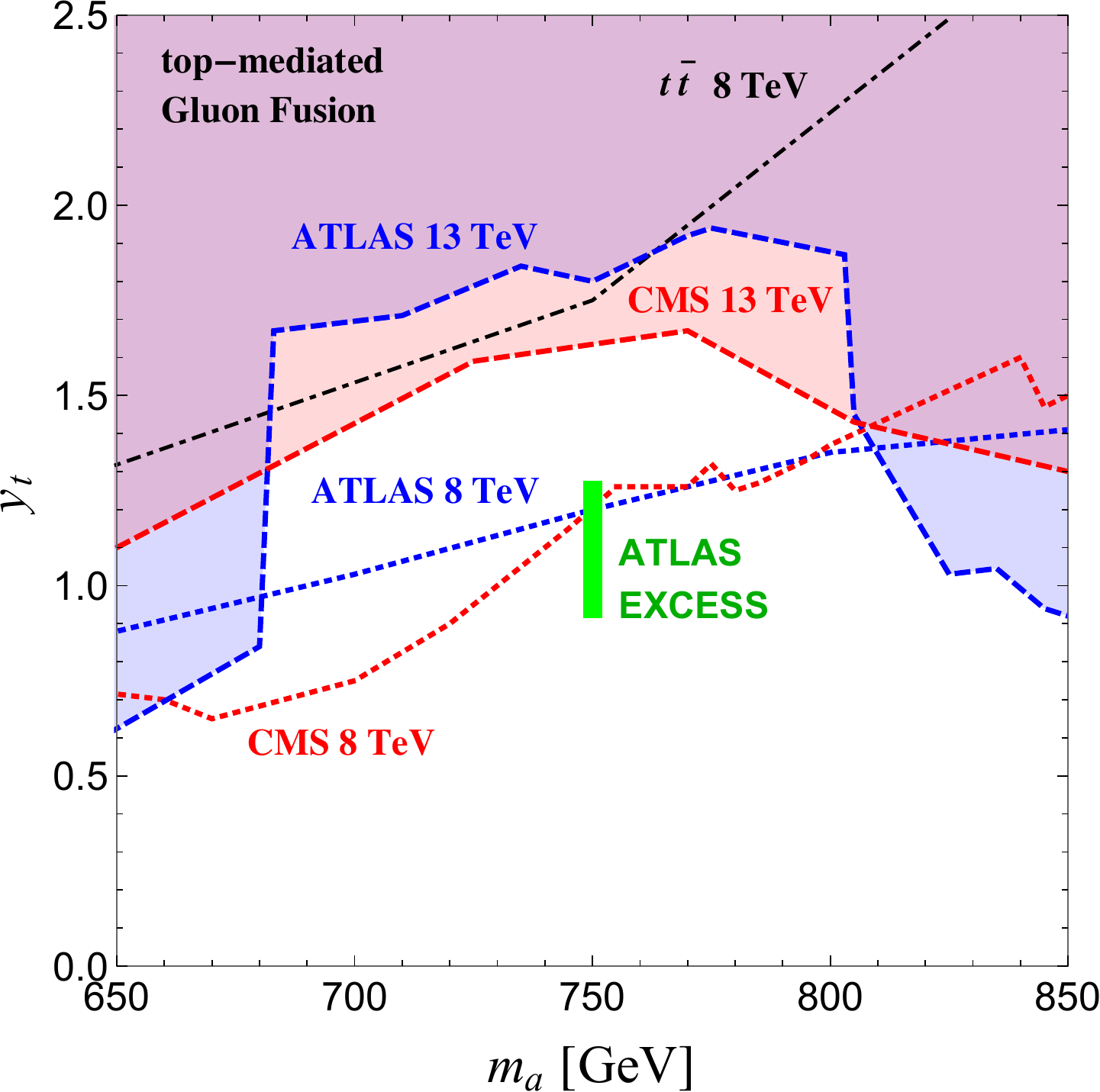}
\end{tabular}
\caption{ \small{\it Left plot}: LHC constraints in the plane ($m_a$, $c_{AA}$). The diphoton resonance $a$ is totally produced via photon fusion and decays to $\gamma\gamma$ with  ${\rm BR}(a\to\gamma\gamma)=60\%$. {\it Right plot:} LHC constraints in the plane ($m_a$, $y_{t}$) for a diphoton resonance produced via top-mediated gluon fusion. We assume $\Gamma_{\rm tot}(a) \simeq \Gamma(a \to t\bar{t})$ and ${\rm BR}(a\to\gamma\gamma)=1\%$. 
The green areas correspond to the region of the parameter space that in both scenarios allow to fit the excess within $1\sigma$. Plots taken from \cite{Molinaro:2016oix}. }
\label{fig-1-molinaro}       \end{figure}

 We report in figure~\ref*{fig-1-molinaro} the constraints from LHC run-1 ($\sqrt{s}=8$ TeV) and run-2 ($\sqrt{s}=13$ TeV) in the case 
 the new resonance is
 either entirely photo-produced (left plot, with $y_t=0$ and $c_{GG}=0$) or it is produced via the effective coupling $c_{GG}$ of $a$ to the gluons (right plot, with $y_t\neq 0$). The latter is 
 generated at one-loop through the coupling of $a$ to the SM top quark. 
 In this case, for the values of $y_t$ that allow to fit the signal (green region in the figure), the total decay width of $a$ is 
 $\Gamma_{\rm tot}(a) \approx 45$ GeV, which is in remarkable agreement with the value preferred by current ATLAS analysis \cite{ATLAS}.
  
 A UV completion of this effective field  theory, which naturally  predicts couplings of $a$ to SM degrees of freedom and a diphoton branching ratio consistent with the values that explain the diphoton excess, as shown in the figure, is realised in a minimal framework of composite dynamics 
 (addressing the SM hierarchy  problem) 
 and is discussed in detail in \cite{Molinaro:2015cwg, Molinaro:2016oix}. 
 In this scenario, a natural candidate for the diphoton resonance, with a mass around 750 GeV, is given by
 the pseudoscalar boson associated with the axial anomaly of the new composite dynamics. This state is the analogous of the $\eta^\prime$ in QCD. Its coupling  to the electromagnetic field is generated via  the topological sector of the theory, i.e. the gauged Wess-Zumino-Witten effective action, which originates from chiral anomalies related to global axial-vector currents. 
 
 It can be shown that, if the SM top mass is generated via four-fermion operators, as in minimal technicolor models, the coupling of $a$ to the top is unambiguously predicted, resulting in $y_t=1$, which  remarkably explains the wide-width resonance reported by ATLAS (right plot of the figure).
 A more detailed discussion is reported in \cite{Molinaro:2016oix}.
  
\newpage
\setcounter{figure}{0}
\setcounter{table}{0}

%% sessionday{monday} 

\subsection*{\hfil Making Sense of LHC Diboson and Diphoton Excesses \hfil}
\label{ssec:MakingSenseofLH}
\vspace*{10mm}

\begin{figure}[h!]
\centering
\includegraphics[width=0.45\textwidth]{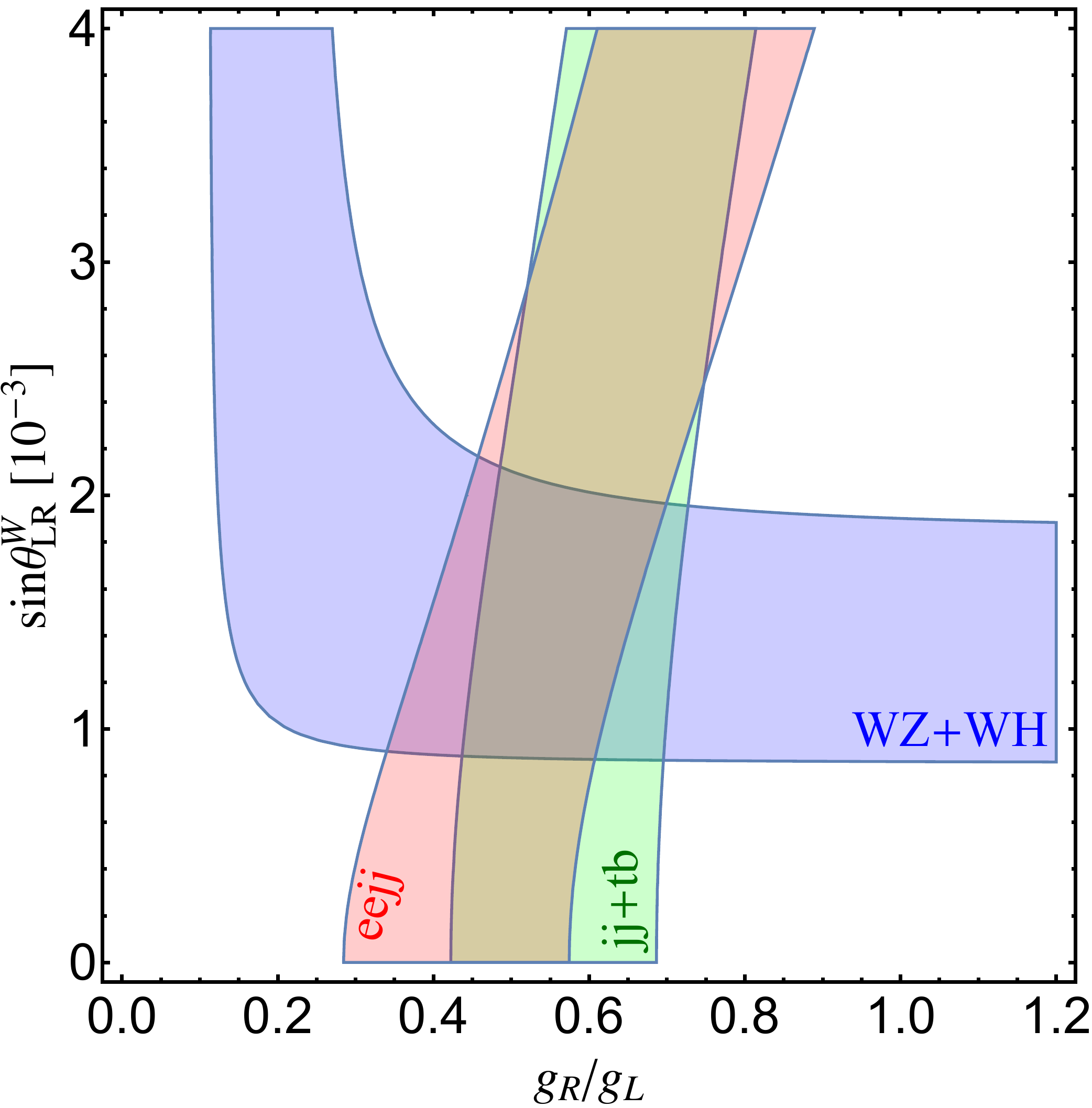}
\includegraphics[width=0.45\textwidth]{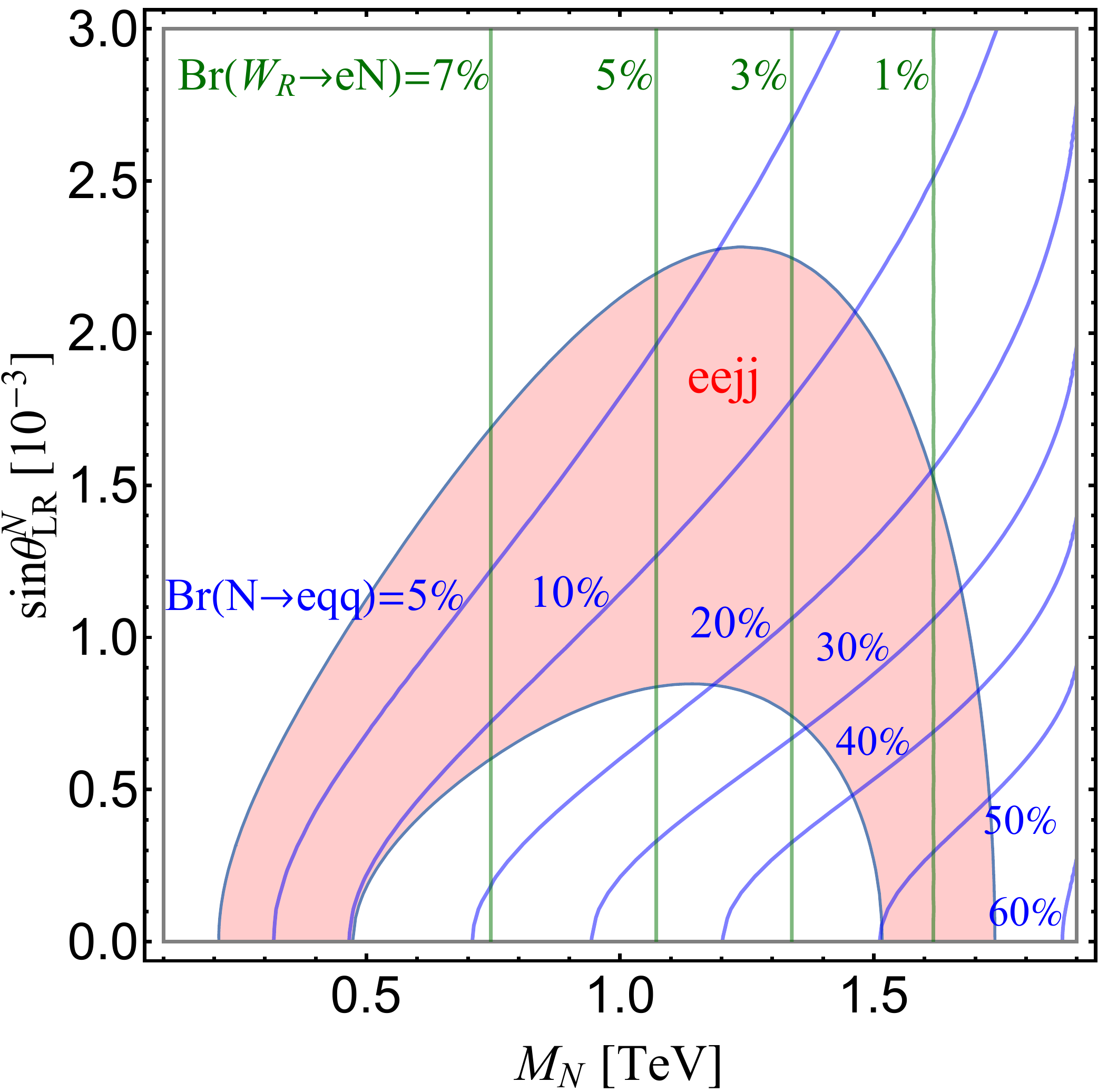}
\caption{Left: Fitting diboson excess in different channels (blue and green band) with the eejj excess (red band) in the plane of the left and right handed sector gauge couplings ratio and the mixing between left and right handed neutrino masses. Right: Fitting the $eejj$ (red band) excess in the ($M_N$-$\sin\theta_{LR}^N$) parameter plane.  The other parameters are chosen as $M_{W_R} = 1.9$~TeV, $g_R / g_L = 0.57$ and $\sin\theta_{LR}^W = 1.5\times 10^{-3}$.  The vertical green lines denote contours of constant $\text{Br}(W_R \to e N_R)$ and the diagonal blue lines of constant  $\text{Br}(N_R \to eqq)$ as denoted.}
\label{chasingexcess:LRSM}       \end{figure}

D.  Barducci$^{1}$, F.  Deppisch$^{2}$, L.  Graf$^{2}$, A.  Goudelis$^{3}$, \underline{S.  Kulkarni}$^{3}$, S.  Patra$^{4}$, W.  Rodejohann$^{5}$, N.  Sahu$^{6}$, U.  Sarkar$^{7}$, D.  Sengupta$^{8}$\vspace*{4mm} 
 \\ $^{1}$ LAPTh, Universit\'e Savoie Mont Blanc, CNRS, B.P. 110, F-74941 Annecy-le-Vieux, France \\  
$^{2}$ Department of Physics and Astronomy, University College London, London WC1E 6BT, United Kingdom \\  
$^{3}$ Institute of High Energy Physics, Austrian Academy of Sciences, Nikolsdorfergasse 18, 1050 Vienna, Austria \\  
$^{4}$ Center of Excellence in Theoretical and Mathematical Sciences, Siksha \textquoteleft O\textquoteright\, Anusandhan University, Bhubaneswar-751030, India \\  
$^{5}$ Max-Planck-Institut f\"ur Kernphysik, Saupfercheckweg 1, 69117 Heidelberg, Germany \\  
$^{6}$ Department of Physics, Indian Institute of Technology, Hyderabad, Yeddumailaram, 502205, Telengana, India \\  
$^{7}$ Physical Research Laboratory, Ahmedabad 380 009, India \\  
$^{8}$ Laboratoire de Physique Subatomique et de Cosmologie, Universit\'e Grenoble-Alpes, CNRS/IN2P3, 53 Avenue des Martyrs, F-38026 Grenoble, France \\  
\newline \noindent 
We explain the 2 TeV diboson excess and the eejj excess~\cite{Khachatryan:2014dka} observed by the CMS collaboration within the framework of Left-Right Symmetric Models(LRSM)~\cite{Deppisch:2015cua}. In these type of models the Standard Model gauge sector is extended by a right handed component with coupling $g_R$, resulting in right handed gauge bosons ($W_R, Z_R$) and right handed neutrinos ($N_R$), which mix with their Standard Model counterparts via mixing angles $\sin\theta_{LR}^W$ and $\sin\theta_{LR}^N$ respectively. The masses of the particles are denoted by $M_{W_R}$ fixed here at 2TeV and $M_N$, which is a free parameter of the theory in this analyis.  As shown in Fig.~\ref*{chasingexcess:LRSM} left panel, we demonstrate that it is possible to have fit all the excesses together pointing to a unique combination of ratio of left-right gauge couplings and mixing in the neutrino sector. In the right panel of the same figure we illustrate the viability of light right handed neutrinos ($ \sim 200$ GeV) with negligible mixing within the left and right handed sector capable of explaining all the excesses together.  We furthermore point out the possibility of existence of large left handed currents denoted by the contours $\text{Br}(N_R \to eqq)$ in Fig.~\ref*{chasingexcess:LRSM} (right panel).

In the second part of the proceedings, we ponder upon the question of the putative 750 GeV diphoton excess resonance $s$ serving as a portal to dark matter~\cite{Barducci:2015gtd}. As a concrete realisation of the portal we consider the following Lagrangian:
\begin{align}\label{chasingexcess:Lcpe}
{\cal{L}}_{\rm NP, CPE} & = \frac{1}{2} (\partial_\mu s)^2 - \frac{\mu_s^2}{2} s^2 + \frac{1}{2} \bar{\psi} (i \slashed{\partial}  - m_\psi) \psi - \frac{y_{\psi}}{2} s \bar{\psi} \psi \\ \nonumber
& - \frac{g_1^2}{4 \pi} \frac{1}{4 \Lambda_1} s ~ B_{\mu\nu} B^{\mu\nu} 
  - \frac{g_2^2}{4 \pi} \frac{1}{4 \Lambda_2} s ~ W_{\mu\nu} W^{\mu\nu} 
  - \frac{g_3^2}{4 \pi} \frac{1}{4 \Lambda_3} s ~ G_{\mu\nu} G^{\mu\nu}
\end{align}
where $B_{\mu\nu}$, $W_{\mu\nu}$ and $G_{\mu\nu}$ are the $U(1)_Y$, $SU(2)_L$ and $SU(3)_c$ field strength tensors respectively and $g_{1,2,3}$ are the corresponding SM coupling constants. The Lagrangian~\eqref{chasingexcess:Lcpe} actually corresponds to the case where $s$ is even under the charge-parity ($CP$) symmetry. It is also possible to write the corresponding CP - odd Lagrangian, we refer the reader to~\cite{Barducci:2015gtd} for more details. 

\begin{figure}[h!]
\centering
\includegraphics[width=0.5\textwidth]{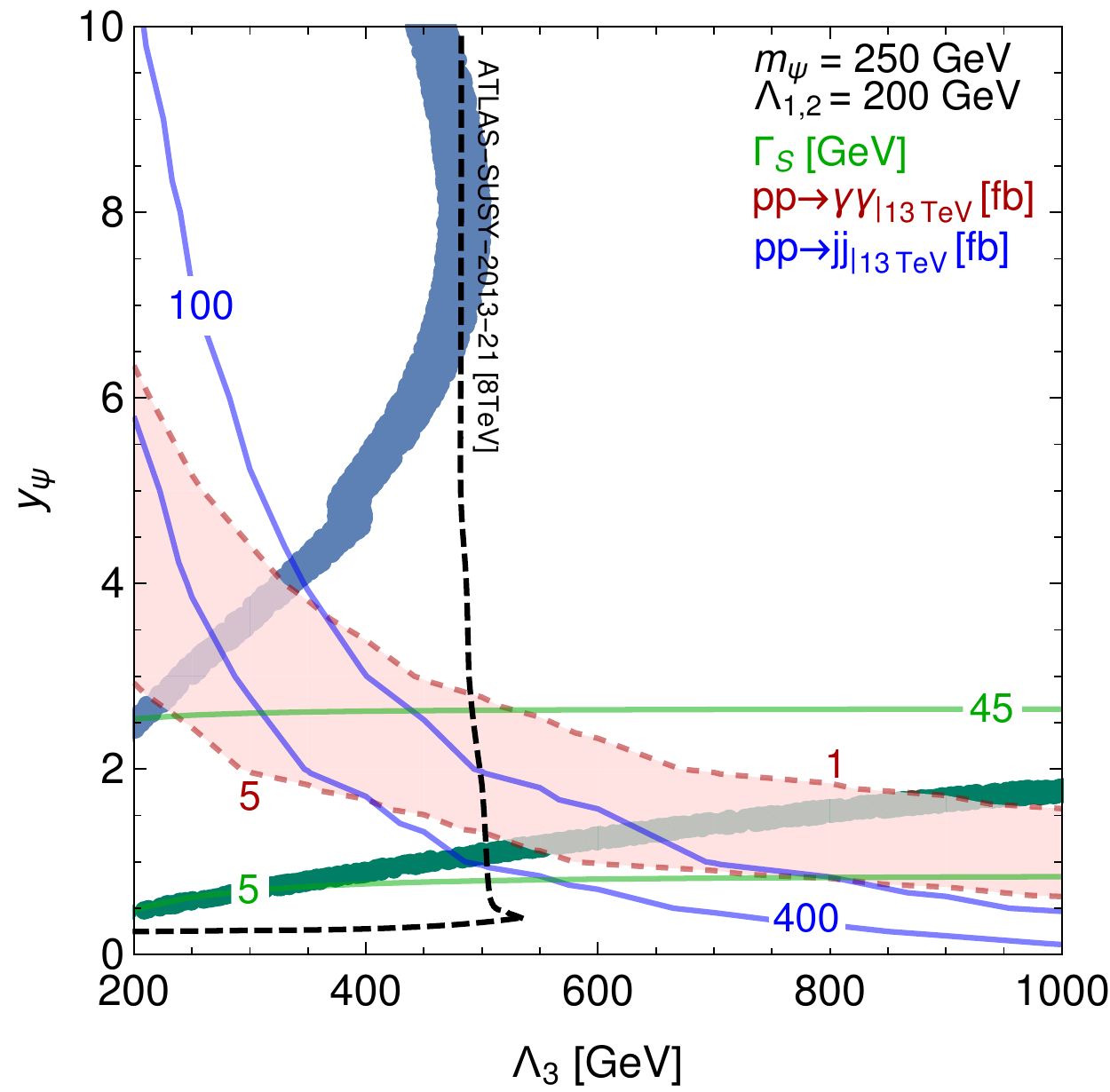}\hfill 
\includegraphics[width=0.5\textwidth]{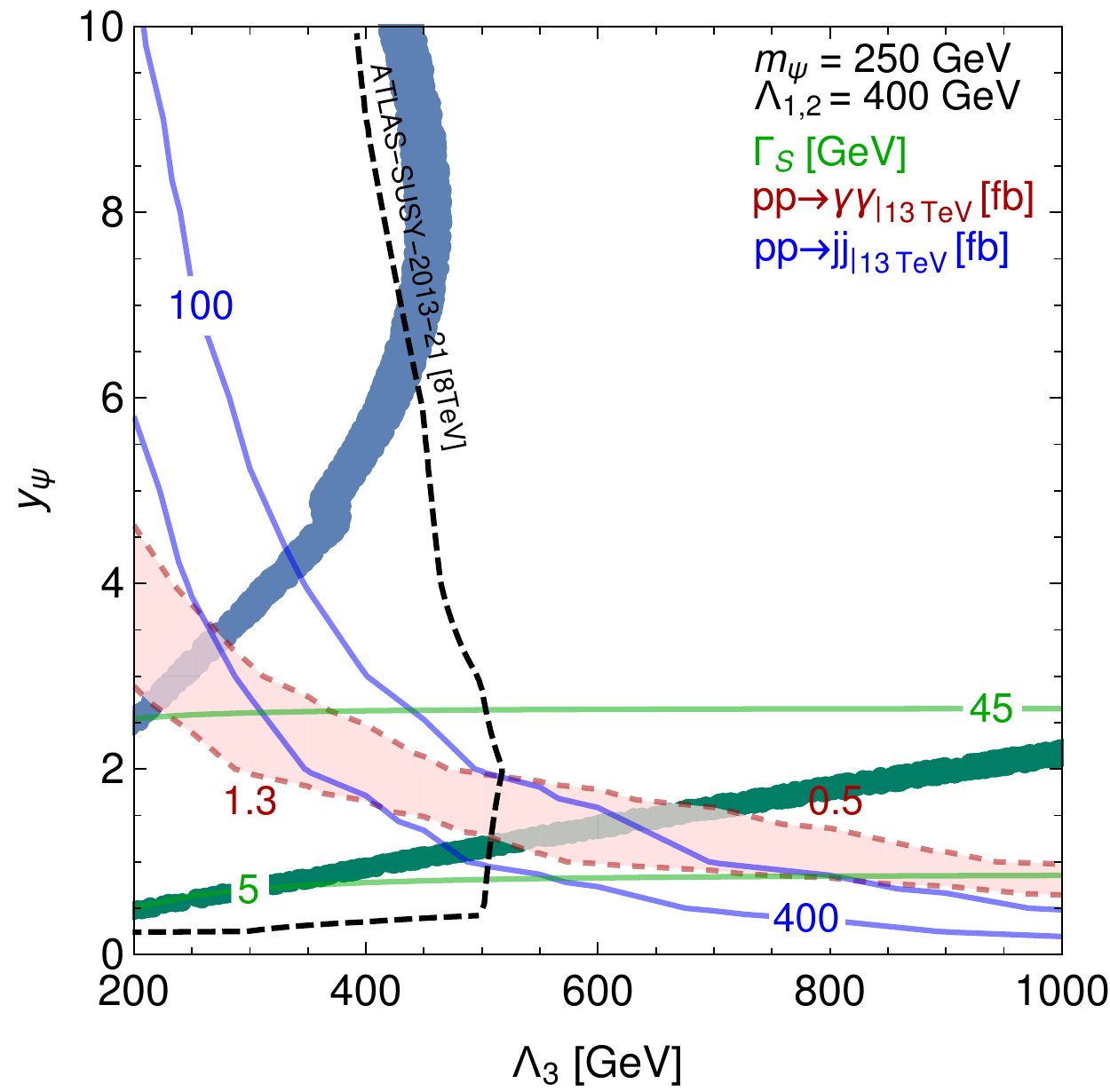} 
\caption[]{Predictions for $p p \rightarrow s \rightarrow \gamma \gamma$ (red band) and $p p \rightarrow s \rightarrow j j$ (blue contours) cross sections at $\sqrt{s} = 13$ TeV, overlaid with 8 TeV monojet constraints (black line) and the width of the resonance $s$ (green contours). The mass of the invisible fermion $\psi$ is fixed at $m_\psi = 250$~GeV and $\Lambda_{1,2} = 200, 400$ GeV in the top left, top right, bottom left and bottom right panels respectively. Monojet constraints are derived at 95\% C.L. The blue (green) band shows regions of parameter space compatible with the observed DM density for a scalar (pseudoscalar) mediator.\label{chasingexcess:monojet250}}
\end{figure}

We explore the constraints from monojet analysis on the production of dark matter at the LHC at 8 TeV. In Fig.~\ref*{chasingexcess:monojet250}, we demonstrate the constraints for a fixed choice of the dark matter mass $m_{\psi}$ and two different scales $\Lambda_1 =  \Lambda_2$. We show the predictions for the width of the resonance, diphoton and dijet cross-sections and highlight the regions of parameter space consistent with the observed relic density of dark matter. Finally, we overlay the contour arising from monojet constraints. The plot demonstrates that monojet searches put strong constraints on the possibility of diphoton resonance serving as a portal to dark matter.

\newpage
\setcounter{figure}{0}
\setcounter{table}{0}

%% sessionday{monday} 

\subsection*{\hfil Studies of Higgs Bosons Decaying to Fermions with CMS \hfil}
\label{ssec:StudiesofHiggsB}
\vspace*{10mm}

\date{}

\underline{D.  Salerno} (for the CMS Collaboration)\vspace*{4mm} 
 \\ University of Zürich \\  
\newline \noindent 
This note reviews the latest results of searches for a Higgs boson decaying to fermions, including both Standard Model (SM) searches from Run I at $\sqrt{s}=8$ TeV and Run II at $\sqrt{s}=13$ TeV and di-Higgs boson production from Run II. The measurements of the Higgs boson coupling to fermions from Run I of the LHC, both in final states and initial states are also presented.

The search for the Higgs boson decaying to two b quarks is performed at CMS in three production channels:
vector boson fusion (VBF), associative production with a vector boson (VH), and associative production with top quarks (ttH). In the case of VBF and VH, a regression method is used to reconstruct the invariant mass of the two b jets \cite{VBF,VHbb}. The final result is presented in terms of the signal strength, $\mu$, defined as the ratio of the cross section to the SM prediction, $\mu = \sigma / \sigma_{\mathrm{SM}}$. In Run I, the VBF analysis reported a best-fit value of $\mu = 2.8 ^{+1.6}_{-1.4}$, which corresponds to an observed (expected) significance of $2.2\sigma$ $(0.8\sigma)$ for a Higgs boson mass of 125 GeV, while the VH analysis published a best fit value of $\mu = 1.0 \pm 0.5$ for $m_H = 125$ GeV, corresponding to an observed (expected) significance of $2.1\sigma$ $(2.1\sigma)$.

The search for ttH production is performed in three broad decay channels: H$\to$bb, multileptons, and H$\to \gamma \gamma$. The ttH, H$\to$bb search uses two different discriminators for signal extraction, a Matrix Element Method and a multivariate Boosted Decision Tree (BDT). In Run I, these two discriminating techniques were used in two separate analyses \cite{BDT8,MEM8}. In Run II, they are used together in the same analysis \cite{ttHbb}. The Run II observed (expected) 95\% confidence level (CL) upper limit on the ttH, H$\to$bb signal strength is 2.6 (3.6) corresponding to a best-fit $\mu = -2.0 \pm 1.8$, which is a 1.7$\sigma$ fluctuation below the SM expectation.

The ttH, multileptons analysis includes three decay channels of the Higgs boson: ZZ, WW and $\tau \tau$. It uses a BDT for the final discriminant in both Run I \cite{multi8} and Run II \cite{multi}. The observed (expected) 95\% CL upper limit on $\mu$ is 3.3 (2.6) in Run II, corresponding to a best-fit value of $\mu = 0.6 ^{+1.4}_{-1.1}$.

The results of three different analyses are used in the ttH combination in Run II \cite{ttHbb,multi,gamma}. The combined result is an observed (expected) limit on $\mu$ of 2.1 (1.9), corresponding to a best-fit value of $\mu = 0.15 ^{+0.95}_{-0.81}$. This compares to an observed (expected) limit of 4.5 (2.7) in Run I, corresponding to a best-fit $\mu = 2.8 ^{+1.0}_{-0.9}$ \cite{ttHcomb8}.

The search for Higgs bosons decaying to two tau leptons employs a likelihood method for the reconstruction of the di-$\tau$ invariant mass \cite{tautau}. The search includes the three main Higgs boson production mechanisms: gluon fusion, VBF and VH. In Run I, the best-fit signal strength was $\mu = 0.78 \pm 0.27$ at $m_H = 125$ GeV, corresponding to an observed (expected) significance of $3.2 \sigma$ $(3.7 \sigma)$.

The analysis of the Higgs boson decaying to two muons searches for a peak in the di-$\mu$ invariant mass over a smoothly falling background \cite{mumu}. In Run I, the observed (expected) 95\% CL upper limit on the signal strength was 7.4 (6.5), corresponding to a best-fit value of $\mu = 0.8 ^{+3.5}_{-3.4}$.

\begin{figure}[h!]
\centering
\hspace*{-0.1\textwidth}
\includegraphics[width=0.44\textwidth]{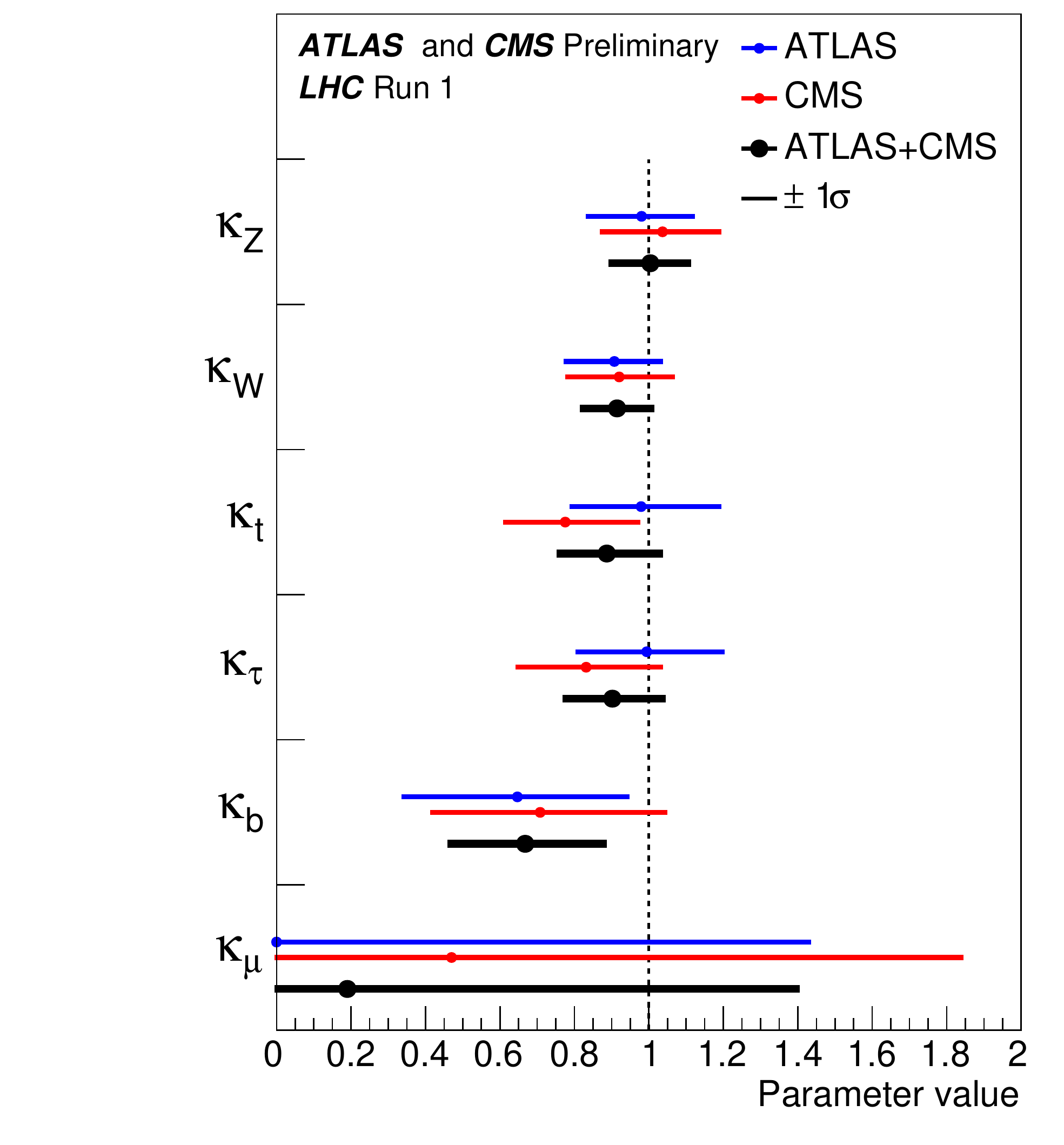}
\hspace*{0.06\textwidth}
\includegraphics[width=0.5\textwidth]{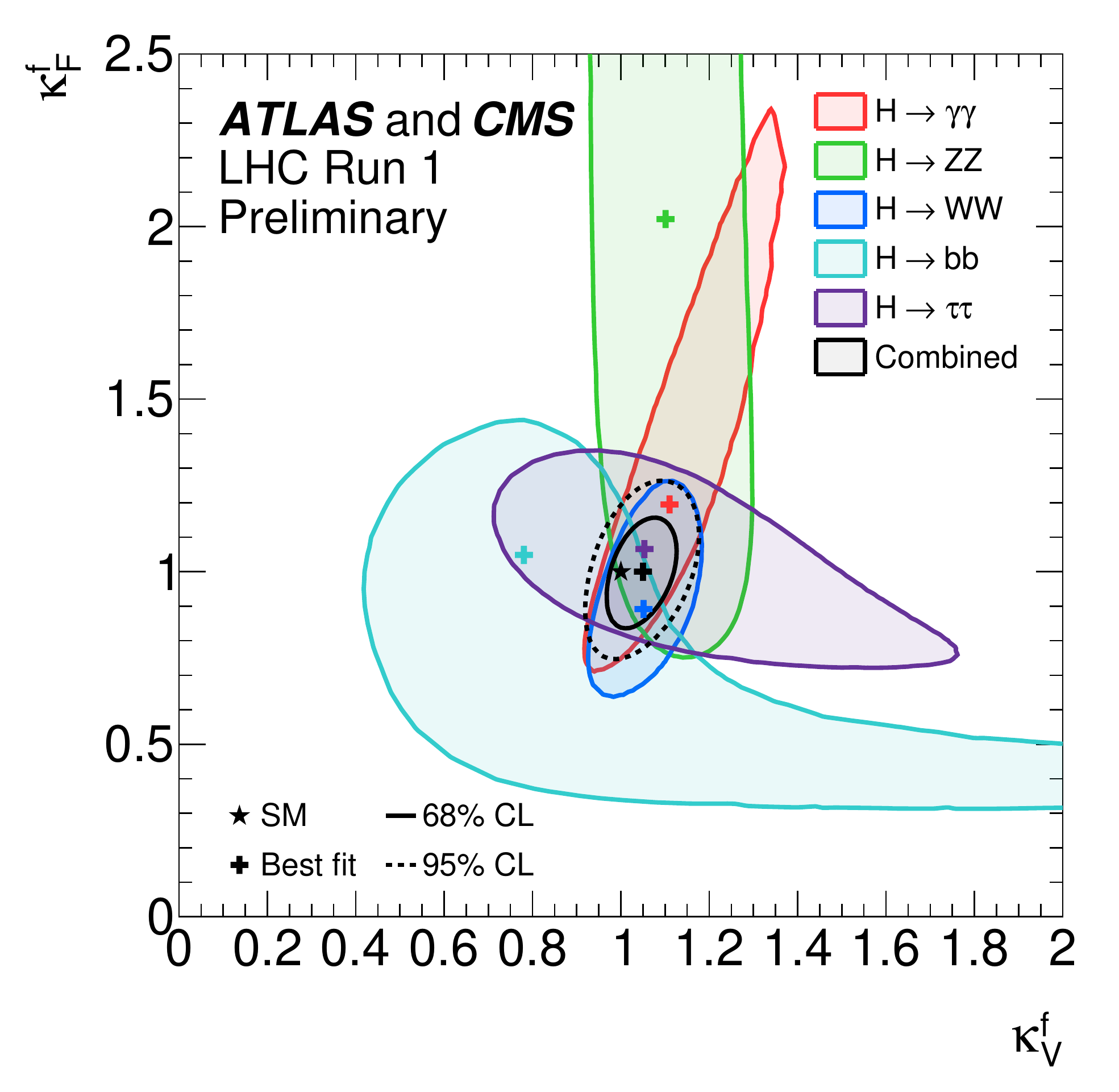}
\caption{Left: Observed coupling modifiers from measurements of all relevant processes for the ATLAS and CMS experiments and combined. Right: 68\% contours of the vector-fermion coupling modifiers, assuming all vector modifiers are equal and all fermion modifiers are equal, as determined from different decay channels and combined.}
\label{fig-1-salerno}       \end{figure}

The Run I results of all SM Higgs boson searches in ATLAS and CMS have been
combined to provide measurements of all accessible parameters \cite{ATLAScomb,
CMScomb}. The coupling modifiers (CMs) for bosons and fermions, defined as the
ratio of the relevant decay width or production cross section to the SM
prediction, are shown on the left of Fig.~\ref*{fig-1-salerno}. The 68\%
contours for the vector-fermion CMs are shown on the right of
Fig.~\ref*{fig-1-salerno}. The significance of ttH production, H$\to$bb and H$\to \tau \tau$ decays are shown in Tab.~\ref*{tab-1}. All observations in Run I are consistent with the SM predictions.

\begin{figure}[h!]
\centering
\includegraphics[width=0.8\textwidth]{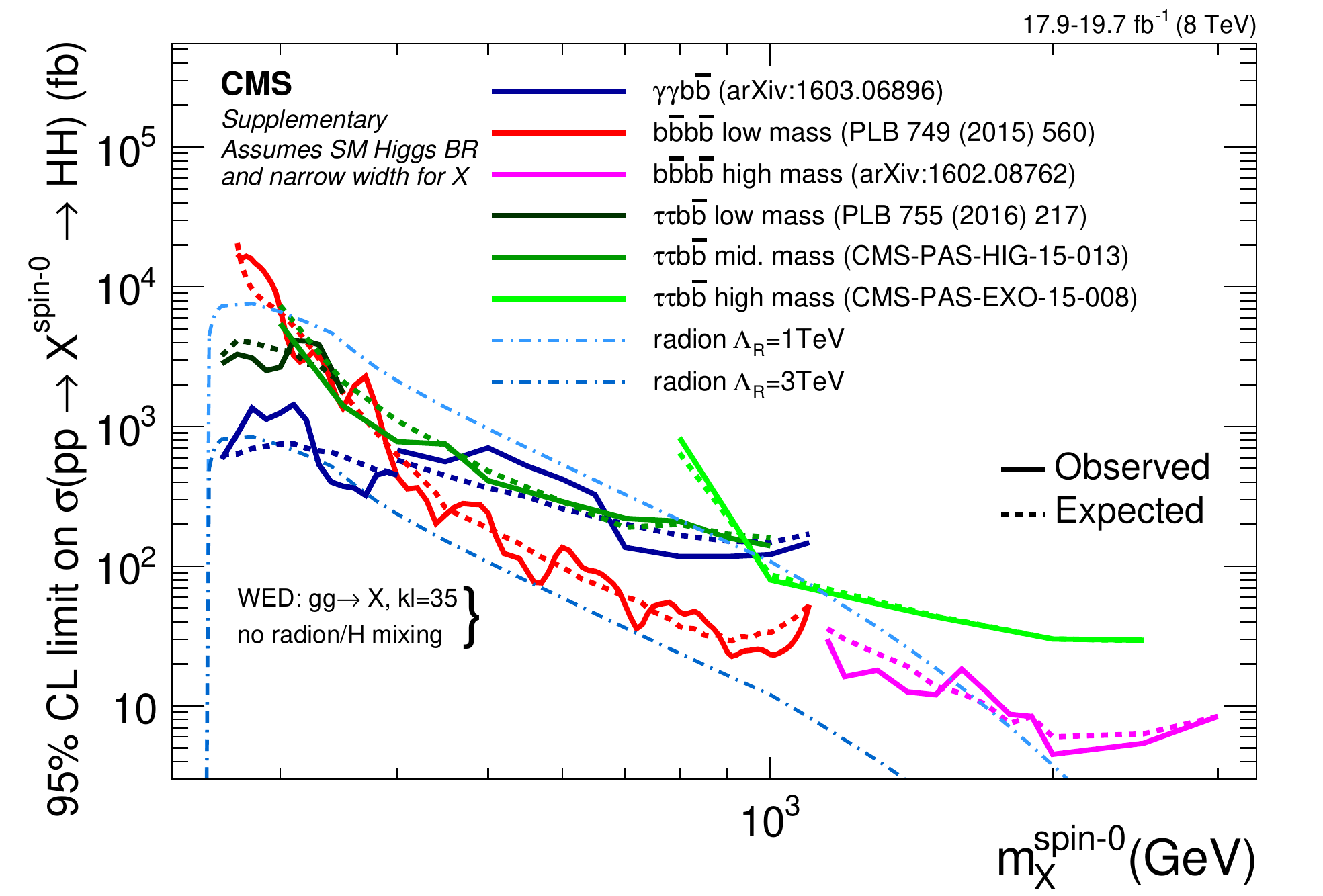}
\caption{Observed and expected 95\% CL upper limits on the cross section times branching fraction, $\sigma(\mathrm{gg} \to \mathrm{X}) \times B(\mathrm{X} \to \mathrm{HH})$, obtained by different analyses assuming a spin-0 hypothesis. Theory lines corresponding to warped extra dimension models with a radion are also shown.}
\label{fig-2}       \end{figure}

In addition to the searches for SM Higgs boson production presented above, this note also includes results of searches for  di-Higgs boson production in which at least one Higgs boson decays to fermions. Fig.~\ref*{fig-2} shows the Run I comparison of several di-Higgs boson resonant searches at CMS, assuming a spin-0 resonance.

The resonant di-Higgs boson search where both Higgs bosons decay to a bb pair is performed in two mass regions in Run II \cite{Xbbbb}. The result is presented in terms of 95\% CL upper limits on $\sigma(\mathrm{pp} \to \mathrm{X}) \times B(\mathrm{X} \to \mathrm{HH} \to \mathrm{bbbb})$ as a function of the resonant mass. For a RS1 KK-Gravition with $kL = 35$ and $k/M_{\mathrm{Pl}} = 0.1$, the observed (expected) exclusion region at 95\% CL is 775-850 (390-790) GeV.

The resonant HH search where one Higgs boson decays to a pair of b quarks and the other decays to a pair of W bosons is performed for the first time at the LHC in the Run II CMS analysis \cite{XbbWW}. The observed (expected) 95\% CL upper limits on the cross section times branching fraction for a spin-0 resonance with a mass between 500 and 900 GeV are between 174 and 101 (135 and 76) fb.

The resonant HH search where one Higgs boson decays to a bb pair and the other decays to a $\tau\tau$ pair employs a likelihood reconstruction for the di-$\tau$ invariant mass and a kinematic fit for the di-Higgs boson invariant mass \cite{Xbbtt}. The Run II observed (expected) 95\% CL upper limit on $\sigma(\mathrm{pp} \to \mathrm{X}) \times B(\mathrm{X} \to \mathrm{HH} \to \mathrm{bb}\tau\tau)$ for a spin-0 resonance with a mass of 300 GeV is 4.0 (3.6) pb.

The non-resonant HH search where one Higgs boson decays to a bb pair and the other decays to a $\tau\tau$ pair also uses a likelihood reconstruction for the di-$\tau$ invariant mass, but uses a simple reconstruction of the di-Higgs boson invariant mass \cite{HHbbtt}. The result is presented in terms of 95\% CL upper limits on the non-resonant HH cross section as a function of the ratio of the anomalous trilinear coupling to the SM coupling, $k_\lambda = \lambda_\mathrm{HHH} / \lambda^\mathrm{SM}_{\mathrm{HHH}}$. The Run II observed (expected) upper limit on $\sigma(\mathrm{pp} \to \mathrm{HH})$ for $k_\lambda = 1$ is 8.8 (7.2) pb.

\begin{table}[h!]
\centering
\begin{tabular}{lll}
\hline
Process  \hspace{15mm} & CMS  \hspace{15mm} & ATLAS+CMS \\\hline
ttH production & $3.6 \sigma$ $(1.3 \sigma)$ & $4.4 \sigma$ $(2.0 \sigma)$ \\
H$\to$bb decay & $2.0 \sigma$ $(2.5 \sigma)$ & $2.6 \sigma$ $(3.7 \sigma)$ \\
H$\to \tau \tau$ decay & $3.2 \sigma$ $(3.7 \sigma)$ & $5.5 \sigma$ $(5.0 \sigma)$ \\\hline
\end{tabular}
\caption{Run I observed (expected) significance for selected Higgs boson processes for CMS and ATLAS+CMS combined.}
\label{tab-1}       \end{table}

\newpage
\setcounter{figure}{0}
\setcounter{table}{0}

%% sessionday{monday} 

\subsection*{\hfil h(125) Boson Measurements in ATLAS: Run-1 Legacy and Early Run-2 Results \hfil}
\label{ssec:h125BosonMeasur}
\vspace*{10mm}

\underline{N.  Venturi}\vspace*{4mm} 
 \\ University of Toronto \\  
\newline \noindent 
Detailed studies of the Higgs boson have been performed by the ATLAS \cite{atlas_detector} Collaboration using $4.7$ fb$^{{-}1}$ of proton-proton collision data collected at a center of mass energy of $\sqrt{s} = 7$ TeV and $20.3$ fb$^{{-}1}$ at $\sqrt{s} = 8$ TeV (run-1).

In 2015, an improved mass measurement \cite{mass} in the $H \rightarrow  \gamma \gamma $ decay channel resulted in a value of $125.98 \pm 0.50$ GeV and in the $ H \rightarrow ZZ^{\ast} \rightarrow 4l $ channel with a value of $124.51 \pm 0.52$  GeV ($2.0 \sigma$ compatibility between the two results). 
Their combination was done with a profile likelihood ratio fit resulting in a mass of $m_H = 125.36 \pm  0.37 (\textrm{stat.}) \pm 0.18 (\textrm{syst.}) = 125.36 \pm 0.41$~GeV. 
Several alternative spin scenarios to the Standard Model (SM) hypothesis of $J^{P}  = 0^{+}$, including non-SM spin-0 and spin-2 models with universal and non-universal couplings to fermions and vector bosons, have been excluded at more than $99.9 \% $ confidence level (CL) \cite{spin_parity}.
Moreover, observations sensitive to non-SM tensor couplings are compatible with the SM predictions and constraints on non-SM couplings have been derived.
The combined yields from the $H \rightarrow \gamma \gamma, ZZ^{\ast}, WW^{\ast}, Z\gamma, b{\bar b}, \tau \tau, \mu \mu$ decay channels relative to their SM prediction is 
$1.18 \pm 0.10 (\textrm{stat.}) \pm 0.07 (\textrm{syst}.) ^{+0.08}_{{-}0.07} (\textrm{theo}.)$ \cite{coupling}.
For the various Higgs boson production modes, the gluon fusion Higgs production mode is confirmed with a significance exceeding $5\sigma$, $4.3\sigma$ evidence is found for vector boson fusion and production in association with a vector boson or a pair of top quarks are compatible with the small SM predictions.

The observed Higgs boson production and decay rates are also interpreted in a leading-order coupling framework.
A wide range of benchmark coupling models both with and without assumptions about the Higgs boson width and the SM particle content of loop processes have been tested.
The observed data are compatible with the SM expectation under a wide range of assumptions (p-values ranging from $29\%$ to $99\%$).
Upper limits on the total width of the Higgs boson are derived directly from fits to the mass spectra under the assumption of no interference with background processes.
The observed (expected) limit at $95\%$ CL in the $H \rightarrow \gamma \gamma$ channel is  5.0 (6.2) GeV and in the $H \rightarrow  ZZ^{\ast} \rightarrow 4l$ channel is 2.6 (6.2) GeV \cite{mass}.
Combining the on-shell and off-shell (for masses above $2m_Z$ and $2m_W$) measurements of the ZZ and WW final states and assuming the Higgs boson couplings are independent of the energy scale of the Higgs production, an upper limit at $95\%$ CL of the $\Gamma_H /  \Gamma_{SM}$ is found in the range $4.5  - 7.5$ depending on the variation of the $gg\rightarrow ZZ, VV$ K-factors \cite{width_indir}.

Combining the $ZZ^{\ast}$ and $\gamma \gamma$ decay channels the total Higgs boson production cross section at $8$ TeV is 
$ 33.0 \pm 5.3 (\textrm{stat.}) \pm 1.6 (\textrm{sys.})$ pb \cite{tot_diff_Xsec}.
Measurements of several differential cross sections have been performed. The total production cross section is larger, the Higgs is produced with larger transverse momentum and more associated jets than predicted by the most advanced SM calculations.
With the 3.2 fb$^{-1}$ of $pp$ collision data at $\sqrt{s} = 13$ TeV collected in 2015 (run-2) ATLAS measured the fiducial cross section in the $ZZ^{\ast}$ \cite{Prod_XSec_ZZ} ($0.6^{+1.3}_{{-}0.9}$ fb) and $\gamma \gamma$ \cite{Prod_Xsec_yy} ($52\pm 34 (\textrm{stat.}) ^{+21}_{{-}13} (\textrm{syst.}) \pm 3  (\textrm{lumi.})$ fb) Higgs decay channels.
The total Higgs cross section is $12^{+25}_{{-}16}$ pb with an observed upper limit of 69 pb at 95$\%$ CL from the $ZZ^{\ast}$ and $40^{+31}_{{-}28}$ pb with an observed upper limit of 106 pb at $95\%$ CL from the $\gamma \gamma$ decay channel.
Combining these results the total Higgs cross section at 13 TeV is $24^{+20}_{-17} (\textrm{stat.}) ^{+7}_{-3} (\textrm{syst.}) $ fb compatible at 1.3 $\sigma$ with the SM prediction, computed under the asymptotic approximation \cite{tot_Xsec}. 
Searches for resonant and non-resonant production of pairs of Higgs bosons have been performed with the 2015 data in the $b\bar{b}\gamma\gamma$ final state resulting in a $95\%$ CL observed limit of $3.9$ pb for non-resonant production \cite{di_higgs}.
In the search for a narrow $X \rightarrow hh$ resonance, the observed limit ranges from 7.0 pb to 4.0 pb for resonance masses between 275 and 400 GeV. Many other ATLAS results not presented here are available online.

\newpage
\setcounter{figure}{0}
\setcounter{table}{0}

%% sessionday{tuesday} 

\subsection*{\hfil Constraining Composite Higgs Models with Direct and Indirect Searches \hfil}
\label{ssec:ConstrainingCom}
\vspace*{10mm}

\underline{P.  Stangl}, C.  Niehoff, D.  Straub\vspace*{4mm} 
 \\ Excellence Cluster Universe, TUM, Boltzmannstr.~2, 85748~Garching, Germany \\  
\newline \noindent 
The naturalness problem of the Standard Model (SM) can be avoided in models where
the Higgs is a pseudo-Nambu-Goldstone boson
(pNGb) of a spontaneously broken global symmetry of some new strongly interacting
sector \cite{Kaplan:1983fs,Dugan:1984hq}. In the effective description, other bound
states like vector and fermion resonances are present in addition to the composite
Higgs and
elementary fields with the same quantum numbers as SM vector bosons and fermions.
A possibility for fulfilling flavour
constraints is to couple the elementary fermions to the Higgs only via a mixing
with the composite fermion resonances. Because the fermion mass eigenstates are
then a mixture of composite and elementary fields, this is called
partial compositeness \cite{Kaplan:1991dc}.

The goal of our work~\cite{Niehoff:2015iaa} is to use a singe framework to constrain the
parameter space of composite Higgs models (CHMs) containing a pNGb Higgs, partial
compositeness and a dynamically generated Higgs potential,
with indirect bounds (from correct radiative electroweak symmetry breaking,
Higgs physics, electroweak precision tests and flavour physics) and direct LHC bounds
on fermion and vector boson resonances.
This is challenging, because the parameter space of such CHMs does not
``factorize'' into a SM and a new physics (NP) part. This has two
main reasons:
First, due to partial compositeness, the fermion masses and the CKM parameters
are complicated functions of many of the model parameters. Second, the dynamically
generated Higgs potential depends on all gauge boson and fermion masses and
couplings.
Due to this complication, a brute-force scan of the parameter space is not
applicable. Our strategy is instead to construct a $\chi^2$-function from all the
observables we have implemented\footnote{In practice, we use a more general
$\chi^2$-function for being able to take into account correlations between
observables.}:
\begin{equation}
  \chi^2 (\vec \theta) \equiv
  \sum  \limits_{i\, \in\, \text{observables}}
  \left(\frac{ \mathcal{O}^\text{th}_i(\vec \theta) - \mathcal{O}^\text{exp}_i}{\sigma^\text{error}_i}\right)^2.
\end{equation}
Given a parameter point $\vec \theta$, this function measures ``how close'' the
theoretical predictions $\mathcal{O}^\text{th}_i(\vec \theta)$ are to the
experimental observations $\mathcal{O}^\text{exp}_i$. Numerically minimizing
the $\chi^2$-function then yields ``good'' points in the sense that for these
points the predictions are close to the observations.
However, depending on the specific model, the parameter space has a dimensionality
between 30 and 50 and thus the minimization is still technically challenging.
Therefore, we employ a numerical minimization consisting of four steps.
To cover as much of parameter space as possible, we start by generating many
different random initial points. Then, for each initial point we use the global
minimization package \texttt{[NLopt]}\cite{johnsonnlopt} to find a viable region
in parameter space, where the value of the $\chi^2$-function is relatively small.
Next, we use the Markov Chain Monte Carlo from the package
\texttt{[pypmc]}\cite{beaujean_2015_20045} to sample the low-$\chi^2$ regions
and to generate good points.
As the last step, we keep only the points satisfying each individual constraint at
the $3 \sigma$~level.

Using the new numerical method described above, it is possible to constrain
composite Higgs models by both direct and indirect bounds in a single
framework. As a first application, we have performed an analysis of the 4D
Composite Higgs Model (4DCHM) \cite{DeCurtis:2011yx}
with two sites, employing the minimal $SO(5)/SO(4)$ coset, custodial protection
of the $Z\, b_L \bar{b}_L$ coupling \cite{Agashe:2006at,Contino:2006qr} and four
different flavour symmetries \cite{Cacciapaglia:2007fw,Redi:2011zi,Barbieri:2012uh}.
While one of the flavour structures is shown to be strongly disfavoured, all
constraints can be passed for the other three cases.
More details on this analysis and our results can be found in \cite{Niehoff:2015iaa}.

\newpage
\setcounter{figure}{0}
\setcounter{table}{0}

%% sessionday{tuesday} 

\subsection*{\hfil Searches for Supersymmetry at CMS in Leptonic Final States with 13 TeV Data \hfil}
\label{ssec:SearchesforSupe}
\vspace*{10mm}

\begin{figure}[!htb]
\begin{center}
\begin{tabular}{cc}
\includegraphics[width=0.4\textwidth]{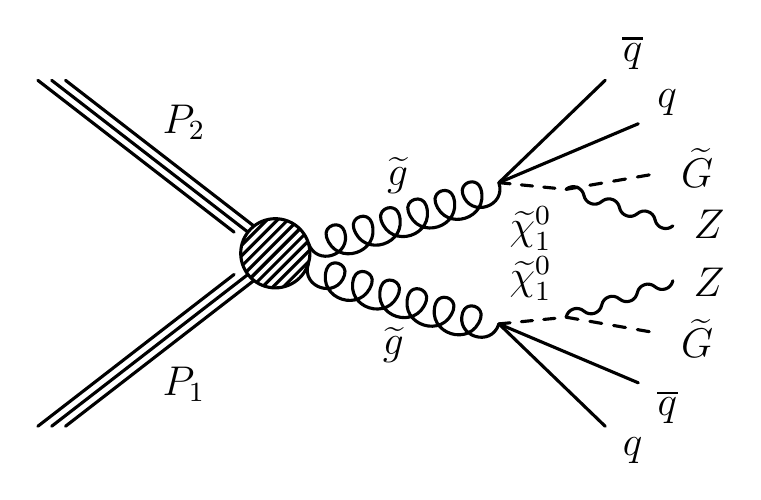} &
\includegraphics[width=0.4\textwidth]{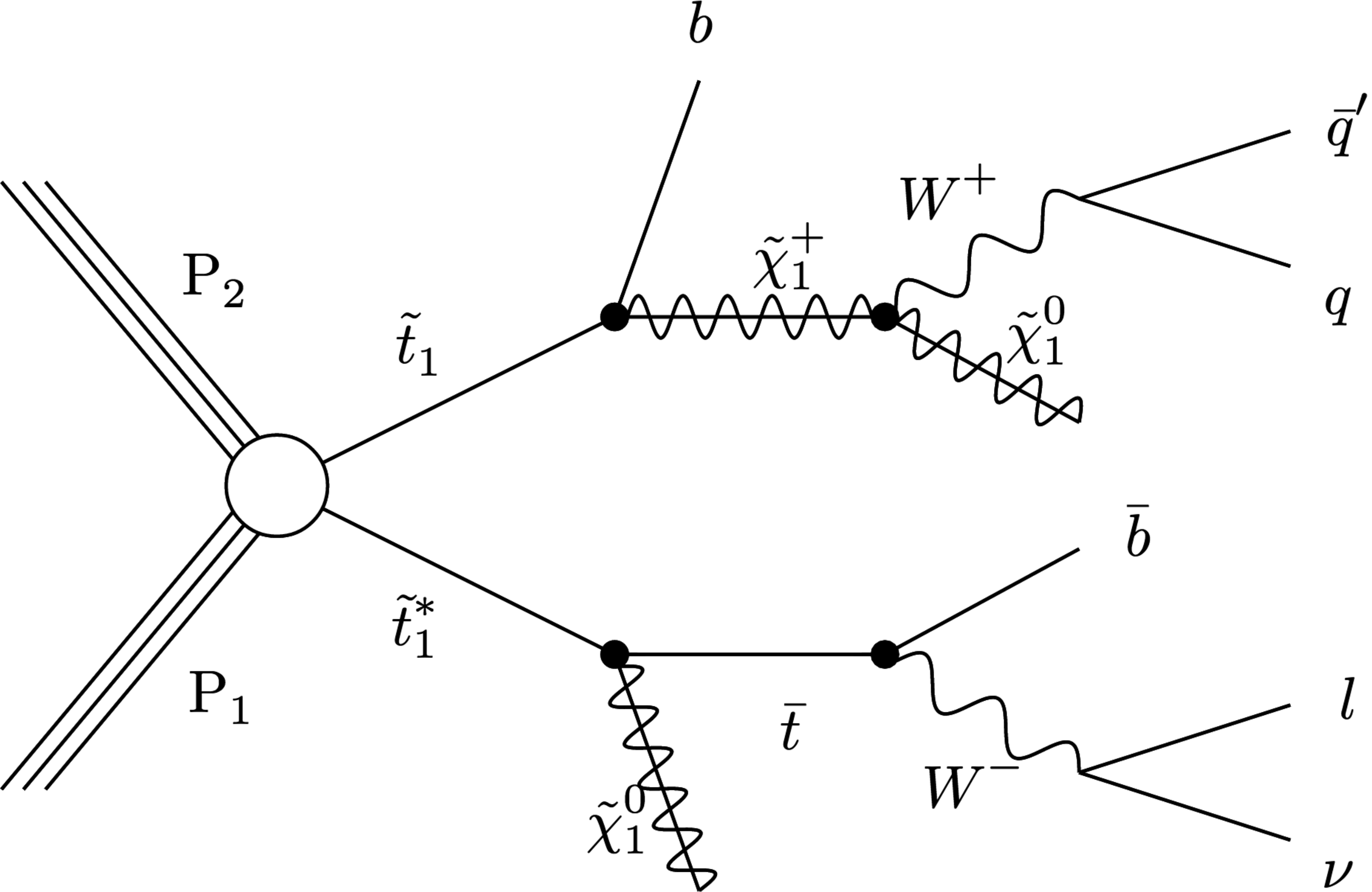}
\end{tabular}
\caption{
\label{fig:SMS}
Diagrams showing different SUSY processes which may contain leptons in the final state are shown in this figure.
On the left, gravitinos are pair-produced, where each eventually decays to a Z boson, two quarks, and a gravitino.
On the right, stops are pair-produced, where one leg eventually decays to a top and a neutralino where the top decays leptonically.
}
\end{center}
\end{figure}

\underline{C.  Welke} (on behalf of the CMS collaboration)\vspace*{4mm} 
 \\ University of California - San Diego, La Jolla, California \\  
\newline \noindent 
  The CMS SUSY program is very active in performing searches with the 13 TeV data including multiple analyses done in regions with leptonic final states.
  The results of these analyses are used to expand the reach of the searches done by CMS at 8 TeV and additionally to investigate two excesses seen in run I,
  namely a 2.6 sigma excess seen by CMS and a 3.0 sigma excess seen by ATLAS.
  These excesses were observed in two separate signal regions both having final states of at least 2 opposite sign same flavor leptons, jets, and \MET.
\\
Supersymmetry (SUSY)~\cite{SUSYPrimer} is an extension to the standard model that can be used to explain some open problems in physics such as
providing a solution the hierarchy problem as well as providing potential candidates for dark matter.
Searches for SUSY are performed by the CMS collaboration in a variety of final states,
including those with leptons (e or $\mu$).
Simplified models~\cite{sms} are used to interpret results of these searches, where two examples of a simplified model are shown in figure~\ref*{fig:SMS}.
On the left is a diagram of a gauge mediated SUSY breaking (GMSB) model with a massless gravitino as the lightest SUSY particle (LSP).
In this model, gluinos are pair produced, and each decays to a pair of quarks and a neutralino, which subsequently decays to a Z boson and a gravitino.
Another model showing direct stop production with one of the stops eventually
decaying to a top that decays leptonically is shown on the right.
The results presented in these proceedings focus on direct squark and gluino production.
In all of the following results, no significant deviation from the SM was observed,
and limits are set on the maximum value of the production cross section of different simplified models at the 95\% level using the CLs method~\cite{Junk:1999kv,Read:2002hq}.

The SUSY analyses performed by CMS share very similar object definitions for electrons, muons, and jets. 
The analyses are then grouped by the number of leptons in the final state,
and for analyses with at least two leptons,
they are categorized according to the charge of the two leptons with the
largest \welkept\
in same-sign, or opposite-sign final states.
Additionally, a large range of topologies are targeted by varying event variables to quantify the visible energy, invisible energy, and jet multiplicity and jet flavor content.
Visible energy in the event can be quantified using the variables \HT, and \MJ,
where \HT\ is the scalar sum of the jet \welkept\ in the event,
and \MJ\ is the sum of reclustered jets; this is explained in more detail in the \MJ\ analysis section.
Invisible energy is quantified using \MET, \MT, \MTtwo, \dphiwl, and \LT.
\MET\ is the magnitude of the vector sum of all the objects in the event, which is corrected to be consistent with the corrections applied to jets.
\MT\ is the transverse mass made using a lepton and the \MET\ vector.
\MTtwo\ is the stransverse mass, which is made from two visible objects and the \MET\ vector; This is explained in more detail in the 1-lepton stop search section.
\dphiwl\ is the difference in $\phi$ between the lepton and the W-candidate in the 1-lepton inclusive analysis.
\LT\ is defined as the vector sum of the lepton and the \MET\ in the 1-lepton inclusive analysis.
Jet multiplicity and flavor content are quantified using \njets\ and \nbtags.

In the analyses with exactly one lepton, three separate searches are performed.
In the first search, a search for direct stop pair production~\cite{1lstop2015}, the signal region is defined by having
exactly 1 lepton, at least 2 jets with at least one of the jets passing the criteria to be tagged as a b-jet, and \MET\ $>$ 250 GeV.
In order to further suppress backgrounds from ttbar, a cut is then made on $\mathrm{M_{T2W}}$,
which is an \MTtwo-like variable that uses the b-jets and leptons as visible objects, and has a kinematic cut-off at the top mass.
After the cuts are applied, the largest background comes from SM \ttbar\ to dilepton where one of the leptons is not reconstructed leading to increased \MET\ in the event.
The results of this analysis increase the sensitivity to models where stop pairs are produced directly to exclude scenarios where the stop mass is up to 750 GeV,
which is an improvement on the previous result which was sensitive to models with a stop mass up to 650 GeV.

The next analysis is a search targeting final states with exactly one lepton~\cite{1lmj2015}, and many jets from hadronic top decay.
In this analysis, events with exactly 1 lepton in the final state, and at least one b-tagged jet are selected.
The jets in the event, which are made using the anti-kt clustering algorithm with a cone size of $\mathrm{R=0.4}$, are reclustered to form large-R jets ($\mathrm{R=1.2}$).
The masses of all the large-R jets in the event are then summed to form the variable \MJ\
which tends to be large for events where large mass particles decay hadronically, for example in SUSY in final states with many top quarks.
In order to reduce backgrounds where a lepton comes from a W not from a top decay, a cut is made of \MT\ $>$ 140 GeV.
The results of this analysis are interpreted in the context of a SUSY model where gluinos are pair-produced then each decays to a \ttbar\ pair and a stable LSP,
and gluinos up to a mass of 1600 GeV are excluded.

The final analysis with exactly one lepton~\cite{1lincl2015} is an inclusive search
which uses the \dphiwl\ and \LT\ variables to suppress SM backgrounds which mostly consist of \ttbar\ and W+jets.
The search is binned in the \HT, \njets, \nbtags\ in order for the search to be inclusive as possible.
The results of the search are interpreted many SUSY scenarios, for example the same model as the \MJ\ analysis,
and this analysis is seen to have similar sensitivity when interpreting the results within the context of this simplified model.

The rest of the analyses all require at least 2 leptons in the final state~\cite{ssdilep2015}.
The first analysis in this category is a search for SUSY in a final state with two same sign leptons.
The baseline selection requires at least two same sign leptons with \welkept\ $>$ 10-15 GeV depending on the trigger, \MET\ $>$ 50 GeV.
The analysis is then binned in \HT, \njets\ and \nbtags.
The largest background in this analysis comes from fake lepton signatures in the detector,
and a data-driven method was developed to predict this background which ends up with an uncertainty of about 40\%.

The next analysis is a search in events with three or more leptons~\cite{multilep2015}.
This analysis is binned in \HT, \njets\ and \nbtags,
and the largest backgrounds in the signal region comes from either SM WZ associated production,
or non-prompt leptons passing all the lepton ID requirements.

The final analysis is a search in final states with at least two opposite-sign same-flavor leptons and at least two jets~\cite{osdilep2015}.
A separate search is performed in events where \mll\ of the two highest \welkept\ leptons is between 81-101 GeV (the on-Z region),
as well as an inclusive search with \mll\ $>$ 20 GeV (the ``edge'' region).
The main backgrounds in this analysis are grouped into three categories, Z+jets, flavor-symmetric (FS), and other SM processes.
Z+jets with no real \MET\ is defined as any background with Z+jets, but no \MET\ from invisble particles, such as neutrinos.
The FS background is defined as any process where ee or $\mu\mu$ pairs are produced at equal rates as e$\mu$,
such as \ttbar, WW, and single top.
Other SM processes include Z+jets with real \MET, multiboson production, and ttV.
In order to predict the Z+jets background, a fully data-driven method was developed where the \MET\ shape is predicted using a $\gamma$+jets control region in data.
The $\gamma$+jets sample is reweighted such that the shape of the \welkept\ of
the photon in the event matches that of the Z boson \welkept\ of the events that pass the baseline selection.
This background is then normalized in a region where Z+jets is the dominant SM background, namely \MET\ $M<$ 50 GeV.
The FS background is predicted using a sample of e$\mu$ data events which is corrected for the difference in reconstruction and trigger efficiencies between electrons and muons.
The largest systematic uncertainty on this background is about 5\% and is mostly due to the trigger uncertainty.
Other SM processes are predicted using simulated monte-carlo datasets which are validated in 3 and 4 lepton control regions.
A conservative systematic uncertainty of 50\% is assigned to these backgrounds.
In order to reduce these backgrounds, cuts are made on \njets, \HT, and \MET.

\begin{figure}[!htb]
\begin{center}
\begin{tabular}{cc}
\includegraphics[width=0.4\textwidth]{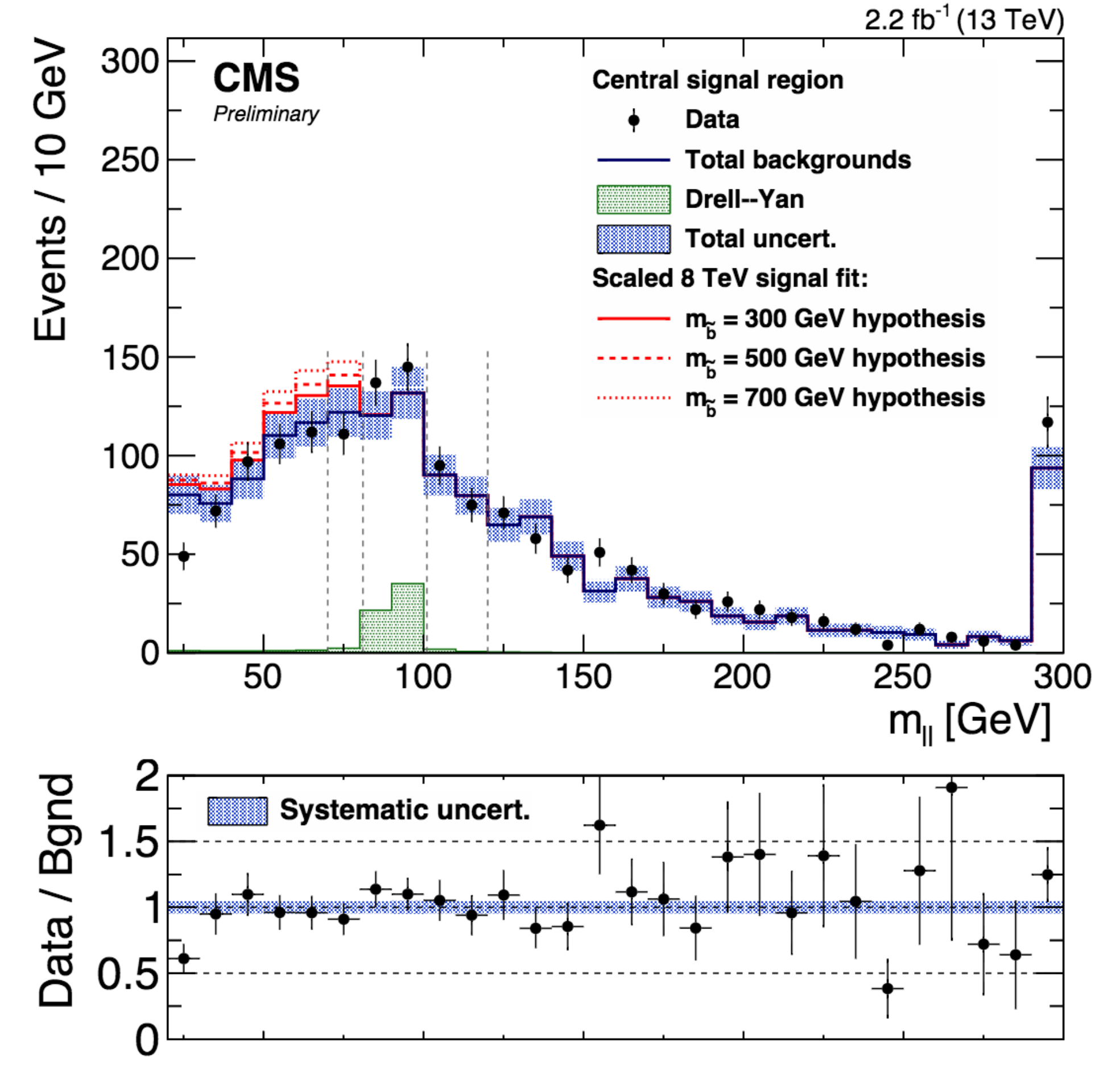} &
\includegraphics[width=0.4\textwidth]{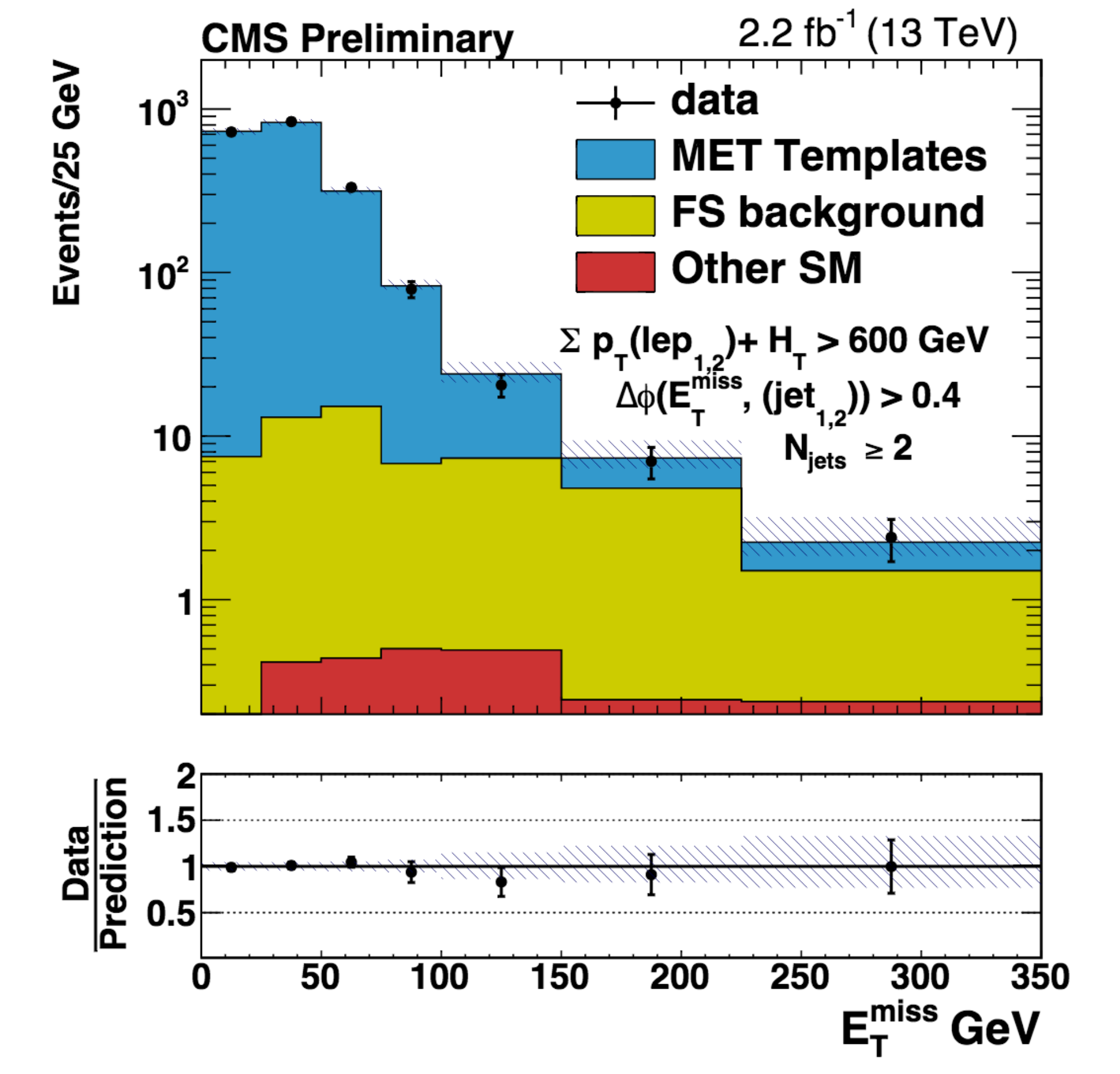}
\end{tabular}
\caption{
\label{fig:2leposresults}
Observed yields and background predictions are shown for the ``edge'' signal regions (left) and ATLAS-like on Z signal region (right).
}
\end{center}
\end{figure}

Two excesses were observed in separate \mll\ regions in run I, one by CMS~\cite{CMSedge} and another seen by ATLAS~\cite{ATLASZPAPER}.
CMS saw an excess in events with \mll\ between 20-70 GeV with a significance of 2.6 $\sigma$ while ATLAS saw no excess in a similar region.
ATLAS observed a 3 $\sigma$ excess in events with two leptons having \mll\ between 81-101, large \HT, and large \MET.
In the CMS analysis performed in run I, a similar signal region was explored and showed no significant deviation from the SM prediction.
In the analysis done by CMS in 2015, and an additional signal region was added in order to search where ATLAS saw an excess at 8 TeV.
No significant excess was seen in either region where an excess was reported in 8 TeV, and the result of this analysis can be seen in figure~\ref*{fig:2leposresults}.

Overall, the SUSY program is well underway in run II, and many exciting results have already been released including many in leptonic final states.

\bibdata{refs}
\bibliographystyle{ieeetr.bst}
\bibliography{refs}

\newpage
\setcounter{figure}{0}
\setcounter{table}{0}

%% sessionday{tuesday} 

\subsection*{\hfil Low Scale Supersymmetry: R.I.P. or Resurrection?  \hfil}
\label{ssec:LowScaleSupersy}
\vspace*{10mm}

\underline{S.  Sekmen} (on behalf of the CMS and ATLAS collaborations)\vspace*{4mm} 
 \\ Kyungpook National University, Daegu, South Korea \\  
\newline \noindent 
Despite all our efforts, supersymmetry still has not revealed itself in any of our detectors.  Yet, we fervently continue the quest, since many physicists still see SUSY as the most legitimate heir to the Standard Model.  This persistent claim is based on the many simultaneous virtues of SUSY: it offers a generic theoretical framework at high scales,  which reproduces SM at low scales; it allows to incorporate gravity; it tames the fine tuning in the corrections to the Higgs mass by introducing partner states with different spin; it unifies gauge couplings at the GUT scale; and it readily offers candidates for dark matter.  However, nowadays SUSY is challenged by many experimental observations, and each challenge bites off a certain piece from its parameter space.  Here, I briefly review the observations that challenge SUSY, and how they effect its feasibility and realization.

The most direct challenge comes from direct SUSY searches at the LHC.  7, 8 and current 13\,TeV searches especially target Natural SUSY, which accommodates O(100\,GeV) Higgsinos, $\sim$TeV stops and $\sim$few TeV gluinos to stabilise the EW scale.  More than 50 analyses in ATLAS and CMS explored a wide range of final states~\cite{lhcsusy}.  When interpreted in terms of simplified model spectra, the inclusive searches exclude gluinos up to $\sim$1.8\,TeV and 1st/2nd generation squarks up to $\sim$1.2\,TeV; 3rd generation searches exclude stops up to $\sim$800\,GeV and sbottoms up to $\sim$900\,GeV; and electroweak gaugino searches exclude light neutralinos and charginos up to $\sim$700\,GeV.  These are complemented by R-parity violating SUSY searches and non-promptly decaying particle searches.  The constraints become looser when results are interpreted in terms of full models such as the phenomenological MSSM~\cite{lhcpmssm}.  Complex decay structures allow cases with much lighter sparticles to survive.  As a result, although they introduce some constraints on the Natural SUSY scenarios, direct LHC searches do not yet constitute a serious threat to SUSY.  

Higgs measurements also have considerable impact on SUSY.  MSSM requires heavy stops to achieve the relatively high measured Higgs mass of 126\,GeV, which creates a tension with Naturalness.  However this tension could be mended by either modifying the definition of fine tuning, or by alternatively considering NMSSM, which introduces a singlet field that generates an effective $\mu$ term and thus naturally augments the Higgs mass.  Searches for heavier SUSY Higgses, or light NMSSM Higgses have achieved no discovery, yet they are far from being fully excluded.

Electroweak precision measurements, especially those on $g - 2$ and FCNC can further constrain SUSY.  So far, SUSY, even the MSSM, is consistent with such measurements, including the recent $b \rightarrow sll$ results.  However $b \rightarrow s\gamma$ imposes a slight tension on Naturalness, as it favours heavier stops.

Next comes the implication of dark matter searches on SUSY.  As LHC drives the sparticle masses higher, the lightest neutralino mass is shifted upwards, and consequently, dark matter relic density is shifted downwards.  Current direct measurements of spin-dependent and spin-independent neutralino-nucleon scattering cross section also exclude a small part of SUSY space, though the bigger impact so far comes from the LHC searches.  However next generation of direct detection experiments like Xenon-1ton, LZ(10) and DARWIN will push the probed cross sections down by a factor of $\sim 10^{-2}$ to $10^{-4}$ and test the bulk of the MSSM phase space.  These will be complemented by the indirect detection experiments which will probe high neutralino masses.  If all experiments disfavour neutralinos, axions can stand as DM candidates, whose feasibility will be tested by the near future experiments such as ADMX.

\begin{figure}[h!]
\centering
$\vcenter{\hbox{\includegraphics[width=0.4375\textwidth]{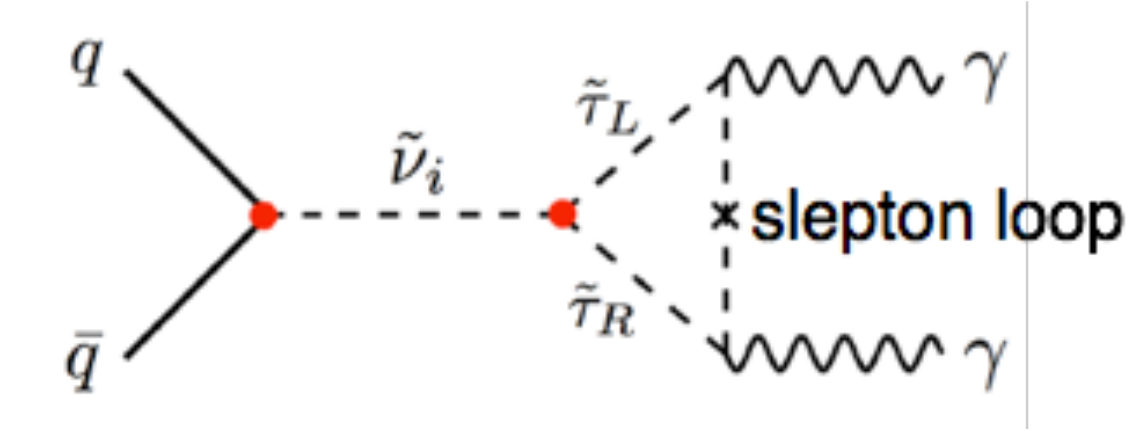}}}
\vcenter{\hbox{\includegraphics[width=0.5625\textwidth]{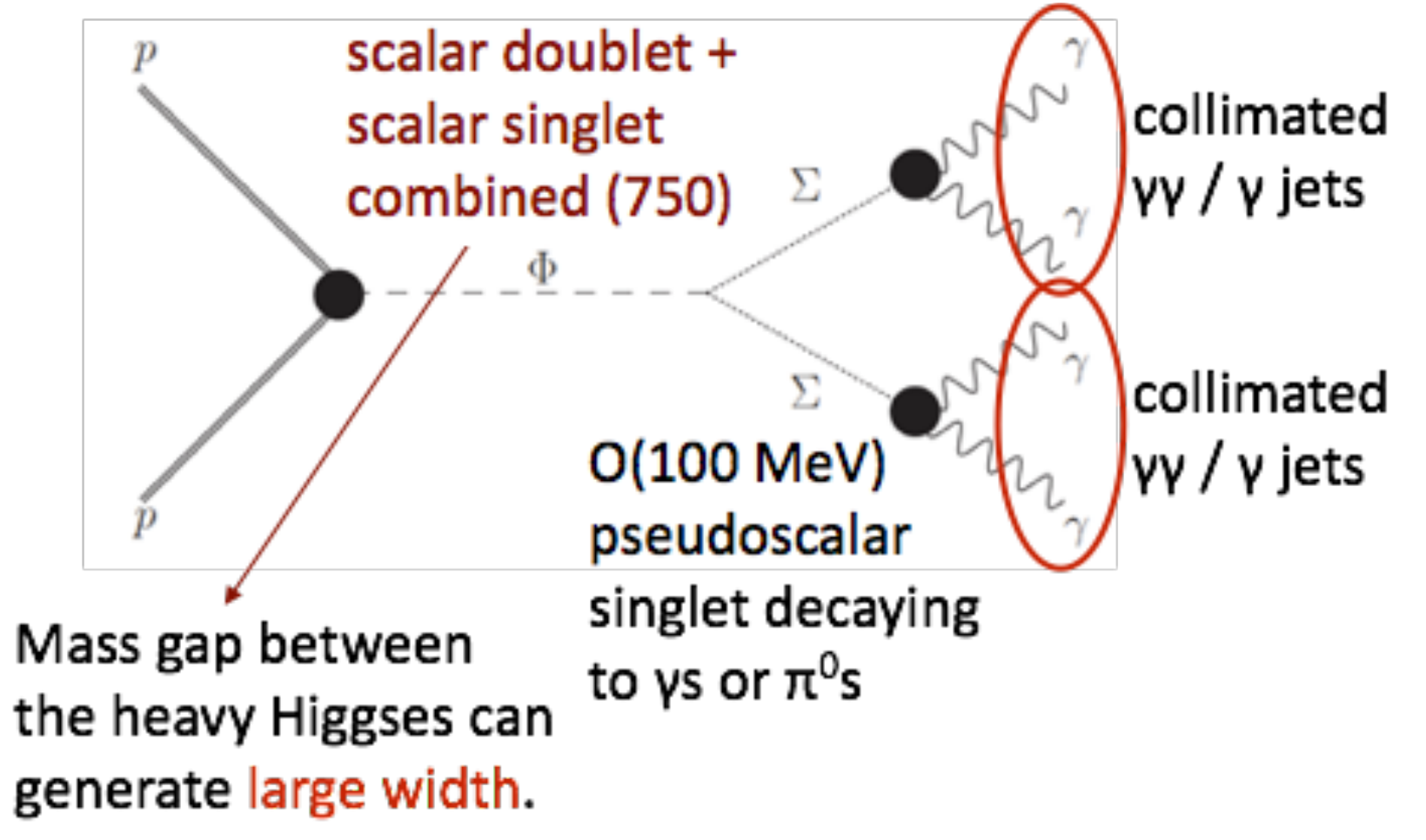}}}
\vspace*{6mm}
\newline
\vcenter{\hbox{\includegraphics[width=0.5625\textwidth]{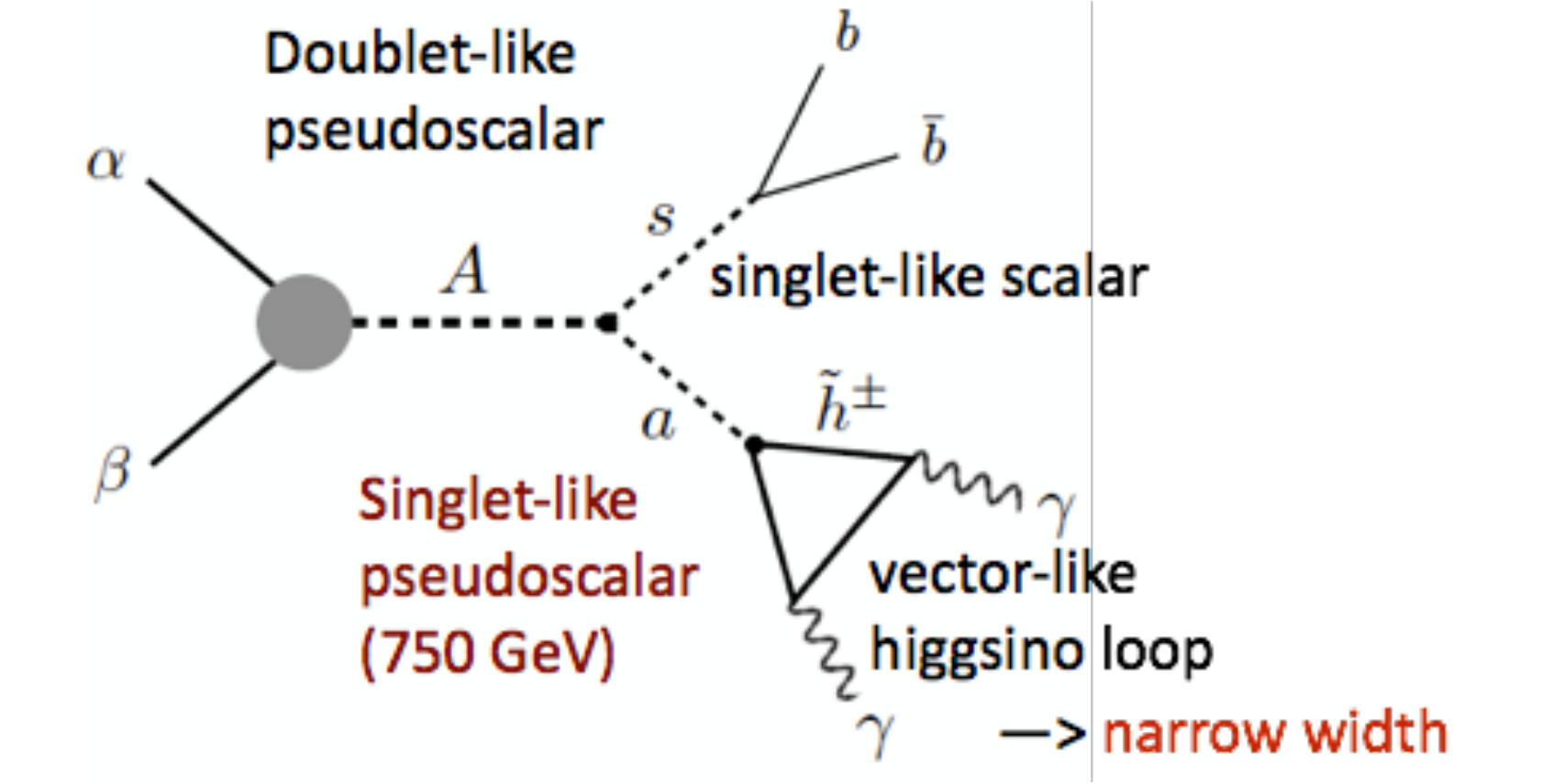}}}$
\caption{Example diagrams realizing diphoton excess in SUSY: sneutrinos in RPV
MSSM (left), NMSSM (right and bottom).}
\label{fig:diphotonsol}       \end{figure}

However, the most intriguing challenge on SUSY comes from the recent diphoton
excess observed by ATLAS and CMS at 750\,GeV~\cite{lhcdiphoton}.  Heavy MSSM
Higgs states fail to reproduce the observed $\gamma\gamma$ rate, which may
signal the end to the MSSM story.  However, the funeral can be postponed by
ideas that enhance the $A \rightarrow \gamma \gamma$ rate by presence of
sfermions with mass $\sim0.5 A$; or propose a sneutrino resonance decaying to
$\gamma \gamma$ through a slepton loop in the R parity violating MSSM (see
Figure~\ref*{fig:diphotonsol} left)~\cite{diphotonsol1}.   NMSSM can also offer
a consistent resonance, and adjust the $gg \rightarrow X \rightarrow
\gamma\gamma$ rate via its additional particle content (see
Figure~\ref*{fig:diphotonsol} right and bottom)~\cite{diphotonsol2}.  Other resonance alternatives are (s)goldstinos responsible for SUSY breaking at low scales; singlet fields in deflected anomaly mediation models that fix the tachyonic slepton problem in minimal AMSB; sbino, the real component of the scalar partner of the field giving Dirac mass to bino in models with Dirac gauginos; or scalar components of superfield systems in MSSM with extra lepton and baryon symmetries~\cite{diphotonsol3}.  Upcoming data has yet to confirm the diphoton exces, but in case it is real, studies to characterize it are already underway~\cite{diphotonchar}.  

SUSY has valiantly endured all observational challanges so far.  Although reconciliation with the diphoton excess requires some creative thinking, the search will still go on with new data and innovative search methods.

\newpage
\setcounter{figure}{0}
\setcounter{table}{0}

%% sessionday{tuesday} 

\subsection*{\hfil Searches for New Physics with Bosons at the ATLAS Detector in LHC Run II \hfil}
\label{ssec:SearchesforNewP}
\vspace*{10mm}

\underline{K.  Iordanidou} (on behalf of the ATLAS Collaboration)\vspace*{4mm} 
 \\ Nevis Laboratory, Columbia University, Irvington NY, USA  \\  
\newline \noindent 
LHC Run II started in 2015 at center of mass energy of $\sqrt{s} = 13$ TeV and the ATLAS detector~\cite{atlas} recorded $3.2$ fb$^{-1}$ of data. This talk focused on early Run II high mass searches with dibosons, $VV$, $Vh$, $\gamma \gamma$, $Z \gamma$ and $hh$, where $V$ denotes an intermediate vector boson and $h$ the Higgs boson.

The basic methodology is based on the search for a mass bump above the smoothly falling background. Since the final decay products can be highly boosted novel techniques had to be developed for boosted objects tagging, which exploits the substructure inside large-R jets with distance parameter of $R = 1.0$ (denoted as $J$) and make use of the jet mass. The resolved regimes, where the two quarks from the boson decays are reconstructed as two $R = 0.4$ jets, are also explored since they are not yet fully excluded.

Searches for $VV$ resonances are performed in the final states of $\ell \ell
J$, $\nu \nu J$, $\ell \nu J$, $JJ$. The results are interpreted in the
context of Heavy Vector Triplet (HVT), Randall-Sundrum Graviton (RS G$^*$) and
scalar large and narrow width resonances. HVT benchmark model A with coupling
constant $g_{V} = 1$ is excluded at $95\%$ confidence level (CL) from the $\nu
\nu J$ channel for $WZ$ masses up to $1.6$ TeV~\cite{VVvvJ}, from the $\ell
\nu J$ channel for $WZ$ and $WW$ masses below $1250$~GeV~\cite{VVlvJ} and from
the $JJ$ channel in the $WZ$ mass region $1.38 - 1.6$ TeV~\cite{VVJJ}.
Additionally, RS G$^{*} \to WW~or~WZ$ resonances with masses below $1060$ GeV
are excluded at $95\%$ CL from the $\ell \nu J$ channel~\cite{VVlvJ}. Upper
limits are placed for the various productions~\cite{VVvvJ, VVlvJ, VVJJ,
VVllJ}.

Searches for heavy $Vh$ ($h \to b \bar{b}$) resonances are very similar to
boosted $VV$ searches. The main difference is the large-R jet mass range and
tagging which requires at least one b-tagged $R = 0.2$ jet. HVT and  $A \to
Zh$ (for the neutral only final states) interpretations are provided and no
significant excess is observed. Specifically, HVT benchmark model A with
coupling constant $g_{V} = 1$ is excluded for $m_{V'^{0}h} < 1480$ GeV and
$m_{V'^{\pm}h} < 1490$ GeV, while model B with coupling constant $g_{V} = 3$
is excluded for  $m_{V'^{0}h} < 1760$ GeV and $m_{V'^{\pm}h} < 1620$~GeV at
$95\%$ CL~\cite{Vh}.

The search for a heavy di-photon resonances can be interpreted as a spin-$0$ particle or RS G$^*$ (spin-$2$). The largest deviation from the background is seen around $750$ GeV and corresponds to $2.0~\sigma$ and $1.8~\sigma$ for the scalar and RS G$^*$ searches respectively. Run I data, after being re-analyzed with the Run II analysis techniques, are compatible with Run II data within $1.2~\sigma$ and $2.1~\sigma$ for the $gg$ and $qq$ scalar productions and $2.7~\sigma$ and $3.3~\sigma$ for the $gg$ and $qq$ RS G$^*$ productions~\cite{yy}.

$X^0 \to Z \gamma$ searches are subdivided, based on the $Z$ decay, into leptonic ($Z \to e^+e^-$ or $\mu^+ \mu^-$) and hadronic ($Z \to qq$) searches. Limits are set separately and only small fluctuations around the expected values are observed. The largest deviations from the background-only hypothesis are found for masses of $350$ GeV and $1.9$ TeV and correspond to a local significance of $2.0~\sigma$ and $1.8~\sigma$ respectively, for a narrow boson produced in a gluon fusion process~\cite{Zy}.

Searches for di-Higgs resonances with early Run II data have been performed for $h \to b \bar{b}$ and $h \to \gamma \gamma$ decays. The $hh \to 4b$ search makes use of the 3 and 4 b-tagged categories in the resolved and boosted regimes. Upper limits on the production cross section times branching ratio to the $b \bar{b} b \bar{b}$ final state are set for RS G$^*$ and scalar resonances with values varying between $30$ and $160$~fb at $95\%$ CL in the mass range between $500$ and $3000$ GeV~\cite{bbbb}. For the non-resonant $hh \to 4b$ production the upper limit is $1.22$ pb at $95\%$ CL~\cite{bbbb}. For the $hh \to b \bar{b} \gamma \gamma$ channel an upper limit of $3.9$ pb is set for the non-resonant production and $7.0 - 4.0$ pb for the narrow resonant production in the mass range of $275-400$ GeV~\cite{bbyy}.

In summary, improved limits are set on various theoretical models and production mechanisms by the ATLAS collaboration from searches with bosons using early Run II data. Small or modest excesses are observed in some channels, however more data are needed in order to be conclusive.

\newpage
\setcounter{figure}{0}
\setcounter{table}{0}

%% sessionday{tuesday} 

\subsection*{\hfil Searches for Beyond-Standard-Model Higgs Bosons in ATLAS \hfil}
\label{ssec:SearchesforBeyo}
\vspace*{10mm}

\underline{Z.  Zinonos}\vspace*{4mm} 
 \\ Georg-August-Universit\"at G\"ottingen \\  
\newline \noindent 
The Minimal Supersymmetric Standard Model (MSSM)~\cite{Fayet:2015sra} is an
extension of the SM, which provides a framework addressing the naturalness
problem, gauge coupling unification, and the existence of dark matter. The
Higgs sector of the MSSM contains two Higgs doublets, which results in five
physical Higgs bosons after electroweak symmetry breaking: two CP-even neutral
(h, H), one CP-odd (A) and two charged ($H^\pm$) Higgs bosons.

A search for neutral MSSM Higgs bosons decaying to a pair of $\tau$ leptons is
performed using a data sample corresponding to an integrated luminosity of
$3.2~\rm{fb}^{-1}$ from $\sqrt{s} = 13~\rm{TeV}$ proton-proton collisions
recorded by the ATLAS detector at the LHC.  The analysis covers the searches
in the $\tau_{\rm{lep}} \tau_{\rm{had}}$ and $\tau_{\rm{lep}} \tau_{\rm{had}}$
channels.  The search finds no indication of an excess over the expected
background in the channels considered.  Hence $95\%$ CL upper limits are set,
which provide tight constraints in the MSSM parameter space.

\begin{figure}
    \centering
		\begin{tabular}{cc}
    \includegraphics[width=0.47\textwidth]{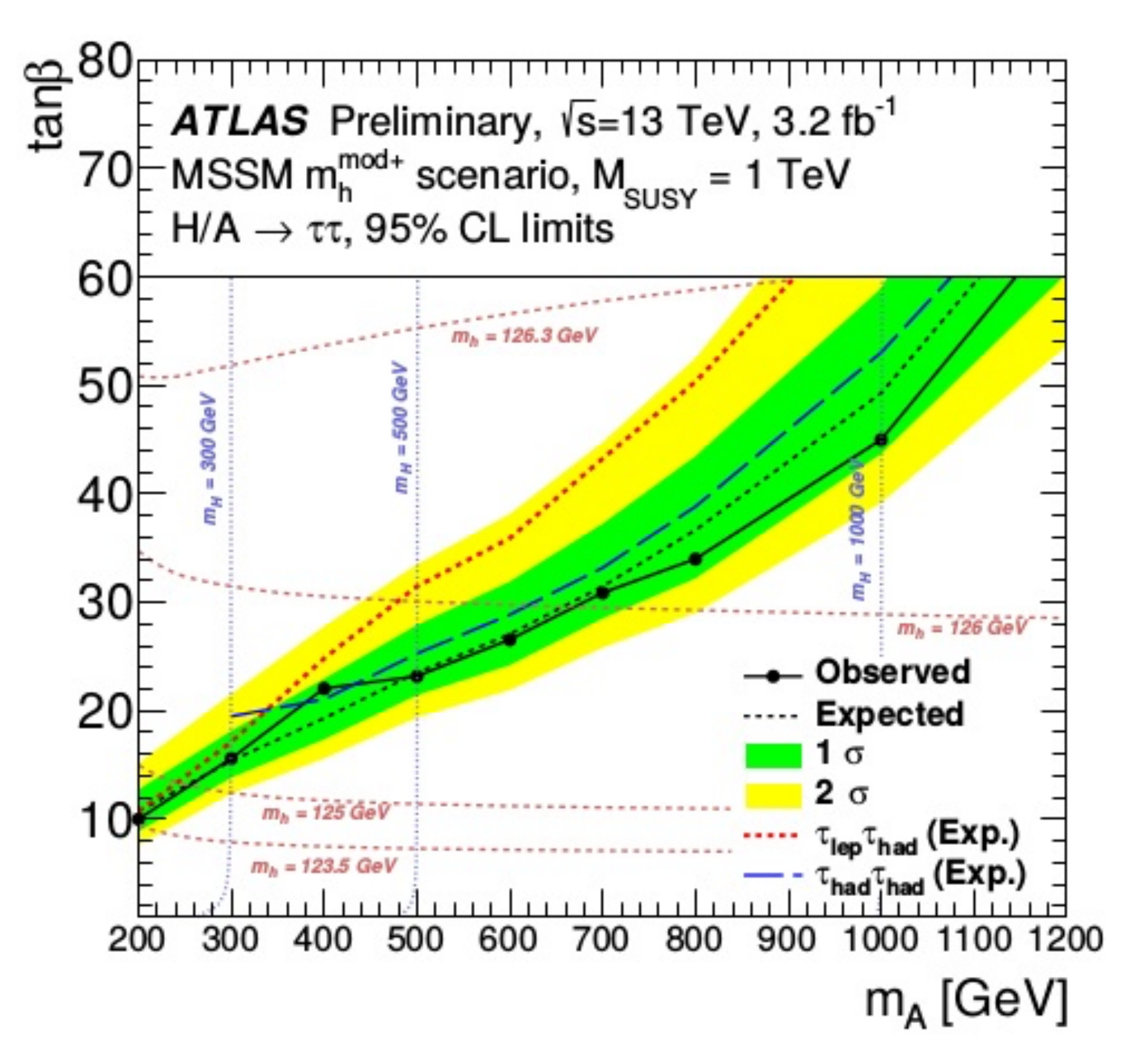} &
	  \includegraphics[width=0.47\textwidth]{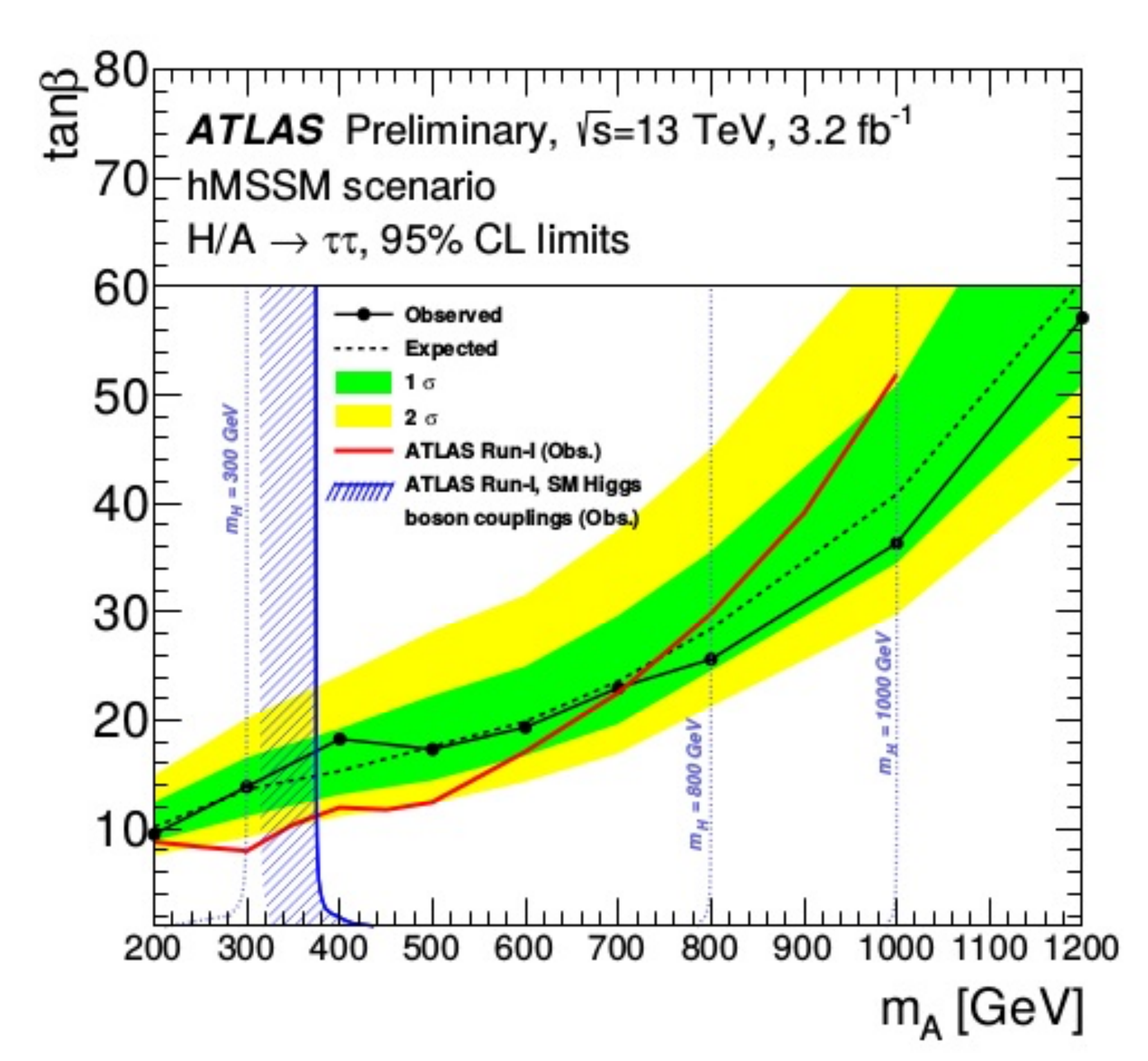} 
		\end{tabular}
    \caption{ The expected and observed $95\%$ CL upper limits on $\tan\beta$ as a function of $m_A$ for the combination of
 $\tau_{\rm{lep}} \tau_{\rm{had}}$ and $\tau_{\rm{lep}} \tau_{\rm{had}}$ channels in the MSSM $m_h^{\rm{mod+}}$ (left) and hMSSM scenario (right).
  {\bf Left:}
 The expected limits of the individual $\tau_{\rm{lep}} \tau_{\rm{had}}$ and $\tau_{\rm{lep}} \tau_{\rm{had}}$ channels are also shown
for comparison. Dashed lines of constant $m_h$ and $m_H$ are shown in red and blue colour, respectively.
{\bf Right:} The observed limits from the ATLAS Run-I analysis from Ref.~\cite{Aad:2014vgg} are
shown as a continuous red line. The continuous blue line denotes the excluded area from the ATLAS Run-I SM
Higgs boson couplings measurement from Ref.~\cite{Aad:2015pla}. Dashed lines of constant $m_H$ are shown in blue colour.
In the hMSSM scenario by construction $m_h = 125~\rm{GeV}$ in the entirety of the parameter space.
	  \label{fig:limits-zinonos}
}
\end{figure}

The expected and observed $95\%$ confidence level upper limits on $\tan\beta$ as a function of $m_A$,
for the combination of $\tau_{\rm{lep}} \tau_{\rm{had}}$ and $\tau_{\rm{lep}} \tau_{\rm{had}}$ channels in various MSSM scenarios
are shown in Figures~\ref*{fig:limits-zinonos} (left) and
\ref*{fig:limits-zinonos} (right).
In both plots, lines of constant $m_h$ and $m_H$ are shown as well with the exception of the
hMSSM\footnote{hMSSM scenario: the mass of the light CP-even Higgs boson in the mass matrix expression is fixed to $125~\rm{GeV}$.}
scenario plot in which only lines of constant $m_H$ are shown, since for this scenario $m_h = 125~\rm{GeV}$
for the entirety of the parameter space. The expected and observed limits of the combination of the
$\tau_{\rm{lep}} \tau_{\rm{had}}$ and $\tau_{\rm{lep}} \tau_{\rm{had}}$ channels in the $m_h^{\rm{mod}+}$
scenario~\footnote{The $m_h^{\rm{mod}\pm}$ scenarios are similar to the $m_h^{\rm{max}}$ scenario,
apart from parameters chosen such that radiative corrections give a light CP-even Higgs boson mass of $\sim 126~\rm{GeV}$.
In the $m_h^{\rm{max}}$ scenario the parameters are chosen such that the radiative corrections maximize $m_h$
for given values of $\tan\beta$ and $M_{\rm{SUSY}}$.
}
are shown in Fig.~\ref*{fig:limits-zinonos} (left) and compared to the expected limits from
the individual $\tau_{\rm{lep}} \tau_{\rm{had}}$ channels.
Similarly, the expected and observed limits for the combination of the $\tau_{\rm{lep}} \tau_{\rm{had}}$
channels on the hMSSM scenario~\cite{Carena:2013ytb, Heinemeyer:2013tqa} are
shown in Fig.~\ref*{fig:limits-zinonos} (right).
For this scenario, the exclusion due to SM Higgs
boson coupling measurements of Ref.~\cite{Aad:2015pla} is also shown, in addition to the ATLAS Run-I
$H/A \to \tau\tau$ search result of Ref.~\cite{Aad:2014vgg}.

In the context of the MSSM $m_h^{\rm{mod +}}$ scenario, the most stringent constraints on $\tan\beta$
for the combined search excludes $\tan\beta > 10$ for $m_A = 200~\rm{GeV}$.
Additionally, this analysis improves the limits of the previous ATLAS analysis, based on data collected at $\sqrt{s} = 8~\rm{TeV}$,
for the mass range $m_A > 700~\rm{GeV}$.
Upper limits for the production cross section times the di-$\tau$ branching fraction of a scalar boson versus its mass,
in both the gluon-fusion and $b$-associated production modes, are also presented in Ref.~\cite{ATLAS-CONF-2015-061}.
The excluded cross section times branching fraction values range from $\sigma \times \rm{BR} > 2.7(2.7)~\rm{pb}$
at $m_\phi = 200~\rm{GeV}$ to $\sigma \times \rm{BR} > 0.030(0.023)~\rm{pb}$ at $m_\phi = 1.2~\rm{TeV}$
for a scalar boson produced via gluon fusion ($b$-associated production), respectively.

\newpage
\setcounter{figure}{0}
\setcounter{table}{0}

%% sessionday{tuesday} 

\subsection*{\hfil Search for New Physics in Z+MET channel at CMS \hfil}
\label{ssec:SearchforNewPhy}
\vspace*{10mm}

\begin{figure}[h!]
\centering
\includegraphics[width=0.36\textwidth]{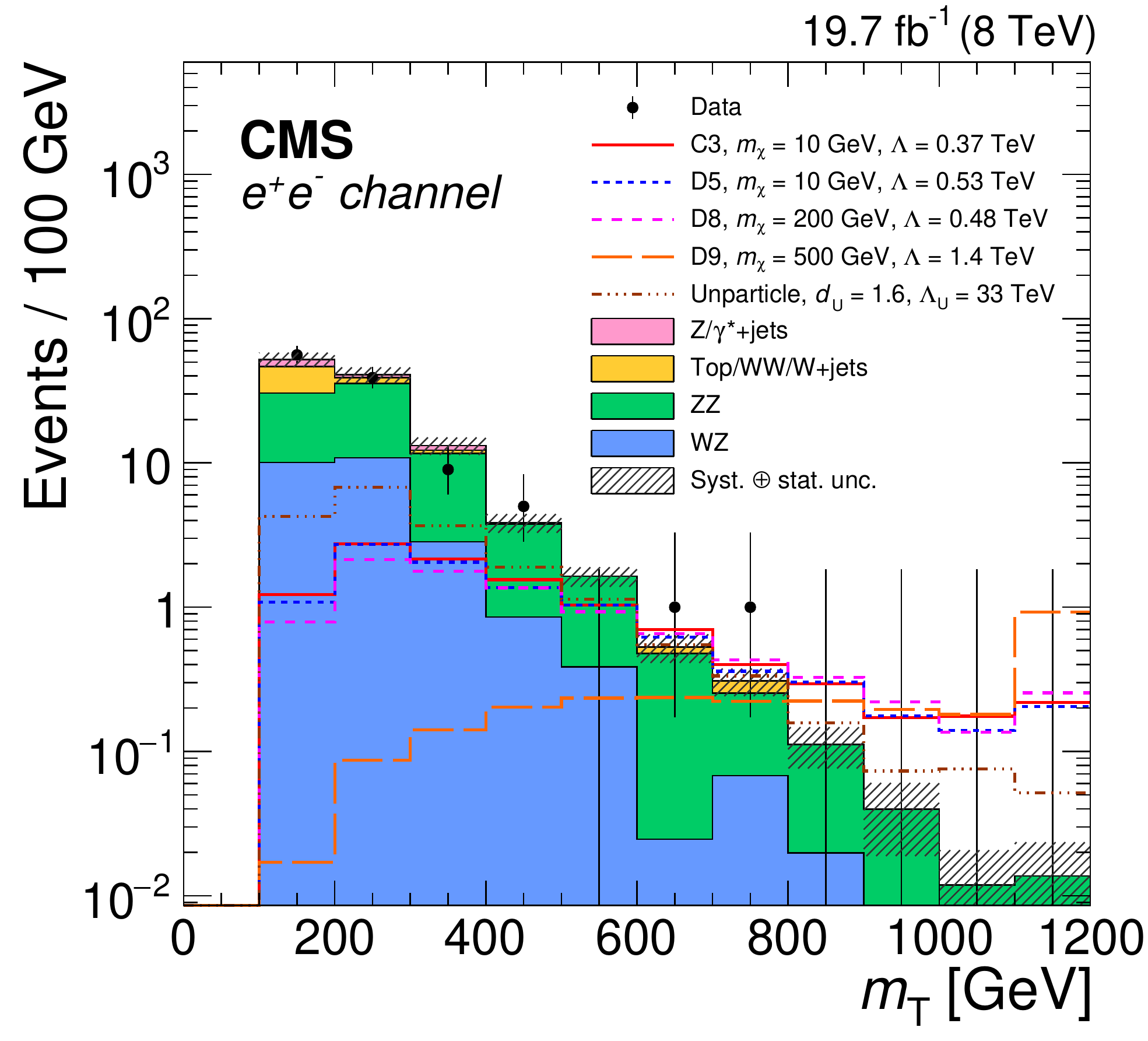}
\includegraphics[width=0.36\textwidth]{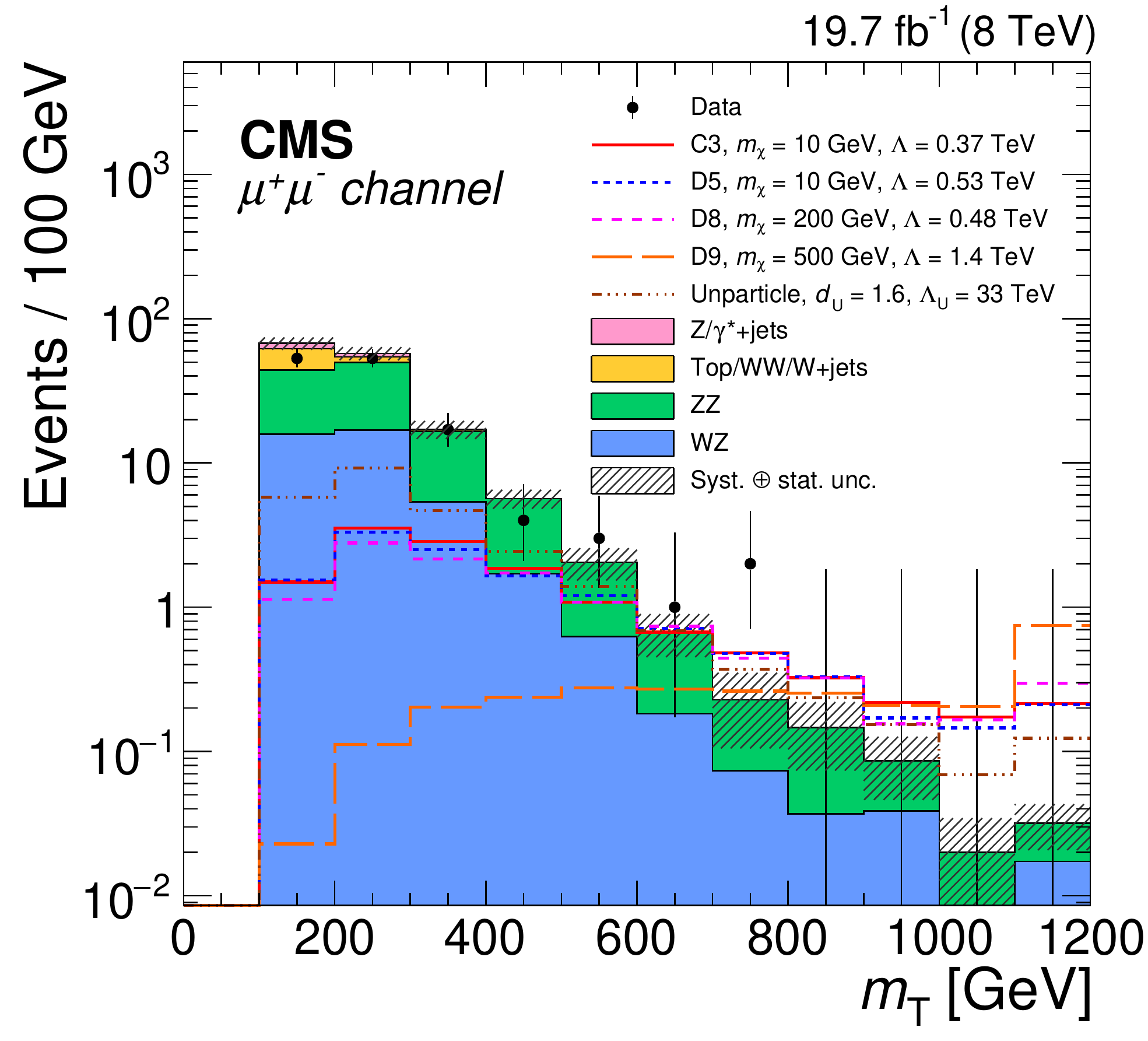}
\caption{Transverse mass distribution for the electron channel (left) and the muon channel (right).
Agreement between data measurement and the standard model prediction is observed~\cite{EXO12054}.}
\label{fig:mt_final}       \end{figure}

\begin{figure}[h!]
\centering
\includegraphics[width=0.36\textwidth]{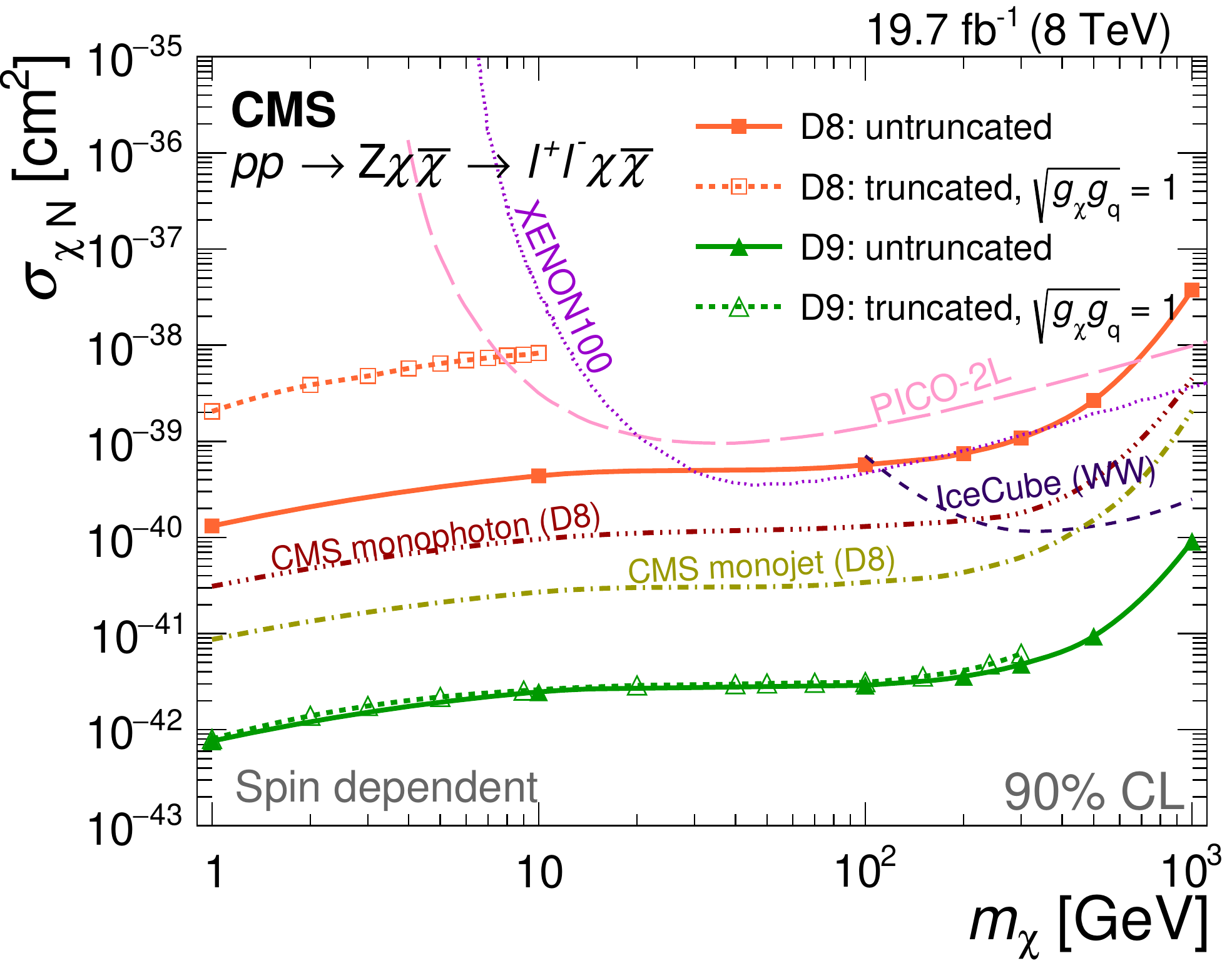}
\includegraphics[width=0.36\textwidth]{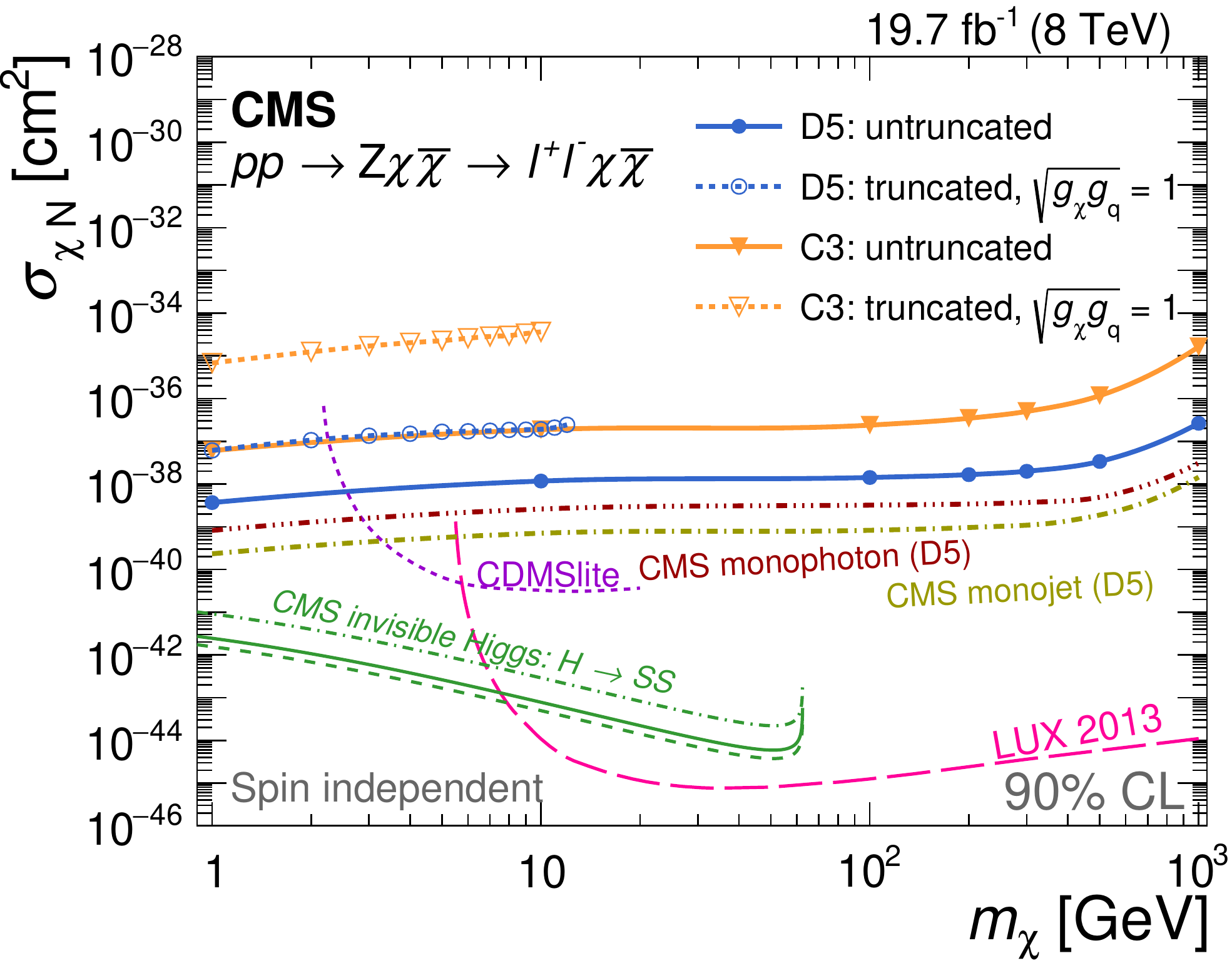}
\caption{Exclusion limits for DM-nucleon cross section for spin dependent (left) and spin independent (right) couplings~\cite{EXO12054}.}
\label{fig:limits_DM}       \end{figure}

\begin{figure}[h!]
\centering
\includegraphics[width=0.36\textwidth]{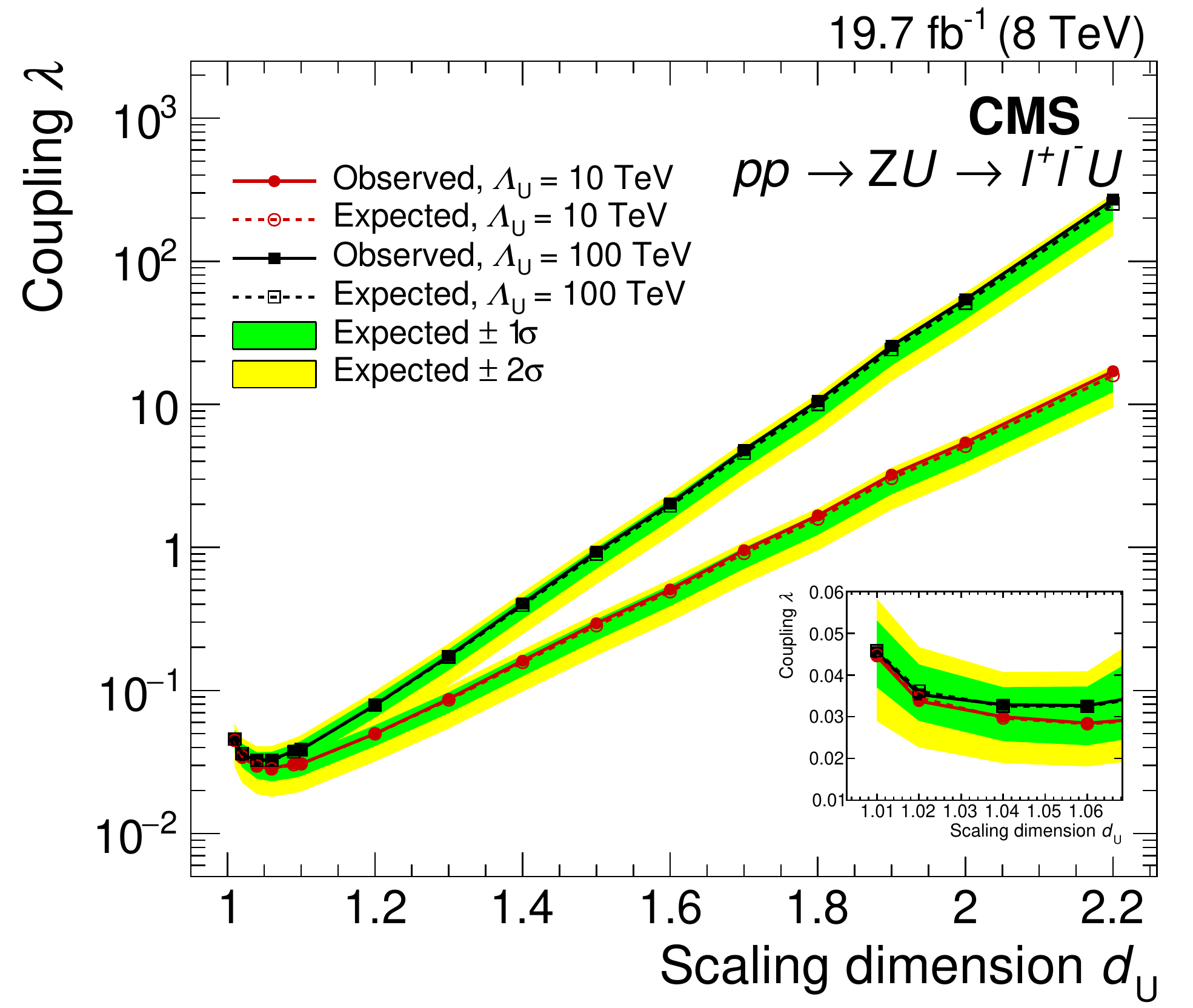}
\includegraphics[width=0.36\textwidth]{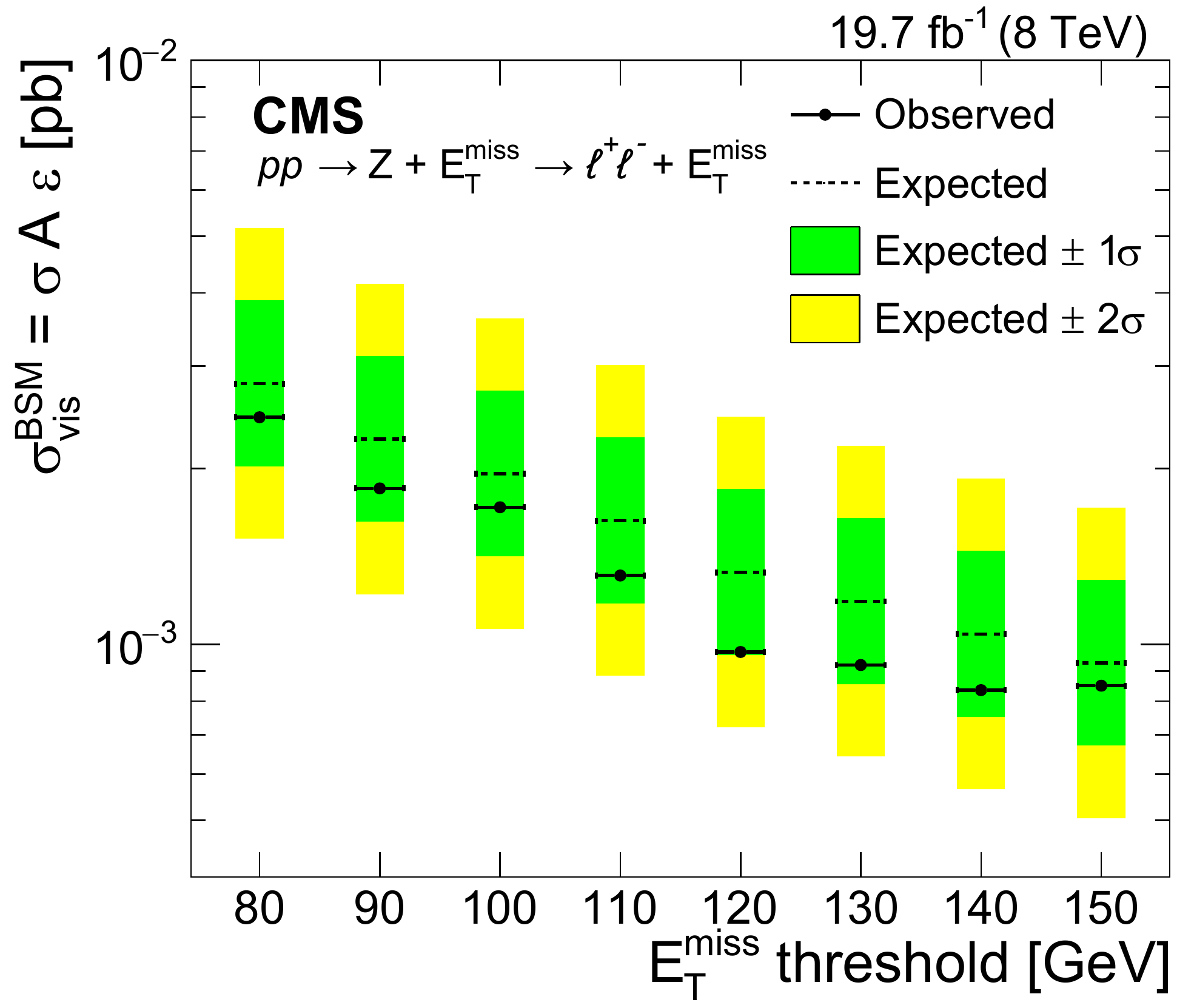}
\caption{Exclusion limits for the unparticle coupling to standard model particles $\lambda$ as function of the unparticle dimension $d_{\mathcal U}$ (left) and model independent exclusion limits for different \ETm signal regions (right)~\cite{EXO12054}.}
\label{fig:limits_unparticle}       \end{figure}

\underline{M.  Brodski} (on behalf of the CMS Collaboration)\vspace*{4mm} 
 \\ RWTH Aachen University \\  
\newline \noindent Searches for New Physics play an important role in the modern research
at collider experiments. In this note, a search for New Physics in final 
states with a Z boson and large missing transverse momentum \ETm at CMS is presented~\cite{CMSDet,EXO12054}.
In this scenario, the Z boson recoils against a new particle which leaves CMS
undetected. Subsequently, the Z boson decays into either a pair of electrons or
a pair of muons. Several possible New Physics signals like dark matter (DM) or unparticles
might be present in this final state. The search is performed in CMS data collected in 2012 using
the full dataset with an integrated luminosity of $19.7~\textrm{fb}^{-1}$ at $\sqrt{s}=8~\textrm{TeV}$.
Leptons with transverse momentum $\pt>20~\textrm{GeV}$ are included in the selection.
The muon pair is required to be reconstructed in both
the tracker and the muon system while the electrons have to be
measured in the tracker and the electromagnetic calorimeter.
Only events where the lepton pair is well isolated against hadronic activity and
activity arising from soft interactions are considered. The invariant mass of 
the lepton pair is required to be within a $10~\textrm{GeV}$ window with respect
to the Z boson mass. The angle between the lepton pair and \ETm
must be larger than $2.7$ and the considered event has to be balanced in the
transverse plane. The \ETm is required to be larger than $80~\textrm{GeV}$. Leading
backgrounds for the signal region are the ZZ and WZ boson production which are estimated
from simulation. Further backgrounds including top quark decays and
WW boson production are estimated using data-driven techniques.
For the final selection, the transverse mass
$\mt = \sqrt{2 \pt^{\ell\ell} \ETm (1-\cos \Delta \phi_{\ell\ell,\ptvecmiss})}$
is considered. The \mt spectra for both channels are shown in Fig.~\ref*{fig:mt_final}.
The data is found to be in agremeent with the standard model prediction.
The observation is interpreted in terms of exclusion limits for possible signatures
of DM as shown in Fig.~\ref*{fig:limits_DM}. Furthermore, exclusion limits for unparticles and 
model independent exclusion limits are set for different signal regions as function
of \ETm lower bound (cf. Fig.~\ref*{fig:limits_unparticle}).

\newpage
\setcounter{figure}{0}
\setcounter{table}{0}

%% sessionday{tuesday} 

\subsection*{\hfil ``Exotica'' - Speaking up for Minorities \hfil}
\label{ssec:``Exotica''Spea}
\vspace*{10mm}

\underline{T.  Berger-Hryn'ova} (on behalf of the ATLAS, CMS and LHCb Collaborations)\vspace*{4mm} 
 \\ LAPP/IN2P3/CNRS, France \\  
\newline \noindent The Standard Model of particle physics (SM) is a comprehensive gauge theory of particle interactions 
which provides accurate predictions for almost all phenomena in high-energy physics.
But if the SM is valid up to the Planck scale ($\Lambda_{Pl}$), 
the SM would have to be fine-tuned such that 
quadratic terms of the Higgs mass of order $\mathcal{O}(\Lambda_{Pl})$ cancel to within 100$\,$GeV. This  fine-tuning would not be necessary if there is new physics at the electroweak scale. 
This hints that the SM is an effective theory, a low energy approximation valid up to $\Lambda\approx 1\,$TeV, 
of this complete theory, so-called ``Theory of Everything''. The nature of this
theory is completely unknown, putting us in a unique
situation which we have not had for over 50 years now, where
the theory can not give us any guidance on what the next discovery should be. ``Exotica'' theoretical landscape is vast, and it is crucial not to focus on 
a single small corner of it, a single theory, however compelling or well-developed 
it seems to be for a ``majority'' of people working on it.  
Only data can provide us answers to our questions.
This summary concentrates on direct searches for new phenomena not related 
to supersymmetry, for other searches please see~\cite{mycitation}.

In the Run-2 of the LHC, the centre-of-mass energy increased to 13$\,$TeV 
with respect to Run 1 ($7-8\,$TeV), leading to potentially 
much higher cross sections for the high-mass new physics processes, making searches a top priority in the 2015 data analysis.
To date, CMS and ATLAS have searched in an extensive range of final states 
consisting of various combinations of leptons, light jets, $b$-jets and top quarks, 
vector bosons (photon, $W$ and $Z$), the Higgs boson and missing 
transverse energy, covering a wide area of signature-based searches~\cite{cms}. 
While we have not yet discovered any new physics, 
both experiments are observing a fascinating excess of events in the diphoton channel
with a global significance of about 2$\sigma$ at
$m_{\gamma\gamma}=750\,$GeV~\cite{atlas2g}. 
The additional data expected in 2016 will either strengthen or weaken this excess. 
Whether the excess persists or not, we will
continue to systematically search in the same broad spectrum of final states as we do at the moment.
Channels, such as $t\bar{t}$, dibosons as well as channels with 
vector-like quarks and leptons, face a lot of scrunity at the moment.
In some of those channels, such as $t\bar{t}$, accounting for the interference effects is crucial, in particular, 
when searching for a spin-0 state.

In addition to searches for final states comprised of conventional objects, there is a broad program of searches for
unconventional signatures, such as highly ionizing particles or monopoles, particles decaying (late) into heavy
neutral particles, 
long-lived particles decaying only in the outer detector components, low-mass (pseudo)scalars, etc.
The means of detecting those processes are very detector specific. 
Fortunately, phase-space coverage of the ATLAS, CMS and LHCb 
detectors is complementary, in particular, in the area of low-mass (pseudo)scalars searches~\cite{llp}. 

From the experimental point of view, now that many analyses are well-established, work is starting in the direction 
of the combinations of the results: 
\begin{itemize}
\item different decays of the same channel (e.g. boosted and resolved $t\bar{t}$ searches, diboson decays);
\item different final states of the same new phenomena signature (e.g. in heavy vector-like quark interpretations);
\item different objects in the same model (e.g. $W'\rightarrow l\nu$ and $Z'\rightarrow ll$~\cite{manuel});
\item different experiments, which would allow to double the effective luminosity of the searches
or increase their phase-space coverage.
\end{itemize}
Communication between different experiments as well as  theoretical input on the potential benchmarks and assumptions
would be crucial to make this work maximally useful. 

The year 2016 is expected to bring 25$\,$fb$^{-1}$ of LHC proton-proton collision data, compared to about $3\,$fb$^{-1}$
collected by each of the ATLAS and CMS experiments in 2015, 
hopefully resolving some outstanding questions which we have at the moment
and giving us new hints to follow up on.

\newpage
\setcounter{figure}{0}
\setcounter{table}{0}

%% sessionday{wednesday} 

\subsection*{\hfil Dark Matter: The Next 5 Years and Beyond \hfil}
\label{ssec:DarkMatterTheNe}
\vspace*{10mm}

\underline{M.  Schumann}\vspace*{4mm} 
 \\ AEC, University of Bern, Switzerland \\  
\newline \noindent 
Dark matter in the Universe outnumbers ordinary, baryonic matter by a factor~$\sim$5. As no known particle can explain the dark matter it is one of the most compelling indications for ``new physics''. A prime dark matter particle candidate is the weakly interacting massive particle (WIMP), predicted by several theories beyond the standard model. Its is assumed to have a mass $m_\chi$\,$\approx$\,1...$10^5$\,GeV/$c^2$ and a ``weak'' coupling to ordinary matter, providing the possibility to detect it with sensitive detectors in underground laboratories.

These ``direct detection'' experiments search for rare WIMP-induced single scatter nuclear recoils (NRs) in a low-background detector. The ${\cal O}$(10)\,keV NRs have a steeply falling exponential spectrum. The most abundant backgrounds for such searches are electronic recoils (ERs) from radioactive $\beta$- and $\gamma$-decays in the detector\footnote{$\alpha$-decays are usually only relevant if most of their energy is lost undetected.} and the shield, and those induced by solar neutrinos. NR backgrounds originate from radiogenic neutrons from $(\alpha,n)$ and fission reactions, and from muon-induced neutrons. As low-energy single-scatter NR interactions are indistinguishable from WIMPs, NR backgrounds are considered as more critical, especially since many experimental techniques are able to efficiently distinguish (``discriminate'') ERs from NRs (see~\cite{ref::dmreview} for an experimental review).

\begin{figure}[t!]
\centering
\includegraphics[width=\textwidth]{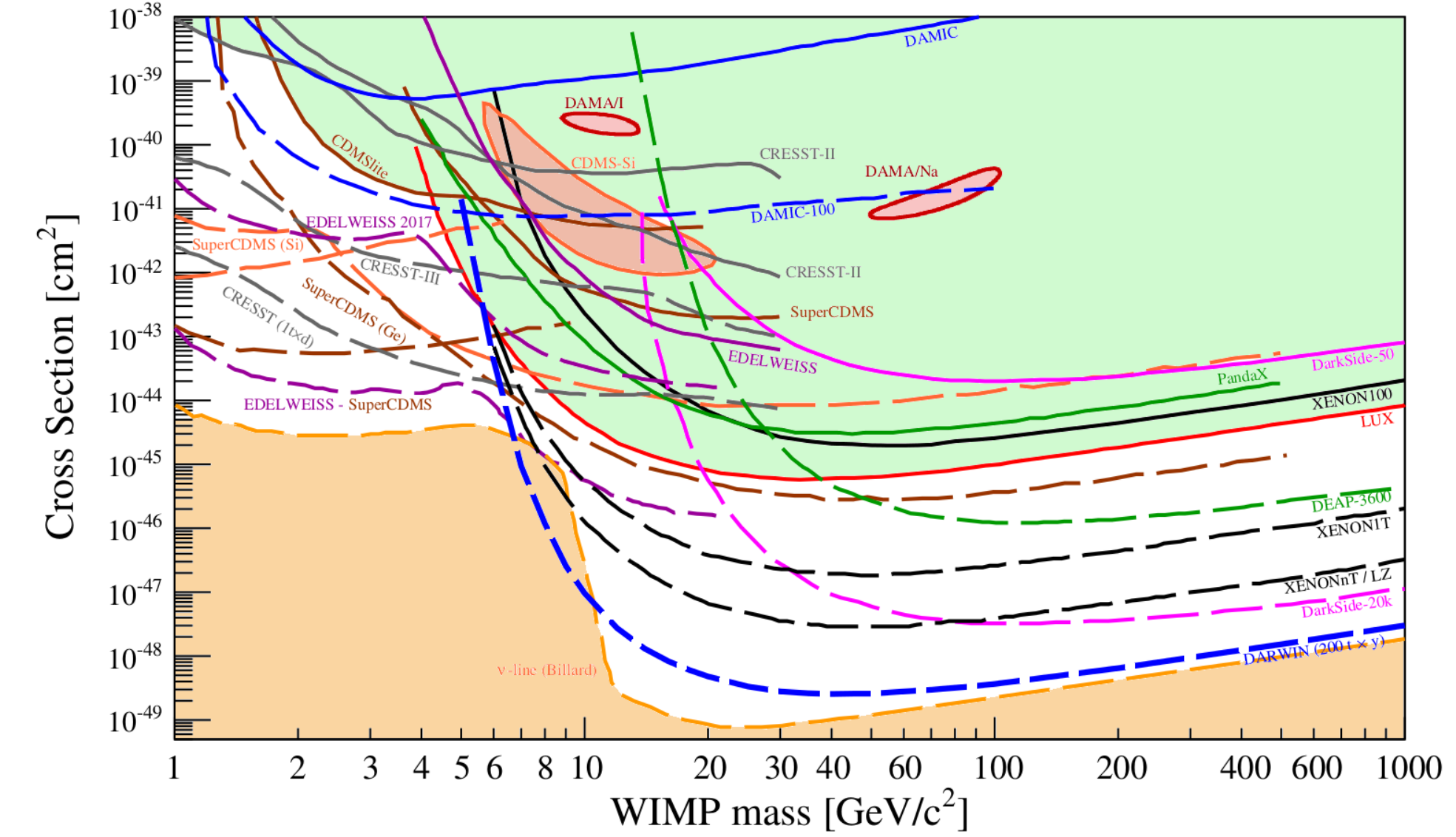}
\caption{Parameter space of spin-independent WIMP-nucleon interactions. The green area is experimentally excluded. Many  future projects (dashed) aim at closing the gap up to the irreducible neutrino background (orange).}
\label{fig:sensitivities}       
\end{figure}

The ``ultimate'', irreducible background, that leads exclusively to single-scatter NRs, is from coherent neutrino-nucleus scattering (CNNS)~\cite{ref::cnns}. At low WIMP masses $m_\chi$, these NRs are induced by $^8$B solar neutrinos, at high $m_\chi$, they are mainly from atmospheric neutrinos. The goal of the direct detection community is to cover the entire parameter space up to the CNNS ``limit'' in the next years, even though this will not be achieved in a 5~years time scale. Fig.~\ref*{fig:sensitivities} provides an illustration for spin-independent (scalar) WIMP-nucleon interactions, together with the current experimental status. (See~\cite{ref::dmreview} for more details and recent references to the various experiments.)

No WIMP signal has been found so far in direct detection experiments. The
long-standing DAMA / LIBRA claim~\cite{ref::dama} has been recently challenged by XENON100, also excluding leptophilic WIMP explanations producing ERs~\cite{ref::xenon}. New experiments also using NaI~crystals are currently being constructed to resolve the situation (Sabre, ANAIS/DM-Ice/KIMS-NaI). At very small $m_\chi$, the best limits are from cryogenic detectors (CDMSlite~\cite{ref::cdmslite}, CRESST-II~\cite{ref::cresst}) with extremely low thresholds  around 50-100\,eV. They employ $\cal O$(1)\,kg crystals (Ge, CaWO$_4$) cooled down to mK-temperatures to measure the tiny heat signal from a particle interaction (see, e.g.,~\cite{ref::cryo}). Future projects such as SuperCDMS and CRESST-III aim at approaching the CNNS line for $m_\chi$\,$<$\,8\,GeV/c$^2$.

Detectors using the cryogenic liquid noble gases xenon (LXe) or argon (LAr) are currently leading the field for $m_\chi \ge 4.5$\,GeV/$c^2$ and can be realized with ton-scale target masses~\cite{ref::dualphase}. With thresholds around 3\,keV and a reasonable ER background rejection, especially dual-phase LXe time projection chambers (TPCs) are very well suited to probe this mass range. At the moment, the most sensitive instrument is LUX (0.25\,t LXe target)~\cite{ref::lux}, which will soon be superseded by the larger XENON1T (2.0\,t LXe target). It is currently under commissioning at LNGS and aims at a sensitivity below $2\times10^{-47}$\,cm$^2$ at 50\,GeV/$c^2$~\cite{ref::xenon1t}. Its upgrade phase XENONnT it is expected to be operational within the next years, together with LZ, the successor of LUX, and DarkSide-20k (LAr), all aiming at a factor~$\sim$10 improvement upon XENON1T. 

Above $m_\chi$\,$\approx$\,6\,GeV/$c^2$, almost the entire parameter space up to the CNNS limit can be probed with a $\sim$40\,t LXe TPC and 200\,t$\times$\,y exposure~\cite{ref::darwinwimp}. Such ``ultimate'' detector is planned within the DARWIN project and would not only be sensitive to WIMP dark matter, but to a plethora of new science channels: solar axions/axion-like-particles, low-energy solar neutrinos (pp, $^7$Be), sterile neutrinos, supernova neutrinos and the CNNS from $^8$B solar neutrinos~\cite{ref::darwinnu}.

\newpage
\setcounter{figure}{0}
\setcounter{table}{0}

%% sessionday{wednesday} 

\subsection*{\hfil Baryonic Dark Matter at the LHC \hfil}
\label{ssec:BaryonicDarkMat}
\vspace*{10mm}

\underline{M.  Duerr}\vspace*{4mm} 
 \\ DESY, Notkestra\ss e 85, D-22607 Hamburg, Germany \\  
\newline \noindent 
In the Standard Model~(SM) of particle physics, baryon and lepton numbers are accidental global symmetries of the renormalizable couplings. The question whether and how these symmetries are broken is connected to such important issues as the generation of the baryon asymmetry of the Universe, the stability of the proton, or the smallness of neutrino masses. In this talk, I discuss gauge theories for baryon (and lepton) number(s) that can be spontaneously broken at the low scale without introducing dangerous baryon-number violating operators generating, e.g., proton decay, and phenomenological aspects of these theories. I focus on the dark matter~(DM) candidate carrying baryon number (hence: baryonic DM), and also discuss how to search for these models at the LHC.

\begin{table}[h!]
\centering
  \begin{tabular}{cccccc}
\hline
Field & $SU(3)$ & $SU(2)$ & $U(1)_Y$ & $U(1)_B$ & $U(1)_L$ \\
\hline \hline
 $\Psi_L$   & $\mathbf{1}$  & $\mathbf{2}$ & $\pm \frac{1}{2}$ & $B_1$ & $L_1$ \\
 $\Psi_R$   & $\mathbf{1}$  & $\mathbf{2}$ & $\pm \frac{1}{2}$ & $B_2$ & $L_2$ \\
 $\eta_R$   & $\mathbf{1}$  & $\mathbf{1}$ & $\pm 1$ & $B_1$ & $L_1$ \\
 $\eta_L$   & $\mathbf{1}$  & $\mathbf{1}$ & $\pm 1$ & $B_2$ & $L_2$ \\
 $\chi_R$   & $\mathbf{1}$  & $\mathbf{1}$ & $0$ & $B_1$ & $L_1$ \\
 $\chi_L$   & $\mathbf{1}$  & $\mathbf{1}$ & $0$ & $B_2$ & $L_2$ \\
\hline
\end{tabular}
\caption{New fermions and corresponding quantum numbers. All relevant anomalies are cancelled if $B_1 - B_2 = -3$ and $L_1 - L_2 = -3$.}
\label{tab:leptobaryons}
\end{table}

A simple model based on the gauge group 
\begin{equation}
 G_{BL} = SU(3) \otimes SU(2) \otimes U(1)_Y \otimes U(1)_B \otimes U(1)_L
\end{equation}
was presented in Ref.~\cite{Duerr:2013dza}. Anomaly cancellation requires the introduction of additional fermions to consistently gauge $G_{BL}$. A simple solution in agreement with all current constraints is provided by lepto-baryons (fields carrying both baryon and lepton numbers) that are vector-like under the SM gauge group, see Tab.~\ref*{tab:leptobaryons}. Various aspects of this model were studied in more detail in Refs.~\cite{Duerr:2013lka,Duerr:2014wra,Duerr:2015vna}. In this talk, for simplicity I focus on a $U(1)_B$ extension of the SM only. 

A new scalar field $S_B \sim (\mathbf{1},\mathbf{1},0,B_1 - B_2)$ can give vector-like masses to the new fermions after spontaneously breaking $U(1)_B$. Due to its quantum numbers, only $\Delta B = 3$ processes will be induced and proton decay is forbidden. A remnant $\mathcal{Z}_2$ symmetry stabilizes the lightest new field after symmetry breaking. If neutral, this is an appealing DM candidate whose stability is an automatic consequence of the gauge symmetry and the particle content. 

At the LHC, the new gauge boson $Z_B$ will be produced via its interactions with quarks, and strongest constraints come from dijet and $t \bar{t}$ searches~\cite{Duerr:2014wra}. The only unsuppressed production for the new Higgs is associated production together with the $Z_B$, which would, e.g., lead to a signal in $t \bar{t}$ plus missing energy.

\begin{figure}[h!]
 \centering
 \includegraphics[width=0.49\linewidth]{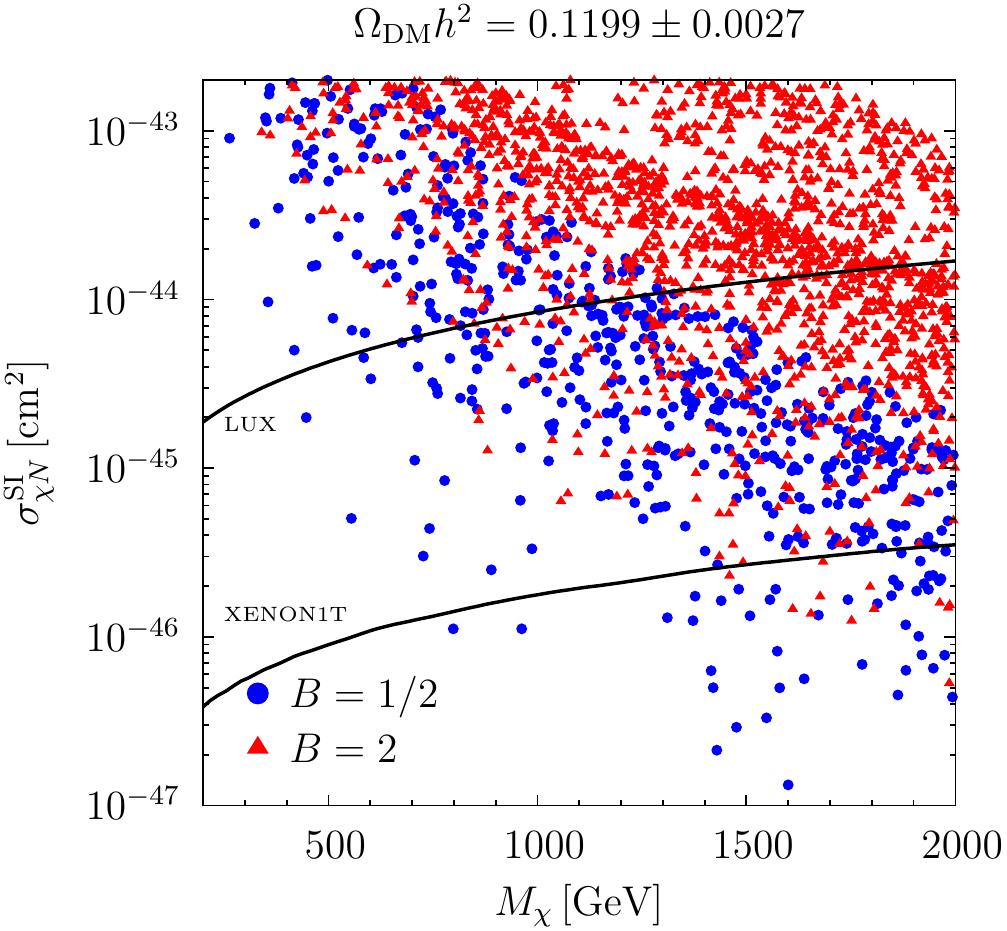}
 \includegraphics[width=0.49\linewidth]{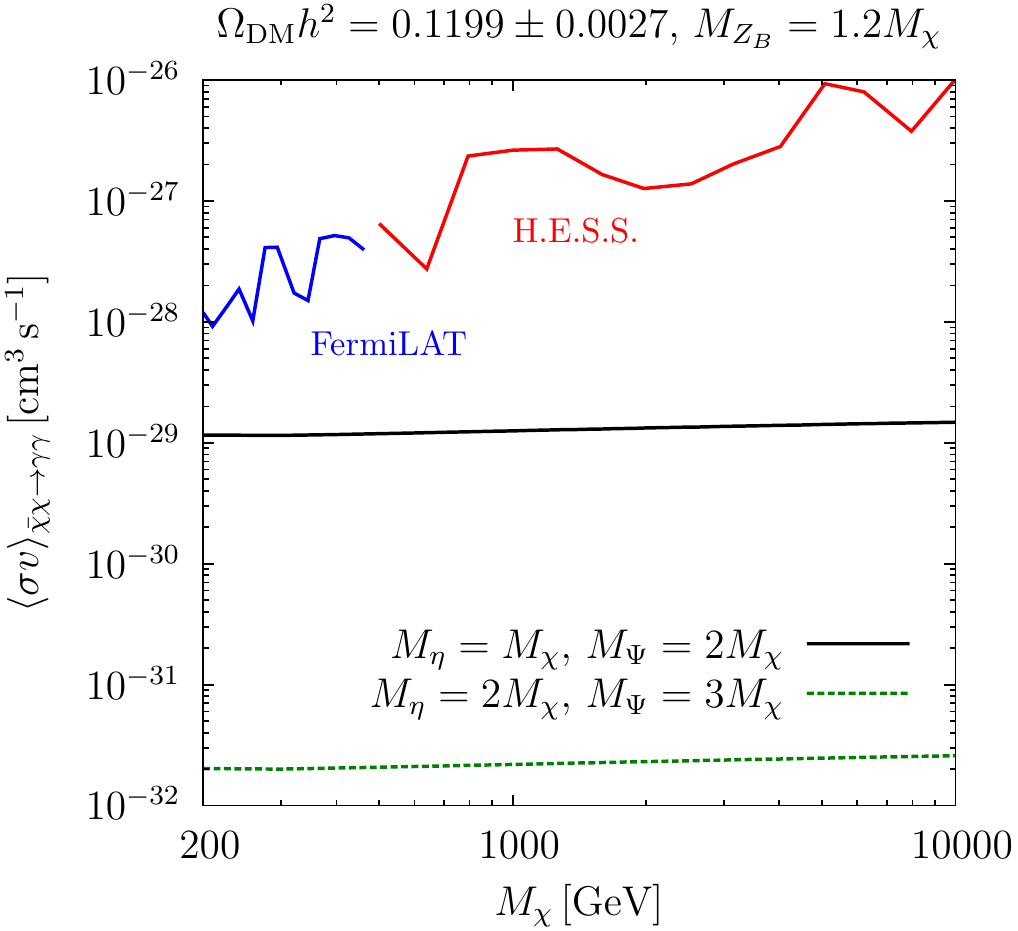}
 \caption{Direct (left) and indirect (right) detection of DM in an extension of the SM with gauged baryon number. The spin-independent DM--nucleon cross section $\sigma^\mathrm{SI}_{\chi N}$ is given for two values of $B \equiv B_1 + B_2$. Figures from Refs.~\cite{Duerr:2014wra,Duerr:2015vna}. }
 \label{fig:results}
\end{figure}

In large part of the parameter space, the properties of the DM $\chi = \chi_L + \chi_R$ are dominated by its interaction with the $Z_B$. This interaction gives the DM annihilation relevant to set the relic density as well as direct detection~\cite{Duerr:2013lka,Duerr:2014wra}. See Fig.~\ref*{fig:results} (left) for the prediction for DM direct detection. However, to discuss indirect DM detection, the additional fermions $\Psi$ and $\eta$ are relevant, since the loop-induced DM annihilation to $\gamma \gamma$ relies on electrically charged particles in the loop that have axial couplings to leptophobic gauge boson ~\cite{Duerr:2015vna,Duerr:2015wfa}. The predicted cross section then depends on their masses; see Fig.~\ref*{fig:results} (right) for more details.

\newpage
\setcounter{figure}{0}
\setcounter{table}{0}

%% sessionday{wednesday} 

\subsection*{\hfil How (not) to Use Simplified Models to Search for DM at the LHC \hfil}
\label{ssec:HownottoUseSimp}
\vspace*{10mm}

\underline{F.  Kahlhoefer}\vspace*{4mm} 
 \\ DESY, Notkestra\ss e 85, D-22607 Hamburg, Germany \\  
\newline \noindent 
Astrophysical observations give almost no indications concerning the particle nature of dark matter (DM). In order to devise experimental search strategies it is therefore necessary to construct specific DM models as a guidance. The top-down approach to this problem aims to obtain a well-motivated candidate for DM from theoretical considerations (concerning for example the hierarchy problem or the strong $CP$ problem). In the bottom-up approach, on the other hand, the foremost aim is to explain the observed DM relic abundance by adding the minimal necessary amount of additional structure to the SM. 

Models containing a DM particle and an additional new particle mediating the interactions of DM (the dark mediator) can be considered the middle ground between these two approaches. One the one hand they can be thought of as a simplification of a UV-complete theory of DM, boiled down to capture the most relevant experimental signatures. On the other hand, such models extend the most minimal models for DM in order to provide all the ingredients necessary to calculate predictions for different experiments in a self-consistent way.

In the context of LHC searches, such models are often referred to as simplified DM models. This name suggests that their main purpose is to provide a tool for generating and studying events with missing energy. As soon as one is interested in comparing results from the LHC with other experimental or observational probes of DM, however, a more ambitious approach is required. It then becomes essential that the models under consideration fulfil certain basic requirements, such as gauge invariance and perturbative unitarity. 

To illustrate this point, let us consider a DM particle coupling to a $Z'$ dark mediator and study the process $\psi \bar{\psi} \rightarrow Z'_L Z'_L$. If the DM particle has non-zero axial coupling to the $Z'$, $g^A_\text{DM} \neq 0$, the matrix element for this process grows with the centre-of-mass energy. 
For
\begin{equation}
\sqrt{s} > \pi \, m_{Z'}^2 / \left[(g^A_\text{DM})^2 \, m_\text{DM}\right]
\end{equation}
new physics must appear to restore unitarity~\cite{Kahlhoefer:2015bea}. This can be accomplished by assuming that the $Z'$ arises from a $U(1)'$, which is spontaneously broken by an additional Higgs boson.

In order for the $Z'$ to mediate the interactions of DM, it is necessary that at least some of the SM particles carry a $U(1)'$ charge. In this case, the charge assignments must respect the structure of the SM Lagrangian. This requirement can be written as 
\begin{equation}
q_H = q_{q_L} - q_{u_R} = q_{d_R} - q_{q_L} = q_{e_R} - q_{\ell_L}.
\end{equation}

\begin{figure}[h!]
\centering
\includegraphics[width=0.45\textwidth]{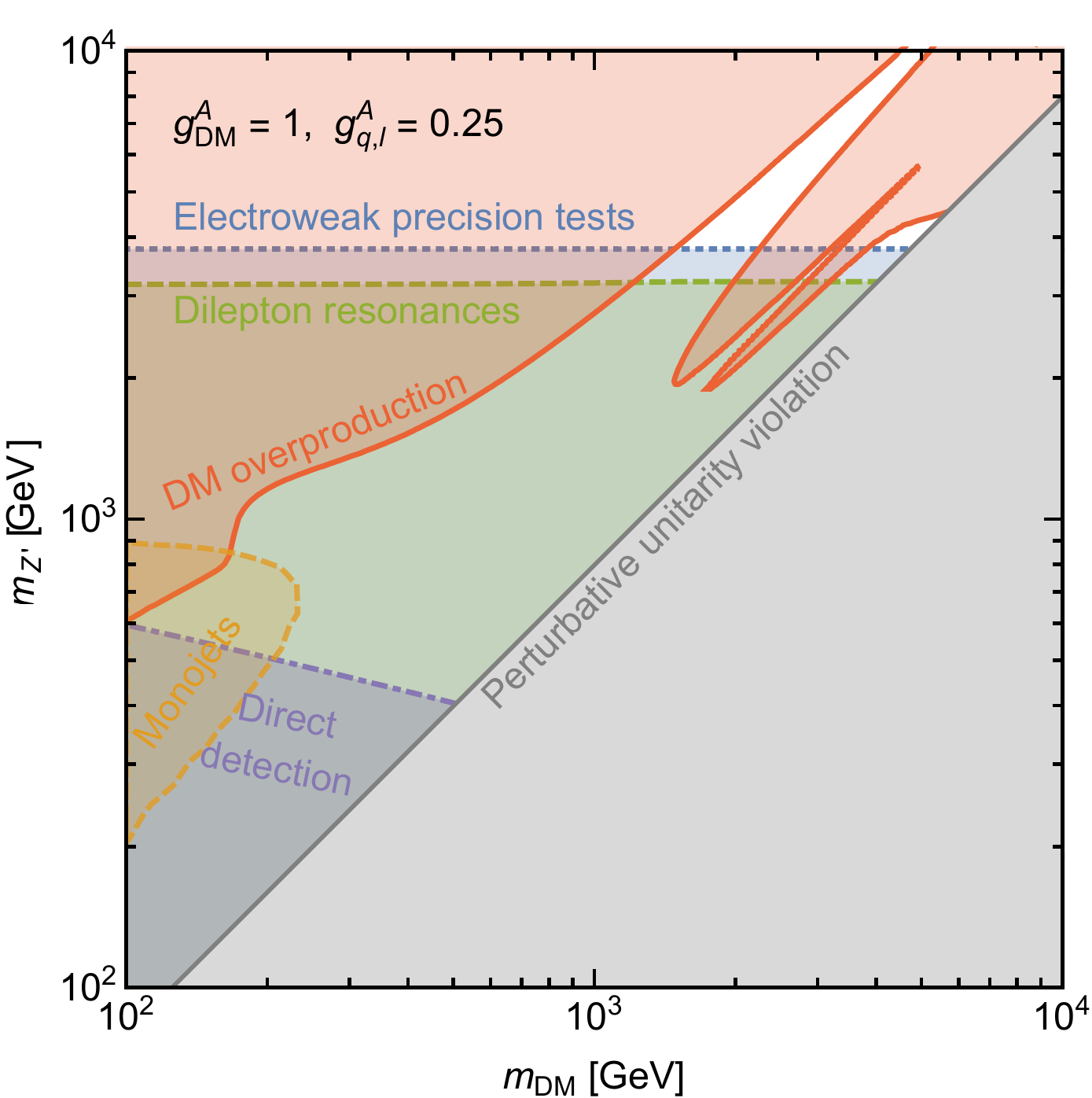}\qquad
\includegraphics[width=0.45\textwidth]{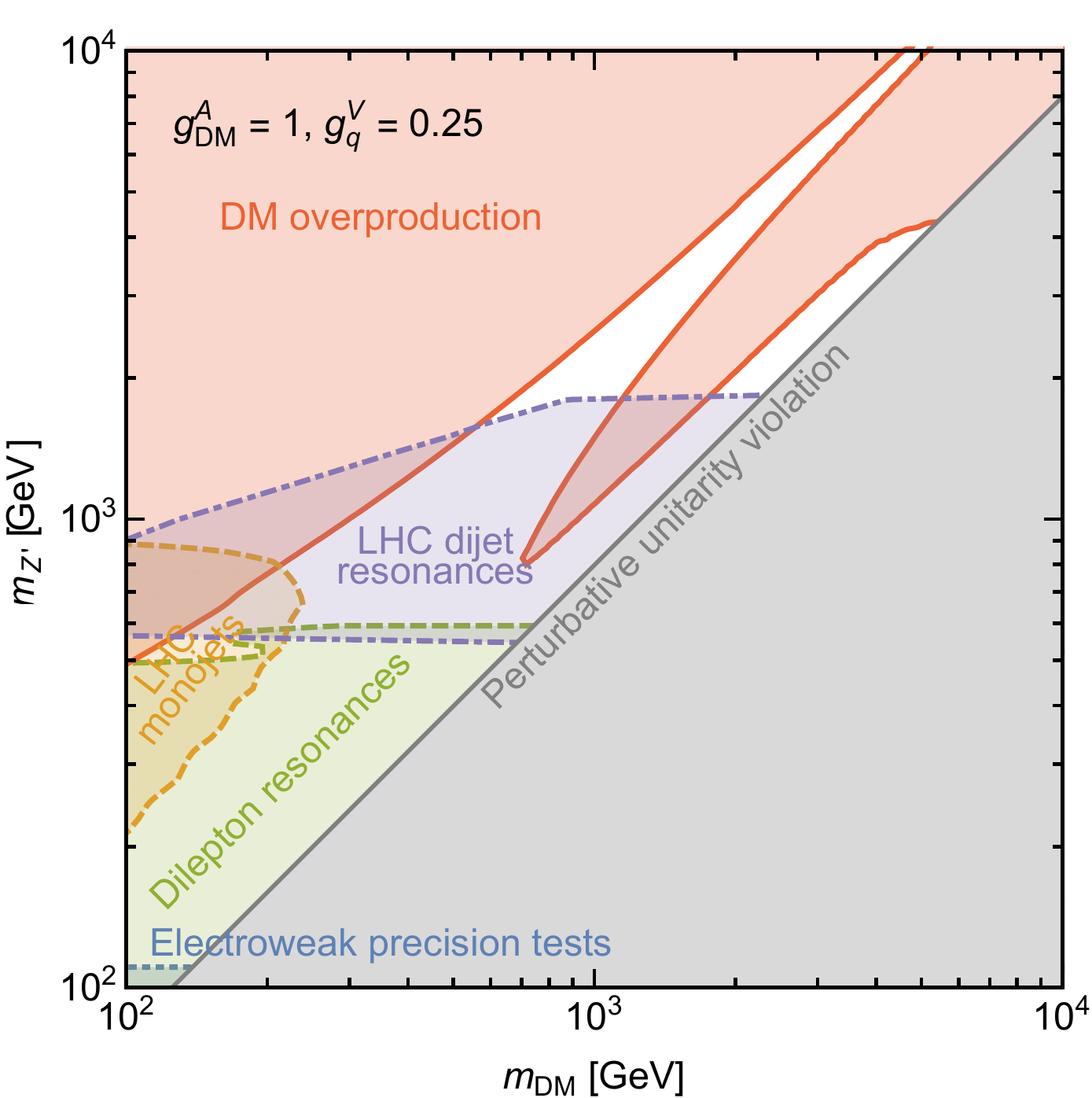}
\caption{Constraints on the parameter space of a dark mediator model with Majorana DM and a $Z'$ with axial (left) and vector (right) couplings to quarks~\cite{Kahlhoefer:2015bea}.}
\label{fig-1}
\end{figure}

For a $Z'$ with axial couplings to quarks, the equation above implies that the $Z'$ must couple with the same strength to all generations of quarks and to leptons, leading to strong constraints from searches for dilepton resonances. Moreover, the coupling of the $Z'$ to the SM Higgs induces a non-diagonal mass term that leads to mixing between the $Z'$ and the SM $Z$ boson and hence strong constraints from electroweak precision tests (EWPT). As shown in the left panel of figure~\ref*{fig-1}, these constraints are typically much more important than the conventional probes of DM in monojet searches or direct detection experiments. The requirement of perturbative unitarity thus leads to various non-trivial relations between the different states and couplings not normally imposed on a simplified model.

Finally, we note that for purely vectorial couplings to quarks these strong constraints are absent. There may still be constraints from searches for dilepton resonances and EWTP due to loop-induced kinetic mixing between the $U(1)'$ and SM hypercharge, but the resulting constraints are much weaker (see the right panel of figure~\ref*{fig-1}). Indeed, in this case we observe a compelling interplay of the constraints due to loop-induced kinetic mixing and bounds from conventional LHC DM searches, probing most of the parameter space where DM overabundance is avoided.

\newpage
\setcounter{figure}{0}
\setcounter{table}{0}

%% sessionday{wednesday} 

\subsection*{\hfil Dark Matter Searches with CMS \hfil}
\label{ssec:DarkMatterSearc}
\vspace*{10mm}

\underline{M.  Jeitler} (for the CMS Collaboration)\vspace*{4mm} 
 \\ Institute of High Energy Physics of the Austrian Academy of Sciences  \\  
\newline \noindent 
The CMS collaboration has performed searches for Dark-Matter production based on data taken in in proton-proton collisions at 7, 8, and 13 TeV with the CMS detector~\cite{CMSdetector} at the LHC. 
They have been interpreted in terms of simplified models with different structures and mediators as well as by using generic effective theory terms.

For spin-independent processes (due to vector or scalar interactions) experimental cross-section limits are currently dominated by direct-detection experiments for Dark-Matter particle masses above about 10 GeV while the lower mass range appears better accessible to collider experiments. Spin-dependent processes (caused by axial-vector or pseudoscalar interactions) are hard to measure by direct detection and limits are dominated by collider results over a wide range of Dark-Matter  particle masses. 

The Run-1 analyses of data taken at collisions energies of 7 and 8 TeV formulated the interpretation of results mostly in terms of Effective Field Theories assuming a contact interaction, which is valid in cases where the mass of the mediator particle is much higher than the momentum transfer in the process. At higher energies this assumption becomes less justified and newer analyses interpret results in terms of explicitly defined mediator particles~\cite{CERNDarkMatterForum1, CERNDarkMatterForum2}. 

As the Dark-Matter  particles themselves would be invisible for the CMS detector, they should show up in terms of missing transverse momentum or missing transverse energy (``MET'' for ``missing $E_T$'') together with visible Standard-Model particles that could arise from initial-state radiation (ISR) or in decay cascades. Many analyses target one such object plus MET and are therefore dubbed ``mono-X'' searches, such as ``mono-jet''~\cite{MonoJet}, ``mono-photon''~\cite{MonoPhoton}, ``mono-Z''~\cite{MonoZ} and ``mono-W''.

Just as in the case of any analysis of hadron collider data, it is essential to correctly define and assess background events. Background is usually measured in control regions (data-driven approach) and applied to signal regions by using Monte-Carlo simulations to calculate the corresponding ``transfer factors''.
So, in the monojet analysis, which requires a high-energy jet plus MET and vetoes leptons, the dominant background channels are $(Z \rightarrow \nu \nu) + \gamma$/jets and 
$(W \rightarrow \ell \nu) + \gamma$/jets (where the lepton is lost). These backgrounds can be best studied by using control regions with
$Z \rightarrow \mu \mu$ and 
$W \rightarrow \mu \nu$ events.  

Another analysis targets heavy-flavor quarks, which could strongly couple to scalar and pseudoscalar mediators. This analysis requires events with one or two b-tagged jets and is also sensitive to top decays. Also in this case, $(Z \rightarrow \nu \nu)$+jets and $(W \rightarrow \ell \nu)$+jets are the dominant sources of background~\cite{DMbottom}. 

Another possibility are the so-called ``Higgs-Portal'' models of Dark Matter, in which the Higgs boson would be the mediator between Standard-Model and Dark-Matter particles~\cite{InvisibleHiggs}. In this case, the Higgs boson would in some cases decay into invisible particles. Such decays can be investigated using Higgs production by vector boson fusion or in associated ZH production, where the final state contains jets or leptons in addition to the Higgs boson. The combined limit from the different analyses on the invisible branching ratio of Higgs decays is 32\%. 

So far, no signals of Dark Matter have been seen at the LHC from the data taken in Run 1 (at $\sqrt{s} = 7$ and 8~TeV) or at the beginning of Run 2.
The start-up of the LHC after ``Long Shutdown 1'' at the increased collision energy of $\sqrt{s} = 13$~TeV required many technical adjustments and only a few fb$^{-1}$ of luminosity could be acquired in 2015. Therefore, in spite of higher cross sections at higher energy in most cases the 2015 data have not allowed to improve limits over LHC's Run 1 but the present fast luminosity build-up of LHC lets expect that this situation will soon change, resulting either in a discovery of Dark-Matter particles or in significantly improved limits.

\newpage
\setcounter{figure}{0}
\setcounter{table}{0}

%% sessionday{thursday} 

\subsection*{\hfil Belle II Studies of Missing Energy Decays and Searches for Dark Photon Production \hfil}
\label{ssec:BelleIIStudieso}
\vspace*{10mm}

\underline{L.  Li Gioi}\vspace*{4mm} 
 \\ Max-Planck-Institut f\"ur Physik, 80805 M\"unchen \\  
\newline \noindent 
SuperKEKB, the massive upgrade of the asymmetric electron positron collider
KEKB in Tsukuba, Japan, aims at an integrated luminosity in excess of $50$
ab${}^{-1}$.
It will deliver an instantaneous luminosity of $8 \cdot 10^{35}$ cm${}^{-2}$
s${}^{-1}$, which is 40 times higher than the world record set by KEKB.
At the same time, the Belle~II detector is under construction. The Belle~II experiment has a broad physics program to search for
new physics beyond the Standard Model in the heavy flavor sector.
The SuperKEKB commissioning has started at the beginning of 2016 and the first data for physics analysis are expected at the end of~2018. \\

{\bf Missing energy decays} can be studied at an $e^{+} e^{-}$  (Super)$B$-factory, reconstructing one of the exclusively produced $B$ mesons in the $\Upsilon(4S)$ decay,
referred as the tag side ($B_{tag}$), either in hadronic or in semileptonic decays. Properties of the remaining particle, referred to as the signal side ($B_{sig}$),
are then compared to those expected for signal and background. This method allows to suppress strongly the combinatorial background from both $B \bar{B}$ and continuum,
$e^{+} e^{-} \to q \bar{q}$ with $q=(u,d,s,c)$, processes.

The precise measurement of $\mathcal{B}(B^{-} \to \tau^{-} \bar{\nu}_{\tau})$ and the search for the decays $B^{-} \to \mu^{-} \bar{\nu}_{\mu}$ and $B^{-} \to e^{-} \bar{\nu}_{e}$
are primary goals of the Belle~II program.
$B^{-} \to \tau^{-} \bar{\nu}_{\tau}$ provides a direct experimental determination of the product of the $B$ meson decay constant and the magnitude
of the CKM matrix element $|V_{ub}|$, but physics beyond the Standard Model could significantly suppress or enhance its branching fraction via exchange of a new charged particle,
such as a charged Higgs boson from supersymmetry or from two-Higgs doublet models~\cite{ref:btnu}.
Another class of decays that will play an important role in leptonic decay measurements is  $B^{-} \to l^{-} \bar{\nu}_{l} \gamma$.
These decays will be crucial for measuring the $B$ meson decay constant. Belle~II will measure $\mathcal{B}(B^{-} \to \tau^{-} \bar{\nu}_{\tau})$ with a precision of 5\% and the branching fraction of the other decay modes with a precision~of~10\%.

Semitauonic $B$ meson decays of the type $b \to c \tau \nu_\tau$ are sensitive probes to search for physics beyond the Standard Model.
Charged Higgs bosons may contribute measurably to the decays due to the large mass of the $\tau$.
Similarly, leptoquarks, which carry both baryon number and lepton number, may also contribute to this process.
The ratio of branching fractions $R(D^{(*)})= \mathcal{B}(B \to D^{(*)} \tau \nu_{\tau})/\mathcal{B}(B \to D^{(*)} l \nu_l)$, with $l=(e,\mu)$,
is typically used, instead of the absolute branching fraction of $B \to D^{(*)} \tau \nu_\tau$, to reduce several systematic uncertainties,
such as those on the experimental efficiency, the magnitude of the CKM matrix element $|V_{cb}|$, and the semileptonic decay form factors.
Including the recent measurement of $R(D^{*})$ performed by Belle~\cite{ref:belleRst}, the combined analysis of $R(D^{*})$ and $R(D)$,
taking into account correlations, finds that the deviation is 4.0$\sigma$ from the Standard Model prediction~\cite{ref:hfagRstR}.
With an expected sensitivity of 2.1\% to $R(D^{*})$ and 3.4\% to $R(D)$, Belle~II will provide a measurement more precise than the present world average.

The expected branching fractions for exclusive $B \to K^{*} \nu \bar{\nu}$ decays have recently been calculated
in the Standard Model~\cite{ref:SMkstNunu}. Various new-physics scenarios exist that could significantly enhance this value.
Based on the Belle results and on the ongoing work on the analysis algorithms, providing better B-Tag efficiency, better $K_L$ identification,
and a $K_S$ efficiency of 30\% better than Belle, The Standard Model expectation of $\mathcal{B}(B^{0} \to K^{*0} \nu \bar{\nu})$
can be probed at 5$\sigma$ by Belle~II. \\

{\bf The dark photon \boldmath{$A'$} and the dark Higgs boson \boldmath{$h'$}} are hypothetical constituents featured in a number of
recently proposed dark sector models. Because of the small coupling to Standard Model particles and the low expected mass of the dark photon, Belle~II at SuperKEKB
will be the ideal tool to discover the dark photon and the dark Higgs boson.

Dark photon can be searched in the reaction $e^+e^- \to \gamma A'$, with $A' \to l^{+} l^{-}$ ($l=e,\mu$) and $A' \to \chi \chi$ ($\chi =$ light dark matter).
Crucial for this analysis are the rate of the low multiplicity and single photon trigger that are still under development in Belle~II.

Given the much higher luminosity, a better invariant mass resolution and an expected higher trigger efficiency,
Belle~II will cover additional regions, compared to Belle, of the parameters space of the dark photon mass and mixing parameter.
Especially, during the first years of data taking, higher low multiplicity and single photon trigger rate are expected.

\newpage
\setcounter{figure}{0}
\setcounter{table}{0}

%% sessionday{thursday} 

\subsection*{\hfil CP Violation in Standard Model \hfil}
\label{ssec:CPViolationinSt}
\vspace*{10mm}

\underline{J.  F. Kamenik}\vspace*{4mm} 
 \\ Jo\v zef Stefan Institute, Jamova 39, 1000 Ljubljana, Slovenia \\  
Faculty of Mathematics and Physics, University of Ljubljana, Jadranska 19, 1000 Ljubljana, Slovenia \\  
\newline \noindent 
The Standard Model (SM) allows for several sources of CP violation. In the gauge sector, the relevant CP violating topological terms can be related to the determinants of corresponding fermion mass matrices ($\theta \sim \arg\det M_f$). The $SU(2)_L$ one is screened by the non-invariant vacuum and can be furthermore related to the anomalous $B+L$ currents. The analogous QCD ($\theta_{\rm QCD}$) term is controlled by the presence or absence of massless chiral quarks. In light of non-vanishing up- and down-quark Yukawas however, existing experimental bounds on the electric dipole moment of the neutron require $\theta_{\rm QCD}<10^{-10}$ signifying the so called strong CP problem. Finally, the strengths of CP violating flavor transitions are dictated by the corresponding Yukawa matrices $J \sim \Im \{ \det (  [ M_u^2, M_d^2 ]  ) \}$. As such they are suppressed by the small intergenerational mixing in the quark sector. During the past three decades they have been well established by precision flavor physics experiments confirming the CKM paradigm at the $\sim 20\%$ level. Unfortunately the size of CPV in the CKM seems to be too small to be relevant for Baryogenesis in the early Universe. The SM with massless neutrinos does not contain analogous sources of CP violation in the lepton sector. Similarly, the gauge invariant potential for a single Higgs doublet is manifestly real. 

This peculiar CP structure is not generically preserved in extensions of the SM. The general agreement with searches and measurements of CP violating phenomena thus leads to potentially very severe constraints on new CP violating sources beyond SM. In particular, for heavy new physics (NP), current experiments are already probing mass scales well beyond the direct reach of particle colliders as well as testing possible flavor structures beyond SM (c.f.~\cite{Kamenik:2014xya}).

One of key predictions of the KM mechanism of CP violation in the quark sector are the CKM matrix ($V_{ij}$) unitarity conditions, where the presence of CP violation emerges as the area of triangles defined in the complex plane, i.e.
${(V_{ud} V^*_{ub})}/{(V_{cd} V^*_{cb})} + {(V_{td} V^*_{tb})}/{(V_{cd} V^*_{cb})} + 1 = 0 \,.
$  Such relations contain an interesting interplay and correlations between the different flavor sectors, e.g. the kaon and $B$-meson systems. Current measurements exhibit  an excellent overall consistency with SM predictions, whose increasing precision has been made possible in part by the continous advances in non-perturbative QCD techniques on the lattice. Currently (and in the foreseeable future), the most precise determination of the CKM phase $\beta \equiv \arg (- V_{td} V^*_{tb} / V_{cs} V^*_{cb})$ is possible through the time dependent measurement of the CP asymmetry in decays of $B^0 \to \psi K_S^0 $, currently reaching the percent level precision. Beyond SM, these observables can be affected by NP contributions to $B^0 - \bar B^0$ mixing amplitudes. On the other hand, the dominant theory uncertainties come from s.c. penguin contributions, which are doubly Cabibbo suppressed in the SM and thus have been usually neglected in the past, also in part because they cannot be reliably estimated without additional experimental input. This in turn is provided through the measurements of the light flavor $SU(3)_F$ related $B_s \to \psi K_S^0$ and $B^0 \to \psi \pi$ decay modes~(c.f.~\cite{DeBruyn:2014oga}). A somewhat similar discussion applies to the corresponding $B_s$ meson observables measuring $\beta_s \equiv \arg (- V_{ts} V^*_{tb} / V_{cs} V^*_{cb})$ albeit with a crucial difference: the experimental precision is only now starting to reach the size of SM predictions.  In the future, such measurements will thus allow for tighter probes of NP in $B_s$ mixing away from the CP conserving limit.
Contrary to $\beta$ and $\beta_s$, the experimental determination of $\gamma \equiv \arg (- V_{ud} V^*_{ub} / V_{cd} V^*_{cb})$ can be done in charged $B^\pm \to D^0 K^\pm$ meson decays  and is under excellent theoretical control. The relevant hadronic parameters as well as possible subleading CP violating effects due to $D^0$ mixing can be extracted directly from data. The residual theory uncertainty in the SM has recently been estimated to be at the level of $\delta \gamma / \gamma \sim 10^{-7}$~\cite{Brod:2013sga}\,. These measurements are also free from sizable NP effects in most popular scenarios beyond SM and thus represent a SM standard candle of CP violation. Finally, CP violation in the kaon and charm sectors is severely suppressed in the SM due to the hierarchical structure of the CKM. This makes CP violating observables in these sectors the most sensitive probes of generic CP- and flavor violating NP. In the Minimal Supersymmetric SM for example, they allow to probe sfermion masses up to several PeV~\cite{Altmannshofer:2013lfa}\,. Thus recent indications of non-standard effects in direct CP violation in kaon~\cite{Buras:2015yba} and earlier in $D$-meson~\cite{Isidori:2011qw} decays could be the early harbingers of an emerging phenomenology of flavor- and CP violating physics beyond SM.

\newpage
\setcounter{figure}{0}
\setcounter{table}{0}

%% sessionday{thursday} 

\subsection*{\hfil Angular Analysis of $\bkllzero$  \hfil}
\label{ssec:AngularAnalysis}
\vspace*{10mm}

\underline{S.  Wehle}\vspace*{4mm} 
 \\ DESY, Deutsches Elektronen Synchrotron, Hamburg, Germany \\  
\newline \noindent 
We present a measurement of angular observables, $P_4'$, $P_5'$, $P_6'$, $P_8'$,  in the decay  $B^0 \to K^\ast(892)^0  \ell^+ \ell^-$, where $\ell^+\ell^-$ is either $e^+e^-$ or $\mu^+\mu^-$.
The analysis is performed on a data sample corresponding to an integrated luminosity of $711~\mathrm{fb}^{-1}$  containing $772\times 10^{6}$ $B\bar B$ pairs, collected at the \yfs resonance with the Belle detector at the asymmetric-energy $e^+e^-$ collider KEKB.
Four angular observables, $P_{4,5,6,8}'$
are extracted in five bins of the invariant mass squared of the lepton system, $q^2$. We compare our results  for  $P_{4,5,6,8}'$ with Standard Model (SM) predictions  including the $q^2$ region in which the LHCb collaboration reported the so-called $P_5'$ anomaly.
Values from DHMV  refer  to the soft-form-factor method of Ref. \cite{Descotes-Genon:2014uoa}, which is also used in the LHCb measurement.

\begin{figure*}[h!]
	\centering
	\subfigure[Result for $P_4'$]{
		\includegraphics[width=\factorh\textwidth]{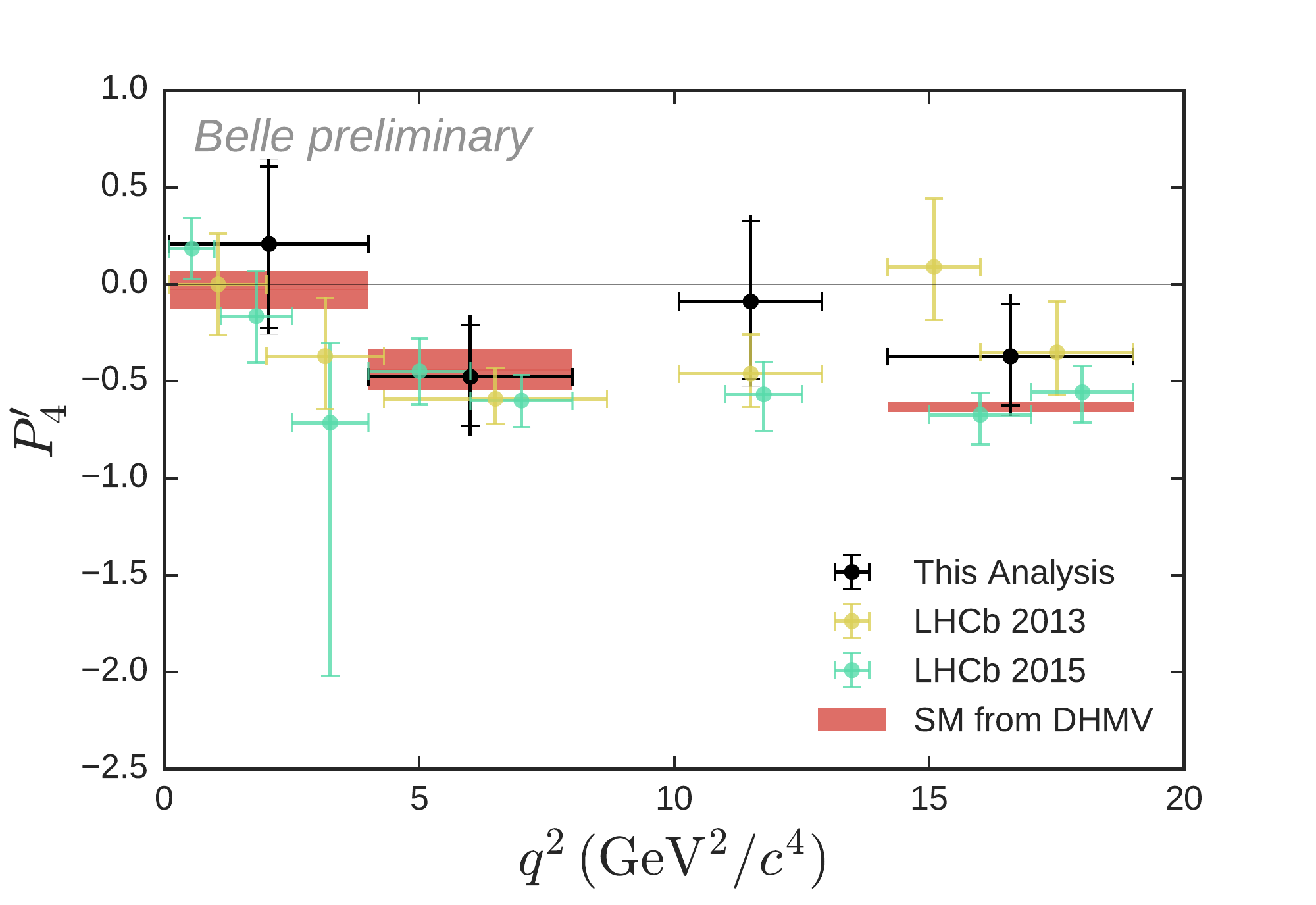}          
	}	\subfigure[Result for $P_5'$]{
		\includegraphics[width=\factorh\textwidth]{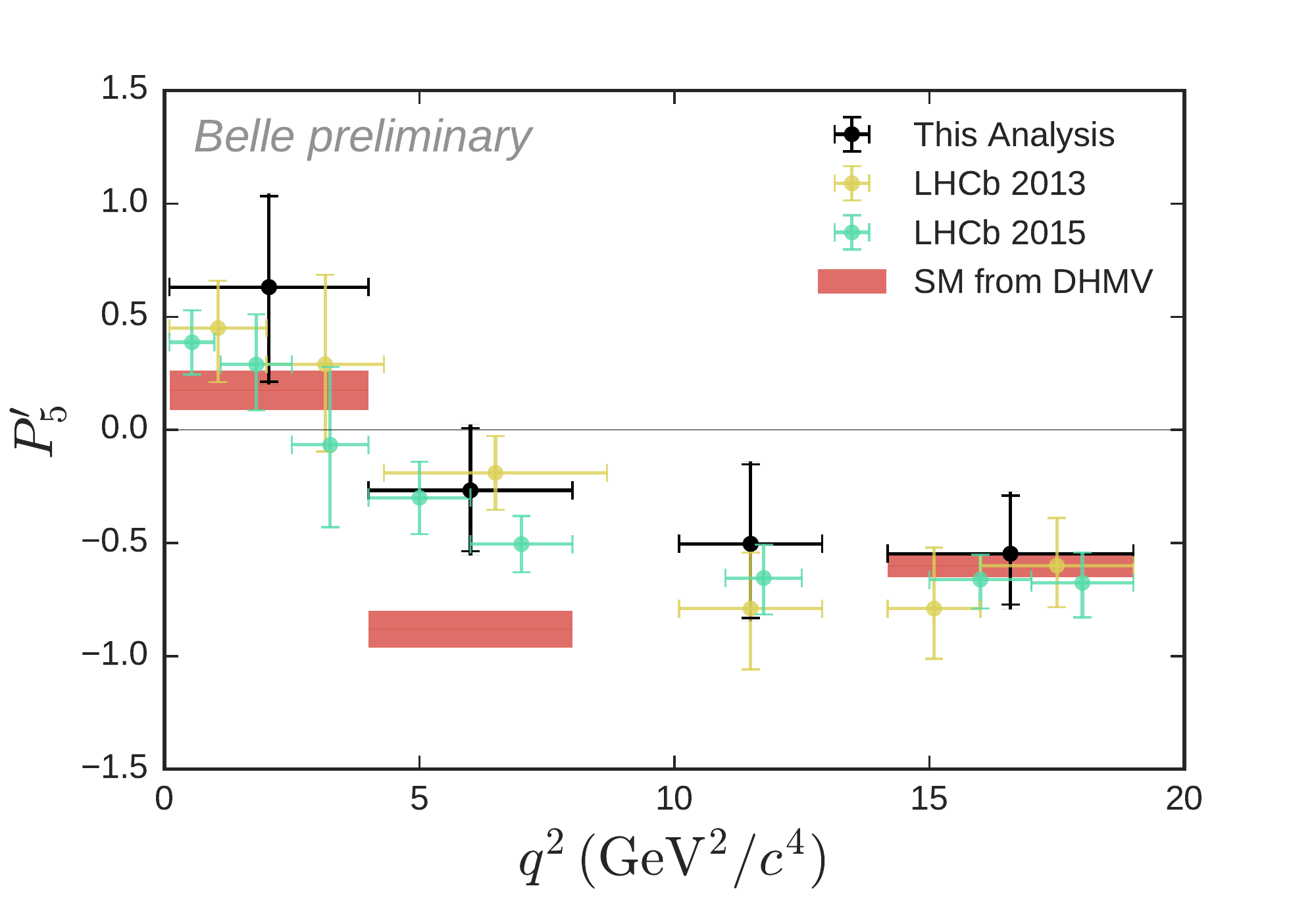}          
	}\\
	\caption{Result for the $P'_{4,5}$ observables compared to SM predictions from \cite{Descotes-Genon:2014uoa}. Results from LHCb \cite{lhcb1,lhcb2} are shown for comparison. }
	\label{fig:res}
\end{figure*}

With the combined data of both channels a full angular analysis in three dimensions in five bins of $q^2$, the di-lepton invariant mass squared, is performed.
A data transformation technique is applied to reduce the dimension of the differential decay rate from eight to three.
By this means the fit is independently sensitive to  observables $P_4'$, $P_5'$, $P_6'$ and $P_8'$, which are optimized regarding theoretical uncertainties from form-factors.
Altogether  20  measurements are performed extracting  $P_{4,5,6}'$ or $P_8'$, the \kast longitudinal polarization     $F_L$ and the transverse polarization asymmetry $A_T^{(2)}$. 

For $P_5'$ a deviation with respect to the DHMV SM prediction is observed with a significance of $2.1\sigma$  in the $q^2$ range $4.0 < q^2 < 8.0 ~\mathrm{GeV}^2/c^4$.
The discrepancy in $P_5'$ supports measurements by LHCb \cite{lhcb1,lhcb2}, where a $3.4\sigma$ deviation was observed in the same $q^2$ region.

The results of the full analysis are published in Ref. \cite{Mypaper}.

\newpage
\setcounter{figure}{0}
\setcounter{table}{0}

%% sessionday{thursday} 

\subsection*{\hfil Precision Measurements in Heavy Flavour Physics \hfil}
\label{ssec:PrecisionMeasur}
\vspace*{10mm}

\underline{T.  Gershon}\vspace*{4mm} 
 \\ Department of Physics, University of Warwick, Coventry, United Kingdom \\  
\newline \noindent 
Studies of $CP$ violation effects and rare decays of heavy flavoured particles provide exciting opportunities to look for physics beyond the Standard Model (SM).
This approach is sensitive to physics at scales higher than those that can be probed through on-shell production of new particles at the LHC.
In-depth reviews can be found, for example, in Refs.~\cite{HFAG,Blake:2015tda,Koppenburg:2015pca}; here only a very brief pointer to the most interesting relevant and recent results is given.

Within the broad area of $CP$ violation and measurements of properties of the Cabibbo-Kobayashi-Maskawa (CKM)~\cite{Cabibbo:1963yz,Kobayashi:1973fv} Unitarity Triangle:
\begin{itemize}
\item 
  The magnitude of the ratio of CKM matrix elements $\left| V^{}_{ub}/V^{}_{cb}\right|$ has been determined by LHCb using $\Lambda_b^0 \to p\mu^- \nu_\mu$ and $\Lambda_c^+ \mu^-\mu_\nu$ decays~\cite{LHCb-PAPER-2015-013}.
\item 
  The ratio $\left| V^{}_{td}/V^{}_{ts}\right|$ has been determined from the rates of $B^{0}$--$\overline{B}{}^{0}$ and $B^{0}_s$--$\overline{B}{}^{0}_s$ oscillations, $\Delta m_d$ and $\Delta m_s$.
  A new result for $\Delta m_d$~\cite{LHCb-PAPER-2015-031} significantly improves the precision.
\item 
  Several new results~\cite{LHCb-PAPER-2016-003,LHCb-PAPER-2016-006,LHCb-PAPER-2016-007,LHCb-PAPER-2015-059} have been released that help to improve the precision with which the angle $\gamma \equiv \arg\left[-V_{ud}^{}V_{ub}^*/(V_{cd}^{}V_{cb}^*)\right]$ is known.
  The combination of LHCb results sensitive to $\gamma$ gives a precision of about $8^\circ$~\cite{LHCb-CONF-2016-001}.
\item
  Charm mixing is now well-established~\cite{HFAG}, and can be used to determine hadronic parameters associated with specific decays~\cite{LHCb-PAPER-2015-057}.
  There is no evidence for $CP$ violation in the charm system~\cite{LHCb-PAPER-2015-055}.
\item
  The phases associated with $B^0$--$\overline{B}{}^0$ and $B_s^0$--$\overline{B}{}^0_s$ mixing, $2\beta$ and $-2\beta_s$ (also known as $\phi_s$) are known to good precision~\cite{LHCb-PAPER-2015-004,LHCb-PAPER-2014-059,LHCb-PAPER-2014-019,Aad:2016tdj,Khachatryan:2015nza}.
  Searches for $CP$ violation in $B_{(s)}^0$--$\overline{B}{}^0_{(s)}$ mixing~\cite{LHCb-PAPER-2014-053,LHCb-PAPER-2016-013} have not confirmed the anomalous result found by D0~\cite{Abazov:2013uma}.
\end{itemize}
All the above results are broadly consistent with the SM predictions, although there are discrepancies between inclusive and exclusive determinations of $\left|V^{}_{ub}\right|$ and $\left|V^{}_{cb}\right|$, and there is a slight tension in the global fit to the Unitarity Triangle (see, for example, Ref.~\cite{Blanke:2015wba}).

Among searches for and studies of rare decays:
\begin{itemize}
\item 
  Exciting progress is being made in the searches for the $K^+ \to \pi^+\nu\overline{\nu}$ and $K_L \to \pi^0\nu\overline{\nu}$ decays by NA62 and KOTO, respectively.  
  New results are anticipated in the next few years.
\item 
  The $B_s^0\to\mu^+\mu^-$ decay has been observed~\cite{LHCb-PAPER-2013-046,Chatrchyan:2013bka,LHCb-PAPER-2014-049,Aaboud:2016ire}.
  All of ATLAS, CMS and LHCb have potential to investigate this highly-interesting channel further in Run~II.
\item
  A full angular analysis of the $B^0 \to K^{*0}\mu^+\mu^-$ decay has been performed by LHCb~\cite{LHCB-PAPER-2015-051}.
  An interesting $3$--$4\,\sigma$ tension with the SM prediction has emerged in the observable $P_5^\prime$, though it remains to be understood if this may be due to underestimated QCD effects.
  Results from other experiments~\cite{Khachatryan:2015isa,Abdesselam:2016llu} are consistent with LHCb, though not as precise.
\item
  Measurements of the differential branching fractions of various $b \to s\mu^+\mu^-$ processes~\cite{LHCb-PAPER-2014-007,LHCb-PAPER-2015-009,LHCb-PAPER-2015-023} differ from their SM prediction, with significances up to about $3\,\sigma$, in what appears to be a consistent manner. 
\item
  Tests of lepton universality in $B^+ \to K^+\ell^+\ell^-$ ($\ell = e,\mu$)~\cite{LHCb-PAPER-2014-024} and $B \to D^{(*)} \ell \nu_\ell$~\cite{Lees:2012xj,Lees:2013uzd,LHCb-PAPER-2015-025,Huschle:2015rga,Abdesselam:2016cgx} decays deviate from their SM expectations, the latter by $4\,\sigma$~\cite{HFAG}.
\end{itemize}
These results all seem to hint at possible contributions from physics beyond the SM.
Numerous theoretical investigations have been and are being carried out, in both model-independent and model-dependent frameworks, to establish if the results may have a self-consistent explanation~\cite{JK,AC}.
More precise measurements are also necessary, and prospects are good with Run~II data-taking proceeding well~\cite{BS}, an LHCb upgrade in preparation~\cite{LHCb-PAPER-2012-031,LHCb-TDR-012,LHCb-TDR-013,LHCb-TDR-014,LHCb-TDR-015,LHCb-TDR-016} before Run~III, and Belle~II to commence operation within the next few years~\cite{FB}.

\bibliographystyle{unsrt}
\bibliography{gershon}

\newpage
\setcounter{figure}{0}
\setcounter{table}{0}

%% sessionday{thursday} 

\subsection*{\hfil A New Class of Family Non-Universal $Z^{\prime}$ models \hfil}
\label{ssec:ANewClassofFami}
\vspace*{10mm}

\underline{A.  Celis}\vspace*{4mm} 
 \\ Ludwig-Maximilians-Universit\"at M\"unchen, Fakult\"at f\"ur Physik, Arnold Sommerfeld Center for Theoretical Physics, 80333 M\"unchen, Germany \\  
\newline \noindent

Extending the scalar sector of the Standard Model (SM) by an additional complex Higgs doublet introduces scalar flavour changing neutral currents (FCNCs) at tree-level.  These are strongly constrained due to precise measurements of $\Delta F =1,2$ transitions in light-quark meson systems, implying a significant tuning in the parameter space of the model~\cite{Branco:2011iw}.  Such suppression of flavour changing effects is \textit{technically natural} when forced by a symmetry, as occurs within the natural flavour conservation paradigm~\cite{Branco:2011iw}.    The latter forbids FCNCs at tree-level by restricting the number of Higgs doublets that couple to a given type of fermion.  This condition is typically enforced via a global $\mathrm{U(1)}$ symmetry or a discrete $\mathrm{Z}_2$.

An alternative suppression mechanism for the FCNCs was proposed in~\cite{Branco:1996bq}.  In this case, specific flavour textures imposed on the Yukawa couplings due to a horizontal global symmetry provide a natural suppression of FCNCs in terms of CKM matrix elements and quark masses~\cite{Branco:1996bq}.      The possibility of implementing this mechanism via a local $\mathrm{U(1)}^{\prime}$ symmetry was explored in~\cite{Celis:2015ara}.     It was found that the suppression of FCNCs proposed in~\cite{Branco:1996bq} can be obtained from an anomaly free $\mathrm{U(1)}^{\prime}$ symmetry if the lepton sector also carries non-trivial quantum numbers under this symmetry.\footnote{A generic solution of the anomaly cancellation conditions involves non-universal charges in the lepton sector.}    Anomaly cancellation in this framework relies on a closure between the quark and lepton sector contributions, as in the SM, but in this case it also involves the three fermion generations.

The flavour structure of the theory is controlled by the CKM matrix due to the underlying Yukawa textures.     As explained in~\cite{Branco:1996bq}, different implementations of the $\mathrm{U(1)}^{\prime}$ symmetry are possible, giving rise to FCNCs in the down-quark sector or in the up-quark sector, see Fig.~\ref*{fig:inZv}.

\begin{figure}[ht!]
\vspace{0.4cm}
\begin{center}
\begin{fmffile}{ZvertexintroI} 
\parbox{10mm}{  \begin{fmfchar*}(20,15)
\fmfleft{i1} \fmfright{o1,o2}  \fmflabel{$d_i$}{o2}  \fmflabel{$\bar d_j$}{o1}
           \fmf{photon,label=$Z^{\prime}$}{i1,v1}    \fmfdot{v1}
\fmf{quark}{o1,v1,o2}  
\end{fmfchar*}  }
$~~~\;\propto\;    (V_{\mbox{\scriptsize{CKM}}})_{q_ui}^*  (V_{\mbox{\scriptsize{CKM}}})_{q_uj}  $ 
\end{fmffile}
~ \quad 
\begin{fmffile}{ZvertexintroII} 
\parbox{10mm}{  \begin{fmfchar*}(20,15)
\fmfleft{i1} \fmfright{o1,o2}  \fmflabel{$u_i$}{o2}  \fmflabel{$\bar u_j$}{o1}
           \fmf{photon,label=$Z^{\prime}$}{i1,v1}    \fmfdot{v1}
\fmf{quark}{o1,v1,o2}  
\end{fmfchar*}  }
$~~~\;\propto\;    (V_{\mbox{\scriptsize{CKM}}})_{iq_d}  (V_{\mbox{\scriptsize{CKM}}})_{jq_d}^* $ 
\end{fmffile}
\end{center}
\caption{Left:  Models with tree-level flavour changing couplings $Z^{\prime} \bar d_i  d_j $ $(i \neq j)$ in the down quark sector, there are three variants corresponding to $q_u=u,c,t$.  Right:  Models with tree-level flavour changing couplings $Z^{\prime} \bar u_i  u_j $ $(i \neq j)$ in the up quark sector, there are three variants corresponding to $q_d=d,s,b$. } \label{fig:inZv}
\end{figure}

Assuming that this horizontal symmetry is spontaneously broken around the TeV scale, one would expect observable effects in low-energy flavour transitions  due to the exchange of heavy scalars and/or the $Z^{\prime}$ boson.  In particular, a violation of lepton universality in $\Delta F =1$ transitions could be expected in these models due to the non-universal $\mathrm{U(1)^{\prime}}$ charges in the lepton sector.    In this regard, it is interesting to note that a series of deviations from the SM have been reported recently by the LHCb collaboration in $b \rightarrow s \ell^+ \ell^-$ transitions~\cite{Aaij:2013qta,Aaij:2014ora}.     Most notably, the ratio
 \begin{equation}
R_K  =  \frac{ \int^{q^2_{\mbox{\scriptsize{max}}}}_{q^2_{\mbox{\scriptsize{min}}}}      \dfrac{d\Gamma(B^+ \rightarrow K^+ \mu^+ \mu^-)}{  dq^2}  dq^2  }{    \int^{q^2_{\mbox{\scriptsize{max}}}}_{q^2_{\mbox{\scriptsize{min}}}}      \dfrac{d\Gamma(B^+ \rightarrow K^+ e^+ e^-)}{  dq^2}  dq^2  } \,,
\end{equation}
has been measured in the range $1<q^2< 6$~GeV$^2$, $R_K =0.745^{+0.090}_{-0.074}\pm0.036$, showing a $2.6\sigma$ deviation from the SM prediction $R_K^{\mbox{\scriptsize{SM}}}\simeq 1$~\cite{Aaij:2014ora,Hiller:2003js}.     The possibility to accommodate such deviations in the class of models discussed here has been discussed in~\cite{Celis:2015ara}.      Further measurements of $b \to s \ell^+ \ell^-$ transitions are clearly needed at this stage, in particular, measurements of $R_{K^*} = \Gamma(B \to K^* \mu^+ \mu^-)/\Gamma(B \to K^* e^+ e^-)$ would shed light on the possible violation of lepton universality.

\newpage
\setcounter{figure}{0}
\setcounter{table}{0}

%% sessionday{thursday} 

\subsection*{\hfil Physics at Future Colliders \hfil}
\label{ssec:PhysicsatFuture}
\vspace*{10mm}

\underline{J.  List}\vspace*{4mm} 
 \\ DESY, Notkestr.~85, 22607 Hamburg, Germany \\  
\newline \noindent 
The LHC, including its high-luminosity phase, is currently foreseen
to operate until 2037. While the LHC will still bring us a wealth of results,
the long preparation and construction times of energy-frontier colliders
require a decision for a successor project very soon, if we expect it 
be up and running by the time LHC data taking ends.
Thus, we have to ask ourselves if we can identify {\itshape already now} 
crucial questions about which we will not learn enough at the LHC,
and which experiments are suited to answer them.

As discussed in depth in this workshop, the Higgs boson discovered at the LHC
not only raises important questions concerning e.g.\ the stabilisation of its mass
and the mechanism of electroweak theory breaking, but might also be our direct portal
to new physics, in particular Dark Matter, inflation, CP violation. Therefore the
model-independent precision characterisation of this particle is of utmost importance.
While LHC offers the opportunty to determine e.g.\ coupling modifiers at the level
of about ten percent, typical extensions of the Standard Model predict deviation of at most
a few percent, depending on the scale of new physics. More importantly, hadron colliders
can only extract ratios of couplings in a model-independent way, since the total production
cross section cannot be measured without assumptions and thus there is no direct access to 
the total width of the Higgs boson. These, however, are at the heart of the physics programme
of future $e^+e^-$ colliders~\cite{ILCTDR, CLICCDR, CEPCpreCDR, FCC}, where the 
knowledge of the initial state allows a model-independent
measurement of the total Higgsstrahlung cross section by counting events where ``anything'' 
recoils against a $Z$ boson. This in turn enables the extraction of the total decay width
and thus the absolute normalisation of all couplings, and provides sensitivity to invisible
decay modes of the Higgs boson. Depending on the assumptions on running scenarios and experimental and
theoretical uncertainties, all proposed $e^+e^-$ projects reach very similar precisions of
1\% or better for most couplings. This level of precision is sufficient to probe
new physics scales significantly beyond the direct reach, covering parameter space complementary to hadron colliders.

As exceptions the top Yukawa coupling and the Higgs self-coupling need to be mentioned,
which require a center-of-mass energy of at least 500\,GeV, thus are not directly accessible at CEPC or FCC-ee. For the top Yukawa coupling, ILC~\cite{Fujii:2015jha} and CLIC~\cite{CLICCDR} reach precisions of a few percent, improving about a factor 4-5 on LHC projections.
The measurement of the Higgs self-coupling $\lambda$ is challenging at any collider since the cross sections for double Higgs production are small and, due to interference effects, have a complicated dependence on $\lambda$. In particular, the cross sections of VBF-like double Higgs processes, which dominate at hadron colliders~\cite{FCC} and at $e^+e^-$ colliders with $\sqrt{s} \ge 1$\,TeV, drop with increasing values of $\lambda$, making any measurement even more difficult.
A unique advantage of $e^+e^-$ collisions at $\sqrt{s}= 500$\,GeV is the access to a process whose cross section increases for $\lambda > \lambda_{\mathrm{SM}}$, which is e.g.\ the region required for electroweak baryogenesis.

Also in top physics, we can identify already now areas where the results still to be expected from the LHC will not be sufficient. In particular, the top mass and
the top electroweak couplings fall into this category~\cite{Vos:2016til}. For the latter, $e^+e^-$ collisions at $\sqrt{s}=500$\,GeV can improve by about two orders of magnitude over LHC projections, which will allow to probe and distinguish various compositeness
models up to scales of typically up to $20$\,TeV, in extreme cases even up to $60$\,TeV. For determining the top quark mass in a theoretically well defined way, the scan of the $t\bar{t}$ production threshold offers a unique opportunity.

Last but not least, there are also well-motivated areas in the BSM landscape where direct searches at an $e^+e^-$ collider with sufficienctly high energy offer unique opportunities for discoveries, fully complementary to the capabilities of hadron colliders.
A famous example for such cases are light, near-degenerate higgsinos as they are required in Natural SUSY models. In $e^+e^-$ collisions, natural SUSY parameter space can be probes unambiguously up to nearly $\mu=\sqrt{s}/2$, independently of all other
SUSY parameters. In such cases, the mass scale of the gauginos can be inferred from precision measurements of the higgsinos,
providing important guidance for the next generation of hadron colliders~\cite{Fujii:2015jha}.

In summary, and coming back to our original question, we can indeed already now identify a very clear and strong physics case for a next collider, namely an $e^+e^-$ collider with center-of-mass energies up to about 1\,TeV.
With the ILC currently under political consideration in Japan, there exists a technologically mature proposal for such a machine, which should be realised in a timely manner.

\newpage
\setcounter{figure}{0}
\setcounter{table}{0}

%% sessionday{thursday} 

\subsection*{\hfil New Physics in the Flavour Sector \hfil}
\label{ssec:NewPhysicsinthe}
\vspace*{10mm}

\begin{figure*}[t]
\centering
\includegraphics[width=0.8\textwidth]{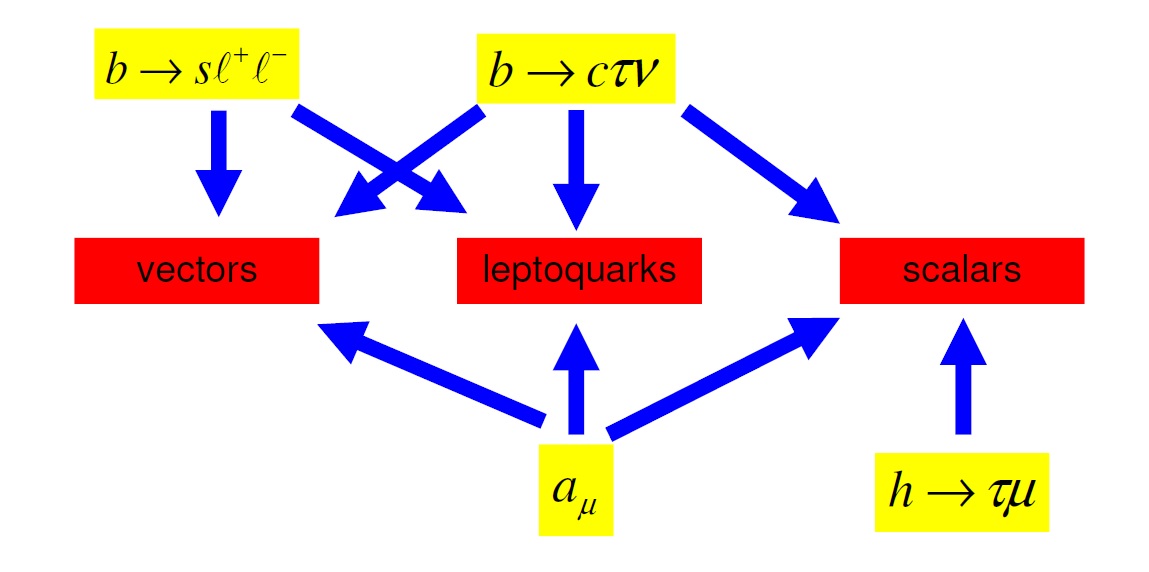}
\caption{Schematic picture of the implications for new particles from the various anomalies.\label{NPplot}}
\end{figure*}

\underline{A.  Crivellin}\vspace*{4mm} 
 \\ Paul Scherrer Institut, CH--5232 Villigen PSI, Switzerland \\  
\newline \noindent 
Several experiments observed deviations from the Standard Model (SM) in the flavour sector: LHCb found a $4-5\,\sigma$ discrepancy compared to the SM in $b\to s\mu^+\mu^-$ transitions (recently supported by an Belle analysis) and CMS reported a non-zero measurement of $h\to\mu\tau$ with a significance of $2.4\,\sigma$. Furthermore, BELLE, BABAR and LHCb founds hints for the violation of flavour universality in $B\to D^{(*)}\tau\nu$. In addition, there is the long-standing discrepancy in the anomalous magnetic moment of the muon. Interestingly, all these anomalies are related to muons and taus, while the corresponding electron channels seem to be SM like. This suggests that these deviations from the SM might be correlated. A schematic view of the additional new particles which can accommodate the anomalies is shown in Fig.~\ref*{NPplot}. 
Let us we briefly review some selected models providing simultaneous explanations:
\smallskip

{\bf Multi Higgs {\boldmath  $L_\mu-L_\tau$} model: {\boldmath $h\to\tau\mu$} and {\boldmath $b\to s\mu^+\mu^-$} \cite{Crivellin:2015mga,Crivellin:2015lwa}}
\newline
Adding to a gauged $L_\mu-L_\tau$ model with vector like quarks~\cite{Altmannshofer:2014cfa} a second Higgs doublet with $L_\mu-L_\tau$ charge 2 can naturally give an effect in $h\to\tau\mu$ via a mixing among the neutral CP-even components of the scalar doublets. In this setup a $Z'$ boson, which can explain the $b\to s\mu^+\mu^-$ anomalies, gives sizable effects in $\tau\to3\mu$ which are potentially observable at LHCb and especially at BELLE II. One can avoid the introduction of vector-like quarks by assigning horizontal charges to quarks as well~\cite{Crivellin:2015lwa}. Then, the effects in $b\to s$, $b\to d$ and $s\to d$ transitions are related in an MFV-like way by CKM elements and the $Z^\prime$ can have an observable cross section at the LHC.

{\bf \boldmath Leptoquarks: $b\to s\mu^+\mu^-$ and $b\to c\tau\nu$~\cite{Calibbi:2015kma}}
\newline 
While in $b\to c\tau\nu$ both leptoquarks and the SM contribute at tree-level, in $b\to s\mu^+\mu^-$ one compares a potential tree-level NP contribution to a loop effect. However, as $b\to c\tau\nu$ involves three times the third generation (assuming that the neutrino is of tau flavour in order to get interference with the SM contribution) but $b\to s\mu^+\mu^-$ only once. Therefore, leptoquarks with hierarchical flavour structure, i.e. predominantly coupling to the third generation, can explain simultaneously $b\to s\mu^+\mu^-$ and $b\to c\tau\nu$ in case of a $C_9=-C_{10}$ (left-handed quark and lepton current) solution for $b\to s\mu^+\mu^-$. In this case one predicts sizable effects in $B\to K^{(*)}\tau\tau$, $B_s\to\tau^+\tau^-$ and $B_s\to\mu^+\mu^-$ below the SM.

{\bf\boldmath 2HDM X: $a_\mu$ and $b\to c\tau\nu$ \cite{Crivellin:2015hha}}
\newline
In a 2HDM of type X, the couplings of the additional Higgses to charged leptons are enhanced by $\tan\beta$. As, unlike for the 2HDM II, this enhancement is not present for quarks, the direct LHC bounds on $H^0,A^0\to\tau^+\tau^-$ are not very stringent and also $b\to s\gamma$ poses quite weak constraints. Therefore, the additional Higgses can be light which, together with the $\tan\beta$ enhanced couplings to muons, allows for an explanation of $a_\mu$. If one adds couplings of the lepton-Higgs-doublet to third generation quarks, one can explain $b\to c\tau\nu$ as well by a charged Higgs exchange. In case of a simultaneous explanation of $a_\mu$ and $b\to c\tau\nu$ (without violating bounds from $\tau\to\mu\nu\nu$) within this model, sizable branching ratios (reaching even the \% level) for $t\to Hc$, with $m_H\approx 50-100\,$GeV and decaying mainly to $\tau\tau$, are predicted.

{\bf\boldmath $L_\mu-L_\tau$ flavon model: $a_\mu$, $h\to\tau\mu$ and $b\to s\mu^+\mu^-$ \cite{Altmannshofer:2016oaq}}
\newline
In this model one adds vector-like leptons to the gauged $L_\mu-L_\tau$ model of Ref.~\cite{Altmannshofer:2014cfa} and one can explain $h\to\tau\mu$ via a mixing of the flavon (the scalar which breaks $L_\mu-L_\tau$) with the SM Higgs. Furthermore, one can account for $a_\mu$ by loops involving the flavon and vector-like leptons without violating the $\tau\to\mu\gamma$ bounds as this decay is protected by the $L_\mu-L_\tau$ symmetry. Despite the effects already present in the model of Ref.~\cite{Crivellin:2015mga}, one expects order one effects in $h\to\mu^+\mu^-$ detectable with the high luminosity LHC.
\smallskip

In conclusion, the models presented here explain different combinations of the anomalies and they predict different signatures in other observables, making them distinguishable and testable with future measurements.

\newpage
\setcounter{figure}{0}
\setcounter{table}{0}

\def\ca{\cos\alpha}
\def\cb{\cos\beta}
\def\sa{\sin\alpha}
\def\sb{\sin\beta}

%% sessionday{thursday} 

\subsection*{\hfil LHC Probe of Leptophilic 2HDM for Muon g-2 \hfil}
\label{ssec:LHCProbeofLepto}
\vspace*{10mm}

\date{}

\underline{E.  Chun}\vspace*{4mm} 
 \\ Korea Institute for Advanced Study, Seoul 02455, Korea \\  
\newline \noindent 
Among four types of two-Higgs-doublet models (2HDMs) with natural minimal flavor violation, 
the type X (lepton-specific) model can only accommodate the observed $3\sigma$ deviation of the muon g-2,
$\delta a_\mu \equiv a^{\rm Exp}_\mu - a^{\rm SM}_\mu = + 262 (85) \times 10^{-11}$,
through two-loop Barr-Zee diagrams in the parameter region of large $\tan\beta$  and a light CP-odd scalar $A$ \cite{broggio1409}. The lepton (quark) Yukawa couplings of the extra Higgs bosons in the type-X 2HDM are proportional to $\tan\beta$ ($\cot\beta$) and thus they become leptophilic (hadrophobic). 

While various constraints from hadronic observables can be easily evaded,  
leptonic precision observables such as lepton universality test in the neutral and charged currents limit stronlgy the allowed parameter space \cite{wang1412,cao0909,abe1504}.
Treating all the lepton universality data in a consistent way, we show in Figure 1 the $1\sigma$ and $2\sigma$ regions allowed by the muon g-2, lepton universality data in $Z$ and $\tau$ decays for given degenerate masses of heavy Higgs bosons $H$ and $H^\pm$, respectively \cite{chun1605}. 
While no overlapping region is found at $1\sigma$, 
sizable parameter space is still viable at $2\sigma$ for $H/H^\pm$ masses at around 200$\sim$400 GeV.
\begin{figure}[!ht]
\centering
\subfigure{\includegraphics[width=0.32\textwidth]{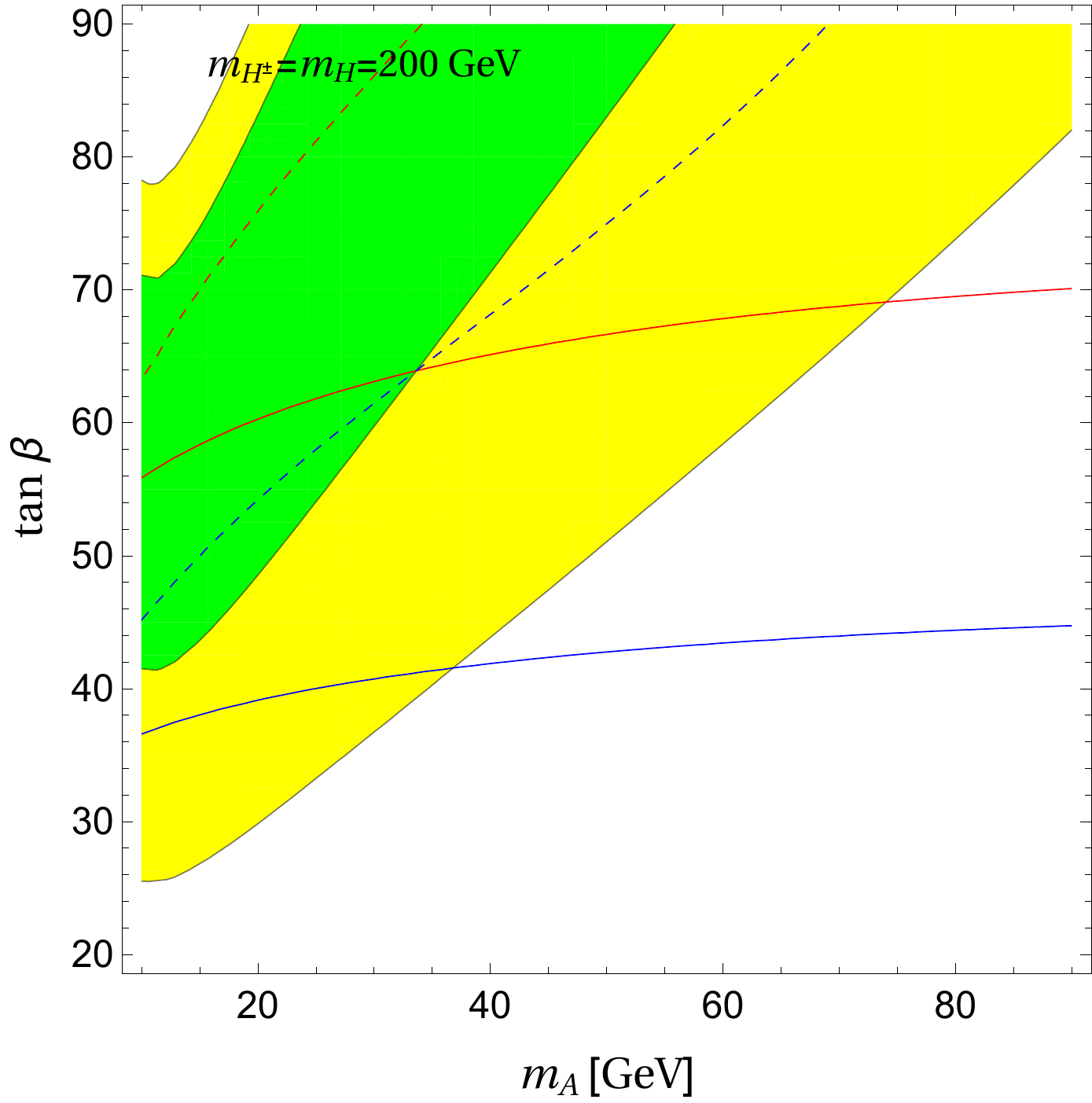}}
\subfigure{\includegraphics[width=0.32\textwidth]{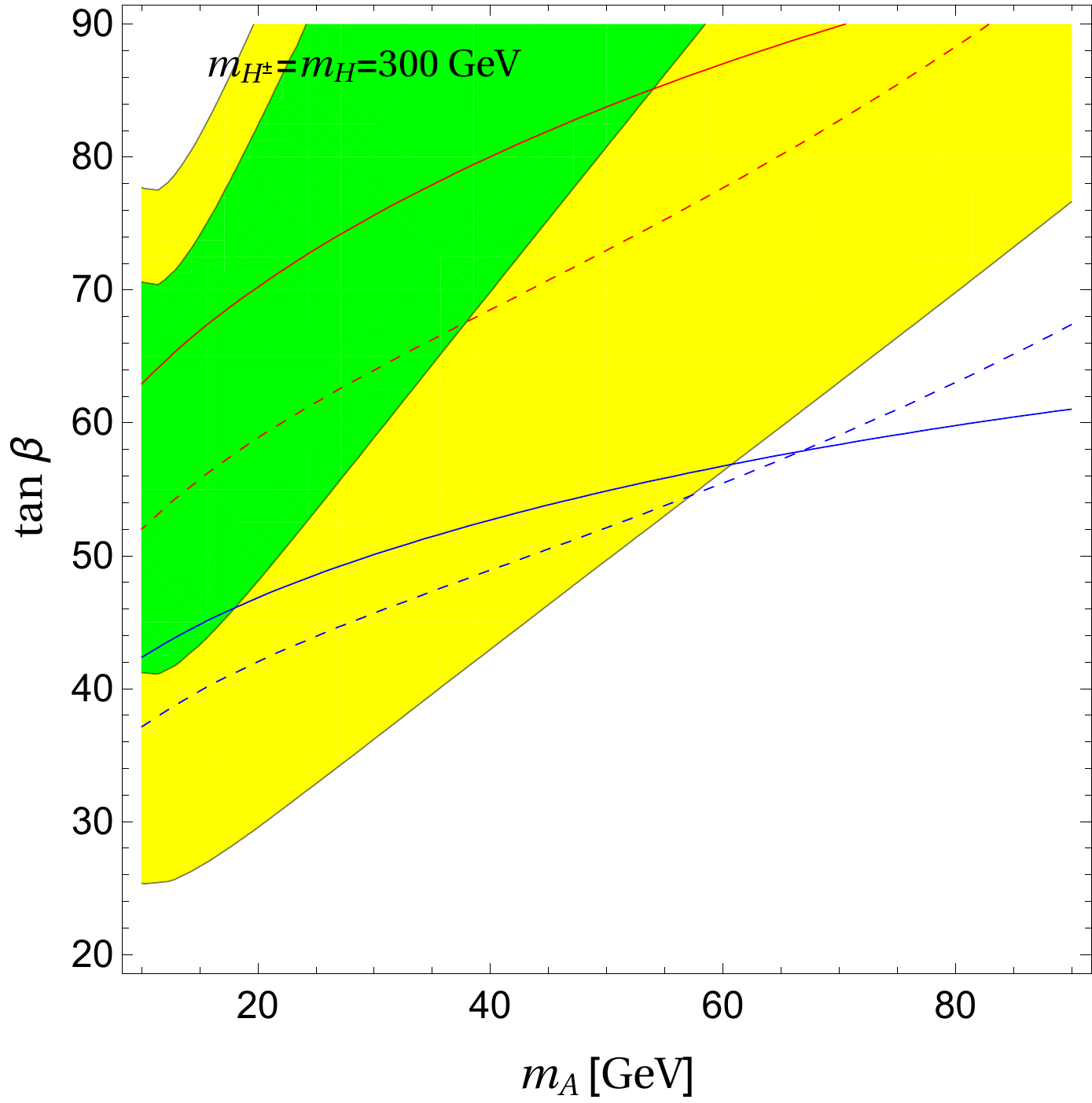}}
\subfigure{\includegraphics[width=0.32\textwidth]{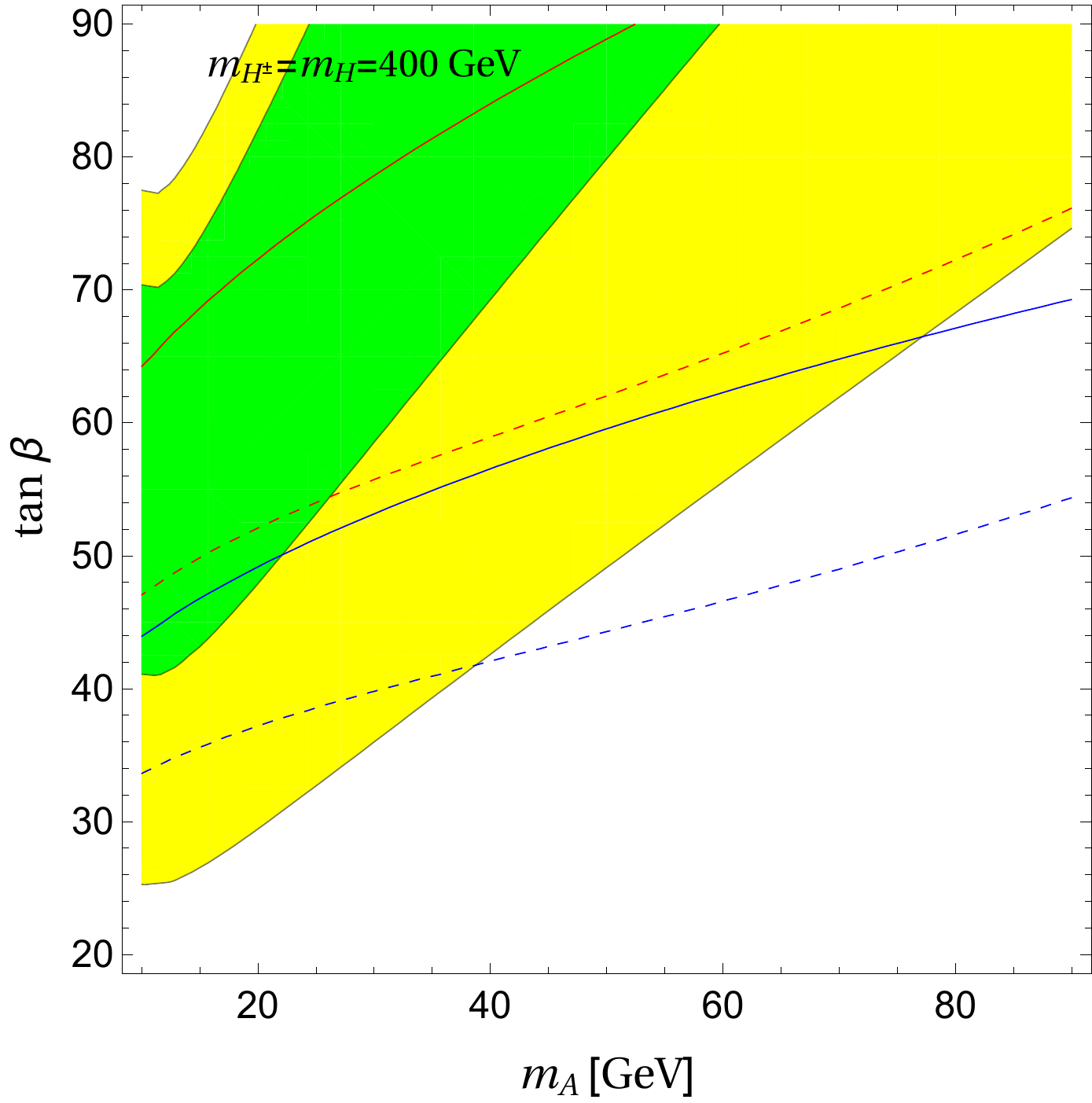}}
\vspace{-3mm}
\caption{The regions allowed (at $1\sigma$ and $2\sigma$) by the muon g-2 
(green (inner) and yellow (outer) shaded areas); by the lepton universality test in $Z$ decays (below the blue (lower) and red (upper) dotted lines); and by the lepton universality test with $\tau$ decays  (below the blue (lower) and red (upper) solid lines).
}
\label{fig:loops}
\end{figure}
It will be an interesting task to search for tau-rich signatures testing the scenario with a light $A$ and heavier $H/ H^\pm$ in the next run of the LHC \cite{chun1507}. The LHC8 bound and LHC14 perspective  are 
presented in Figure 2. For the detailed LHC study, six benchmark points are selected as in Table 1. 
It shows that high-luminosity LHC is needed to probe the favored region (or points A,B,D) as it
leads to soft or boosted $\tau$ pairs.
\begin{figure}[!ht]
\centering
\subfigure{\includegraphics[width=0.4\textwidth]{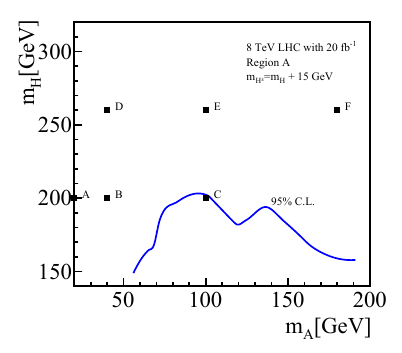}}
\subfigure{\includegraphics[width=0.4\textwidth]{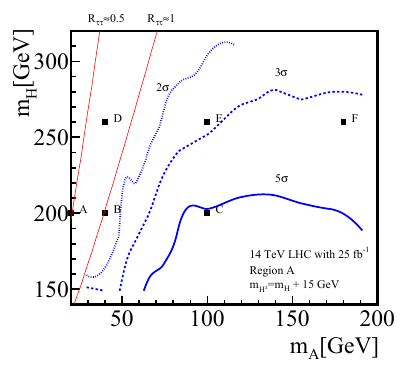}}
\vspace{-4mm}
\caption{The 95\% exclusion contour from LHC8  (left) and the discovery reach countours at LHC14 (right) in the $m_A$--$m_H$ plane. }
\label{fig:lhc-regionA}
\end{figure}
\begin{table}[h!]
\begin{tabular}{l|rrrrrr}
 & point A & point B & point C & point D & point E & point F \cr
\hline
$m_A$~[GeV] & 20 & 40 & 100 & 40 & 100 & 180  \cr
$m_H$~[GeV] & 200 & 200 & 200 & 260 & 260 & 260  \cr
\hline
\hline
total $\sigma_{\rm gen}$ [fb] & 270.980 & 241.830 & 153.580 & 100.430 & 71.271 & 44.163\cr
\hline
$n_{\ell} \ge 3$  & 6.606 & 16.681 & 21.713 & 7.110 & 11.962 & 8.822\cr
$n_{\tau} \ge 3$  & 0.894 & 2.602& 4.386 & 0.888 & 2.346 & 1.971\cr
$E\!\!\!/_T > 100$~GeV    & 0.201 & 0.547 & 1.179 & 0.209 & 0.765 & 0.926\cr
$n_b=n_{j}=0$& 0.098 & 0.314  & 0.857 & 0.121  & 0.479 & 0.631\cr
\hline
\hline
$S/B$ & 0.1  & 0.5 & 1.2 & 0.2 &0.7 & 0.9 \cr
$S/\sqrt{B}_{25{\rm fb}^{-1}}$ & 0.6 & 1.9 & 5.2 & 0.7 & 2.9 & 3.8\cr
\hline
\end{tabular}
\vspace{-2mm}
\caption{The number of events after applying successive cuts for 14 TeV LHC. }
\label{tab:selectioncut2}
\end{table}

\begin{table}[h!]
\begin{center}
\begin{tabular}{|ccc|ccc|ccc|}
\hline
$y_u^A$ & $y_d^A$ & $y_l^A$  & $y_u^H$ & $y_d^H$  & $y_l^H$ & $y_u^h$ & $y_d^h$ & $y_l^h$\\
\hline
$\cot\beta$ & $-\cot\beta$ & $\tan\beta$ & $\frac{\sa}{\sb}$ & $\frac{\sa}{\sb}$ & $\frac{\ca}{\cb}$ & $\frac{\ca}{\sb}$ & $\frac{\ca}{\sb}$ & $\frac{\sa}{\cb}$\\
\hline
\end{tabular}
\caption{Yukawa couplings of neutral Higgs bosons in the type-X 2HDM}
\end{center}
\end{table}

\begin{figure}[!ht]
\centering
\includegraphics[width=1.0\textwidth]{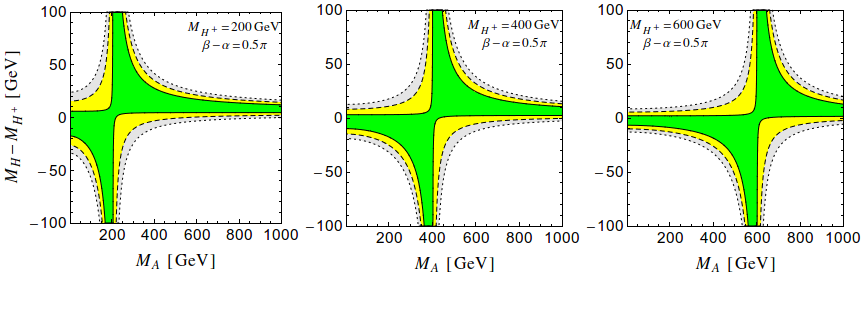}
\caption{The parameter space allowed in the $M_A$ vs.\ $\Delta M_H =M_H-M_{H^\pm}$ plane by the electroweak precision constraints. The green, yellow, gray regions satisfy $\Delta \chi_{\mbox{$\scriptscriptstyle{\rm EW}$}}^2 (M_A, \Delta M)< 2.3, 6.2, 11.8$, corresponding to 68.3, 95.4, and 99.7\% confidence intervals, respectively.}
\label{fig:ewpc}
\end{figure}

\newpage
\setcounter{figure}{0}
\setcounter{table}{0}

%% sessionday{friday} 

\subsection*{\hfil The Rise of Effective Lagrangians at the LHC \hfil}
\label{ssec:TheRiseofEffect}
\vspace*{10mm}

\underline{T.  Plehn}\vspace*{4mm} 
 \\ Institut f\"ur Theoretische Physik, Universit\"at Heidelberg, Germany \\  
\newline \noindent 
During Run~I of the LHC the field of particle theory has entered a
data-driven era. While it remains interesting and relevant to ask what
kind of ultraviolet completions of the Standard Model the LHC
experiments (and related other experiments) could be searching for, we
need to find ways to describe, communicate, and exploit experimental
results in a theoretically well-defined and accessible fashion. Three
distinct approaches have been established to date:
\begin{enumerate}
\item effective Lagrangians for a bottom-up and systematic description
  of LHC observables, obviously including a cutoff or matching scale
  towards the ultraviolet UV;
\item simplified models representing certain theoretically or
  experimentally relevant features without attempting to be be general
  or UV-complete. Consistency questions are secondary
  in their construction;
\item full models completing the Standard Model in the ultraviolet,
  largely motivated by theory and links to other
  fields, like cosmology.
\end{enumerate}
Two aspects are crucial: first, the
theoretical predictions have to be reproducible. An interpretation
framework which is not available as part of standard Monte Carlo codes
is useless. Second, by construction each of these 
approaches has limitations; rather than requiring from
an approach what it is not constructed for, we need to choose
the approach which suits a given question best.\medskip

\begin{figure}[h!]
\begin{center}
\includegraphics[width=0.8\textwidth]{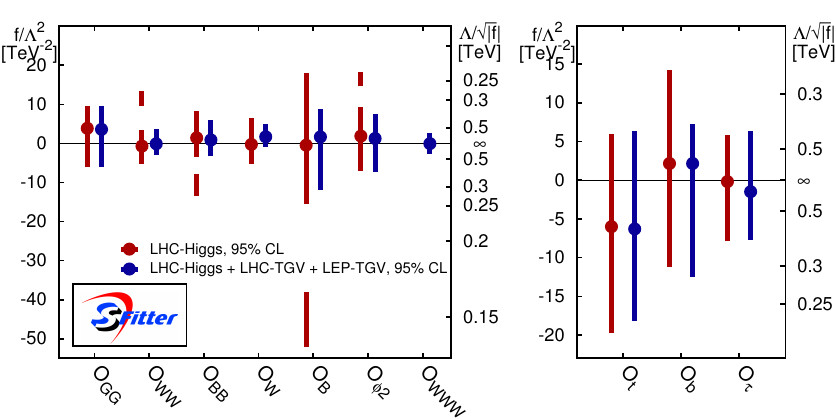}
\end{center}
\vspace*{-6mm}
\caption{Allowed 95\% CL ranges for individual Wilson coefficients
  $f_x/\Lambda^2$. We show results from Higgs observables only (red)
  and after adding di-boson measurements (blue). Figure from
  Ref.~\cite{higgs}.}
\label{fig:higgs}
\end{figure}

In this brief writeup I will discuss effective Lagrangians as a way to
link LHC measurements to theory predictions and to quantify the
agreement between data and Standard Model predictions. As a personal
side remark, I will not use this approach to study an established new
physics signal. In that lucky situation simplified models or even full
models should be well-suited to understand the origin of such a
signal. One application of an effective, dimension-6 Lagrangian to LHC
physics is the \underline{electroweak Higgs gauge sector}. One basis of operators
we can use to interpret Higgs signatures as well as weak boson pair
production is
\begin{alignat}{9}
\mathcal{O}_{WW} &= \phi^{\dagger} \hat{W}_{\mu \nu} \hat{W}^{\mu \nu} \phi  \qquad \qquad \qquad 
&\mathcal{O}_{BB} &= \phi^{\dagger} \hat{B}_{\mu \nu} \hat{B}^{\mu \nu} \phi \notag \\
\mathcal{O}_W &= (D_{\mu} \phi)^{\dagger}  \hat{W}^{\mu \nu}  (D_{\nu} \phi)
& \mathcal{O}_B &=  (D_{\mu} \phi)^{\dagger}  \hat{B}^{\mu \nu}  (D_{\nu} \phi) \notag \\
\mathcal{O}_{GG} &= \phi^\dagger \phi \; G_{\mu\nu}^a G^{a\mu\nu}  \qquad \qquad 
&\mathcal{O}_{\phi,2} &= \frac{1}{2} \partial^\mu\left ( \phi^\dagger \phi \right)
                            \partial_\mu\left ( \phi^\dagger \phi \right) \notag \\
\mathcal{O}_{WWW} &= \text{Tr} \left( \hat{W}_{\mu \nu} \hat{W}^{\nu \rho} 
\hat{W}_\rho^\mu \right)  \; .
\label{eq:operators}  
\end{alignat}
The Higgs-fermion
couplings are limited to shifted Yukawas. In Fig.~\ref*{fig:higgs} we
show the Run~I limits on the individual Wilson coefficients.
Some of them
correspond to the usual $\kappa$ or $\Delta$
modifications of Higgs couplings, others represent new Lorentz
structures leading to a modified kinematics. Unlike for the usual
Higgs couplings analysis we can assign a typical new physics energy
scale to these limits. For example assuming a reasonably weakly
interacting theory with $f_j = 1$ the Higgs sector measurements are
sensitive to scales $\Lambda = 300~...~500$~GeV. Given what we know
from other LHC searches for new physics it might not be surprising
that at this level we see no significant deviation from the Standard
Model in the Higgs operators. In the limits for example on $\mathcal{O}_B$
and $\mathcal{O}_W$ we observe is a significant improvement from adding
di-boson channels. This is because this new set of observables removes
strong non-Gaussian correlations for example in $\mathcal{O}_B$ vs $\mathcal{O}_W$
from the Higgs sector alone.\medskip

\begin{figure}[h!]
\begin{center}
\includegraphics[width=0.5\textwidth]{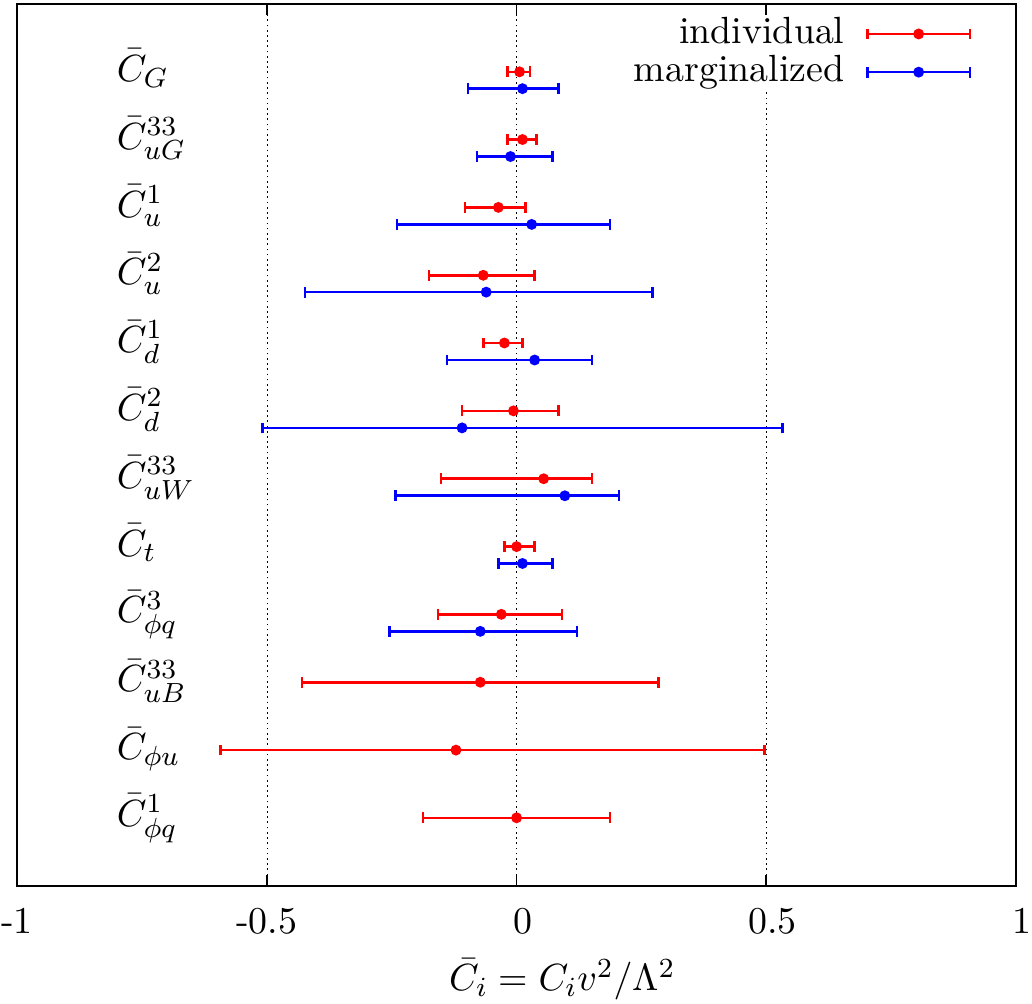}
\end{center}
\vspace*{-6mm}
\caption{Allowed 95\% CL ranges for individual Wilson coefficients
  $C_i v^2/\Lambda^2$, both for individual operators (red) and from a
  multi-dimensional analysis (blue). Figure from Ref.~\cite{top}.}
\label{fig:top}
\end{figure}

A second application of effective Lagrangians is the \underline{top sector},
combining measurements from top pair production, single top
production, and top-associated production. The dimension-6 operator
basis now includes operators of the kind
\begin{alignat}{9}
\mathcal{O}_{qq} &= \left( \bar{q} \gamma_\mu q \right) \left( \bar{q} \gamma^\mu q \right) \qquad \qquad 
&\mathcal{O}_{uG} &= \left( \bar{q} \sigma^{\mu \nu} T^A u \right) \tilde{\phi} G^A_{\mu \nu} \notag \\
\mathcal{O}_{GGG} &= f_{ABC} G_\mu^{A \nu} G_\nu^{B \lambda} G_\lambda^{C \mu}
\qquad \cdots
\end{alignat}
As for the Higgs-gauge sector, these operators describe total rates as
well as kinematic features. The last operator in the above list is a
purely gluonic Yang-Mills operator which is suspected to be best
constrained in top pair production --- a proven Tevatron assumption
which might or might not be true at the LHC. Typical limits in the
top sector, shown in Fig.~\ref*{fig:top}, also probe energy scales
around 500~GeV, similar to the Higgs-gauge sector. Obviously,
resonance searches for example in $t\bar{t}$ production have a much
larger mass reach, reflecting the fact that $s$-channel resonance
searches at the LHC are much more powerful than searches for new
particles in the $t$-channel.\medskip

\begin{figure}[h!]
\begin{center}
\includegraphics[width=0.7\textwidth]{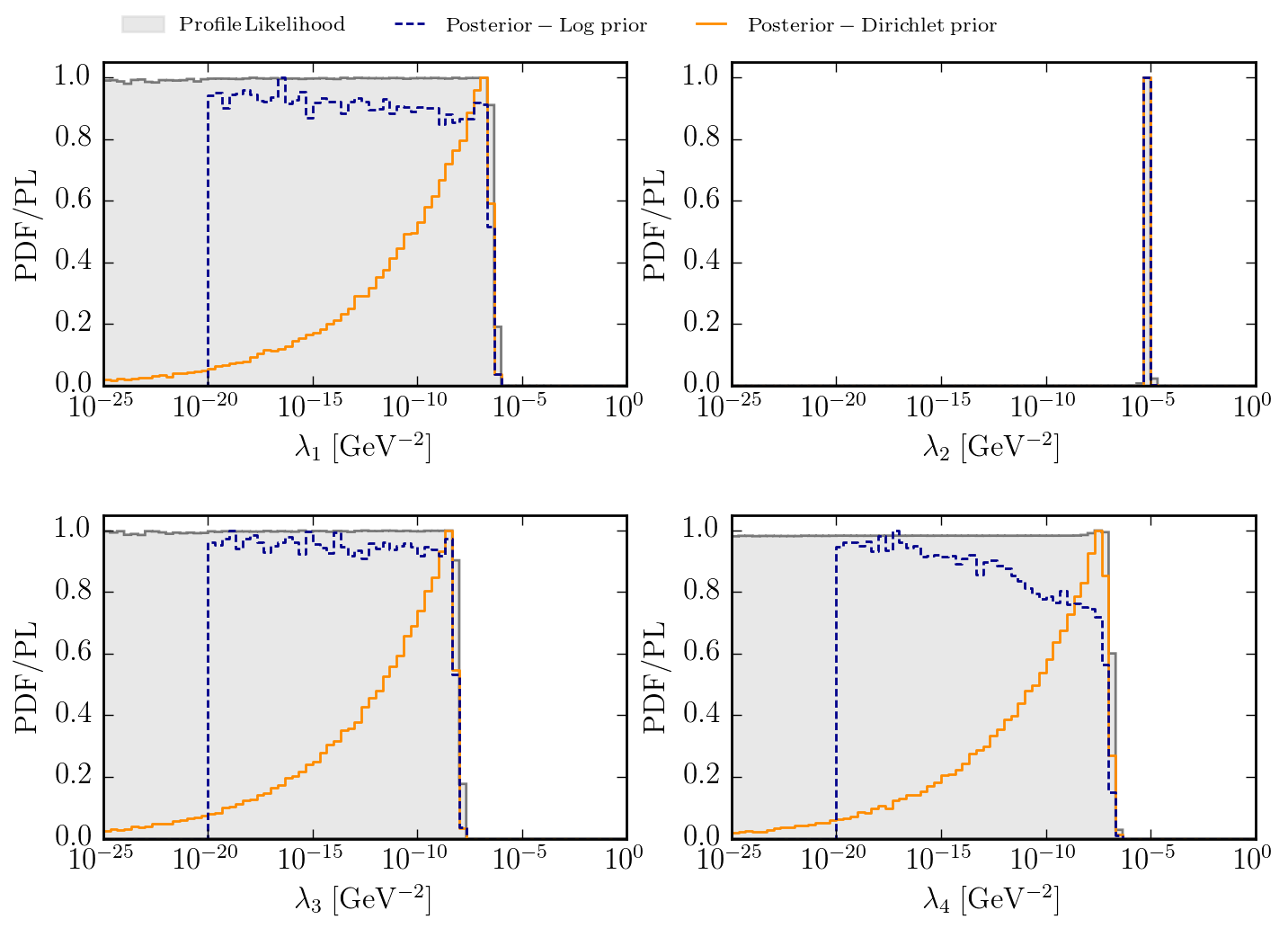}
\end{center}
\vspace*{-6mm}
\caption{One-dimensional posterior distributions of individual Wilson
  coefficients $\lambda_i$ for different priors. Figure from
  Ref.~\cite{dm}.}
\label{fig:dm}
\end{figure}

Finally, we can describe searches for \underline{dark matter} based on
the observed relic density, direct searches, indirect searches, and
even LHC searches in terms of effective operators.  The difference to
the above cases is that now we have to assume the field properties of
the dark matter agent (and of the mediator, if it is too light to be
integrated out). For a real scalar dark matter particle $\chi$ the
relevant operators include
\begin{align}
\mathcal{O}_1 &= m_q \chi^2 \bar{q} q \qquad \qquad 
&\mathcal{O}_2 &= m_q  \chi^2 \bar{q} \gamma^5 q \notag \\
\mathcal{O}_3 &= \chi^2 G_{\mu\nu}G^{\mu\nu} 
&\mathcal{O}_4 &= \chi^2 G_{\mu\nu}\tilde{G}^{\mu\nu} \; .
\end{align}
In Fig.~\ref*{fig:dm} we show the constraints on these operators from
the current dark matter data, including the Hooperon galactic center
excess from Fermi. In the absence of the galactic center excess the
only conclusive measurement is the relic density, strongly correlating
one of these couplings with the dark matter mass. For the profile
likelihood this means that none of the Wilson coefficients will show a
peak. Adding the galactic center excess resolves this degeneracy and
allows for a proper measurement of $\mathcal{O}_2$ and the dark matter mass
$m_\chi \sim 50$~GeV.\medskip

In summary, we see that analyses in different fields of LHC physics
are successfully described in terms of effective Lagrangians. On the
other hand, the fact that these Lagrangians are a powerful means of
communication between theory and experiment does not mean that they
are guaranteed to represent a rapidly converging effective field
theory. The benefits and limitations of this approach is an open
question which does not benefit from matter-of-principle
statements. It requires careful analyses for individual classes of
observables as well as for individual classes of models. Possible
shortcomings can be solved by moving to simplified models, but at the
expense of many more degrees of freedom in setting up the
interpretation framework and on their own limitations. I would very
much like to thank the organizers of the LHCSki2016 conference for
giving us the opportunity of discussing these issues at the level of
detail which they deserve.

\newpage
\setcounter{figure}{0}
\setcounter{table}{0}

%% sessionday{friday} 

\subsection*{\hfil Towards the Next Standard Model - Experimental Challenges \hfil}
\label{ssec:TowardstheNextS}
\vspace*{10mm}

\underline{C.  Kiesling}\vspace*{4mm} 
 \\ Max-Planck-Institute for Physics, and Ludwig-Maximilians-Universit\"at, Munich, Germany \\  
\newline \noindent 
Despite the overwhelming success of the Standard Model of strong, weak and electromagnetic interactions, culminating in the recent discovery\cite{HiggsDisc} of the last missing particle, the Higgs boson, there are inescapable arguments for demanding new concepts in particle physics beyond the present wisdom. Among those arguments one may find theoretical reasoning, such as the hierarchy problem related to the vacuum stability of the Higgs boson mass. Some others are purely experimental, such as the observed non-zero masses of the neutrinos, or the unexplained disappearance of the antimatter which was necessarily generated together with the matter we observe today in the Universe. 

With their ingenious quark mixing scheme, requiring three generations of quark doublets (i.e.~six quarks where only three of them where known at that time) Kobayashi and Maskawa\cite{KobMas} were able to quantitatively describe the matter-antimatter asymmetry (``CP violation'') observed in the neutral kaon system. The physical reality of quark mixing according to the ``Cabibbo-Kobayashi-Maskawa'' (CKM) matrix with its three mixing angles and one complex phase was finally proven in a beautiful series of experiments\cite{BaBarBelle} with quantum-entangled $B^0-\bar{B^0}$-pairs at the previous B-factories at SLAC (USA) and KEK (Japan). Kobayashi and Maskawa had indeed found a mechanism to allow for CP violation within the SM via the weak interaction, which could, at least in principle, provide an explanation for the baryon asymmetry in the Universe. However, the observed astrophysical nucleon-to-photon ratio, which in a direct way yields the excess of matter over antimatter, is by many orders of magnitude larger than expected in the SM with the Kobayashi-Maskawa scheme. 

Another very severe flaw of the SM is the lack of understanding how the energy density in the Universe comes about. Astrophysical observations lead to the conclusion that only about four percent of the total energy density in the Universe, characterized by ``luminous'' baryonic matter, is described by the SM, the rest being attributed to Dark Matter (DM, more than five times the baryonic matter), and, most mysteriously, to Dark Energy (DE), which takes the lion's share of about 70 percent. While DM is needed to explain the star rotation curves within the galaxies and the dynamics of galaxy clusters, the DE is responsible for the (observed) accelerated expansion of the Universe.   

What could this ``New Physics'' (NP) be? There is a great wealth of theoretical models on the market, just to mention Supersymmetry\cite{susy} as one example, which is a good candidate to cure the vacuum instability of the Higgs mentioned above, and to provide in addition some prospect of unification of all the fundamental forces in Nature, although only at very high energy. Despite the fact that none of these models provides conclusive guidance to specific experimental searches they all have in common that the NP is most likely to appear at energy scales barely accessible at present. In any case, the search for NP requires broad experimental strategies, with new hardware tools and data analysis methods. Especially in the field of accelerator-based research, new experimental challenges are visible on the horizon on the way to the ``Next Standard Model''.

Throughout the history of particle physics research two important guiding principles were the keys to success: On the one hand it is the push for ever higher collision energies (``high energy frontier''), which were enabling the discoveries of, for example, the heavy weak bosons $W^\pm$ and $Z^0$, the top quark or the Higgs particle. On the other hand the drive for ever higher precision opened up the window to search for small deviations from the ``standard'' expectation, an early example being the discovery of CP violation in the neutral kaon system. In this type of experimental physics (``precision frontier'', sometimes also called ``intensity frontier'') one is taking advantage of the fact that quantum (loop) contributions from NP propagators will alter the process amplitudes and thus the physical observables. Such contributions are not limited in the NP mass scale. One has, however, to observe that only the ratio of coupling constant over mass can be measured and theory is needed in order to get an estimate of the new mass scale. But this has been exactly the way the SM was uncovered.          

Clearly, the LHC is the paramount representative of the former class of search directions. However, we have to conclude that at present there is no significant evidence of any new scale in the reach of the LHC. Strong efforts are launched to increase the luminosity of the LHC within the coming years. But signals of direct production of new particles, if they exist, should show up nevertheless within the next two years of runs at 13 TeV center of mass energy.       

Given the not unlikely possibility that Nature has prepared for us energy scales for NP way beyond the reach of the present (or even future) accelerator generations, the only way to make progress is to search for small deviations in high statistics experiments. The new SuperKEKB B-factory\cite{skekb} (CP violation studies, rare $B$ and $\tau$  decays) or the International Linear Collider ILC\cite{ilc} (Higgs Factory via the Higgs-strahlung process) are such machines. Technologies are either mature, such as ultra-thin pixel sensors for the Belle~II detector at SuperKEKB, or under development for the ILC, such as high granularity calorimeters for particle flow analyses. Already within the coming two years the SuperKEKB collider will provide luminosity and after a few years outperform the previous KEKB collider by almost two orders of magnitude in instantaneous luminosity. An exciting period of physics with high discovery potential is ahead of us.

\newpage

\end{document}